\DocumentMetadata{}
\documentclass[nonacm]{acmart}

\setcopyright{none}
\renewcommand\footnotetextcopyrightpermission[1]{} 
 
\usepackage{listings}
\usepackage{xcolor}

\lstdefinestyle{prompt}{
  backgroundcolor=\color{gray!10}, 
  basicstyle=\ttfamily\footnotesize, 
  frame=single,
  breaklines=true
}
\usepackage{tikz} 
\usetikzlibrary{arrows.meta, positioning}
\usepackage{threeparttable}
\usepackage{tabularx}
\usepackage{graphicx}
\usepackage{tabularray}
\usepackage{makecell}
\usepackage{soulutf8} 
\usepackage[normalem]{ulem} 
\usepackage{adjustbox}
\usepackage{multirow}
\usepackage{array}
\usepackage{titlesec} 
\usepackage{caption}
\usepackage{cleveref}
\usepackage{underscore}
\usepackage[english]{babel}
\usepackage{subcaption}
\usepackage{multirow}
\usepackage{float}
\usepackage{blindtext}
\usepackage{amsmath}
\usepackage{amsfonts}
\usepackage{booktabs}

\usepackage{colortbl}
\usepackage{booktabs}
\newcolumntype{P}[1]{>{\centering\arraybackslash}p{#1}}

\def\RQone{What factors contribute to delays throughout the lifecycles of the PCVEs?}
\def\RQtwo{To what extent do the SOTA vulnerability detection methods demonstrate efficacy in identifying PCVEs?}
\def\RQthree{How can the SOTA be improved to identify PCVEs that currently remain undetected?}

\def\projectname{\emph{DeeptraVul}}
\def\tintroBOLD{$\boldsymbol{\mathsf{T_{Intro}}}$}
\def\treportBOLD{$\boldsymbol{\mathsf{T_{Report}}}$}
\def\tpatchBOLD{$\boldsymbol{\mathsf{T_{Patch}}}$}
\def\tdiscloseBOLD{$\boldsymbol{\mathsf{T_{Disclose}}}$}

\def\treport{$\mathsf{T_{Report}}$}
\def\tpatch{$\mathsf{T_{Patch}}$}
\def\tdisclose{$\mathsf{T_{Disclose}}$}
\def\tearliest{$\mathsf{T_{Earliest}}$}

\usepackage{enumitem}
\setitemize{noitemsep,topsep=0pt,parsep=0pt,partopsep=0pt,leftmargin=10pt}

\usepackage{tcolorbox}
\tcbset{mybox/.style={colback=white, colframe=blue, left=1mm, right=1mm, 
  fonttitle=\bfseries}, fontupper=\small,
  before upper=\setlength{\parindent}{1em}\everypar{{\setbox0\lastbox}\everypar{}},
}

\newcommand\figuretimeline[0]{
\begin{wrapfigure}{r}{0.3\textwidth}
  \begin{center}
    \includegraphics[trim=0cm 5cm 0cm 9cm, width=0.3\textwidth]{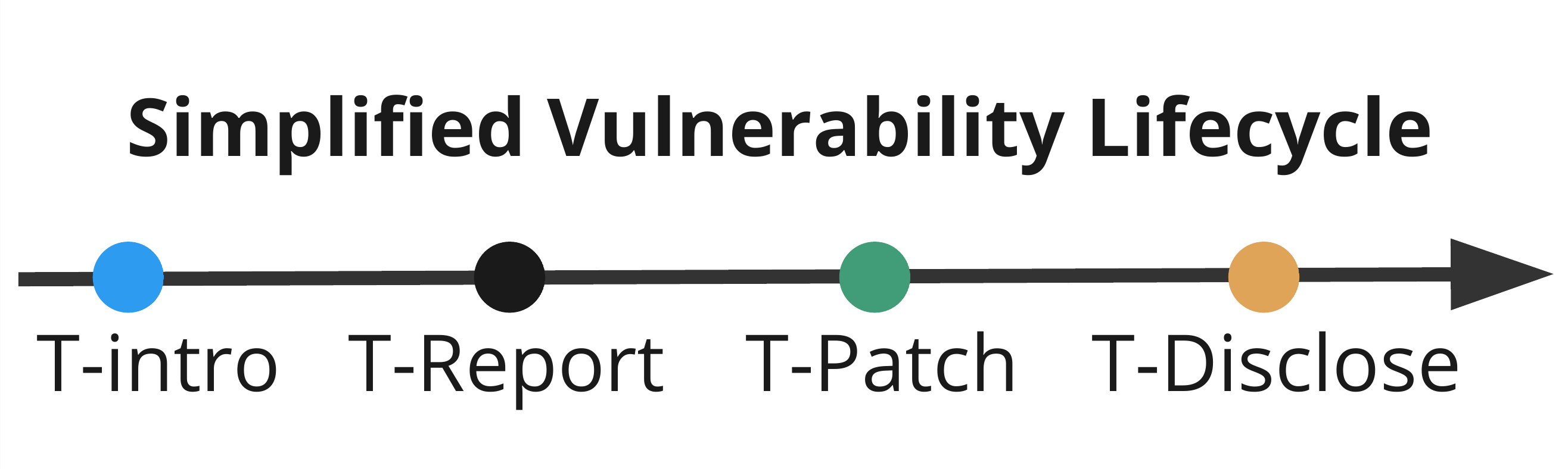}
  \end{center}
\end{wrapfigure}
}

\newcommand\figurecveintro[0]{
\begin{wrapfigure}{r}{0.42\textwidth}
  \begin{center}
    \includegraphics[trim=0cm 4.5cm 0cm 3cm, width=0.42\textwidth]{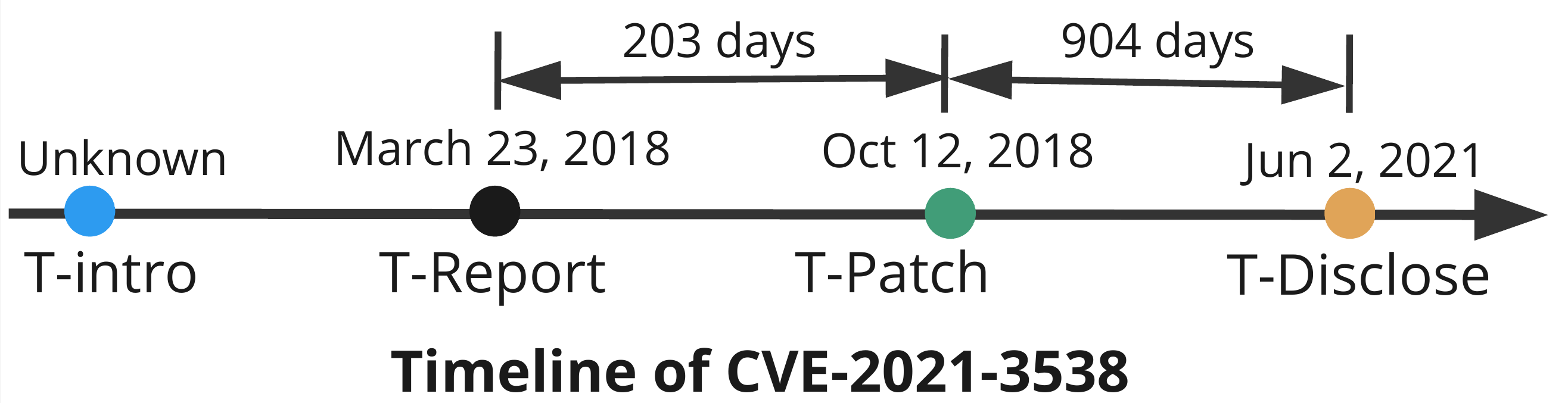}
  \end{center}
\end{wrapfigure}
}
\newcommand\figureRepoisnotActive[0]{
\begin{wrapfigure}{r}{0.32\textwidth}
  \begin{center}
    \includegraphics[trim=0cm 5cm 0cm 5cm, width=0.32\textwidth]
    {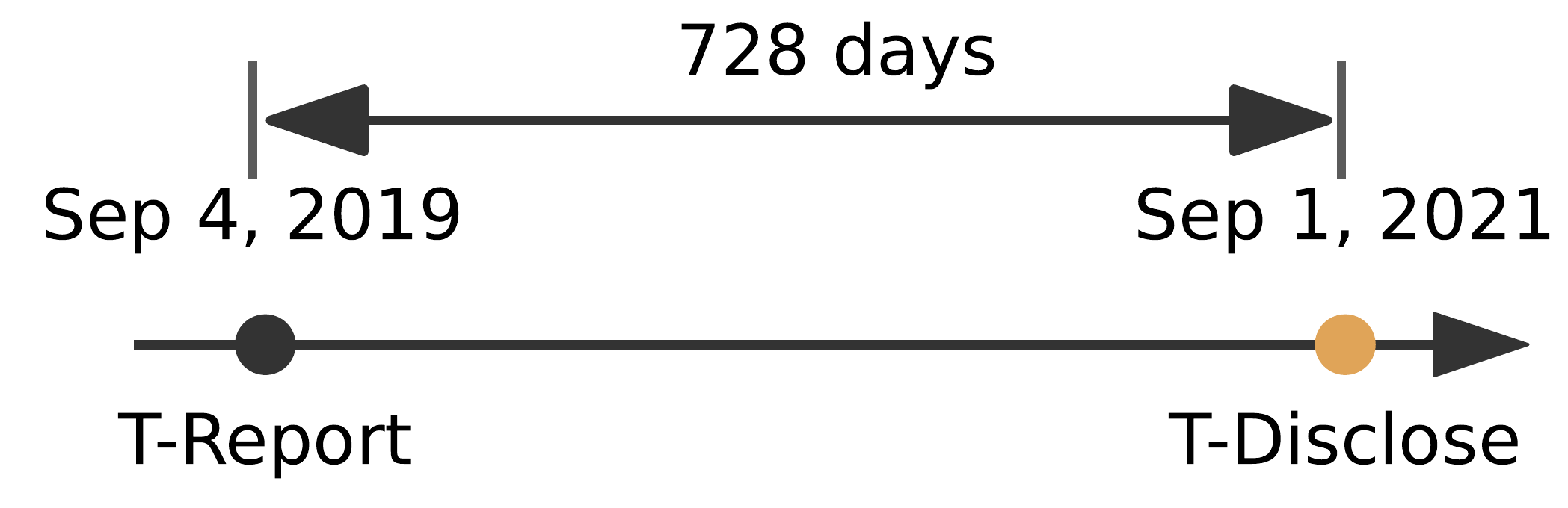}
  \end{center}
\end{wrapfigure}
}
 
\newcommand\figureLowprioritysecondcase[0]{
\begin{wrapfigure}{r}{0.32\textwidth}
  \begin{center}
    \includegraphics[trim=0cm 5cm 0cm 5cm, width=0.32\textwidth]
    {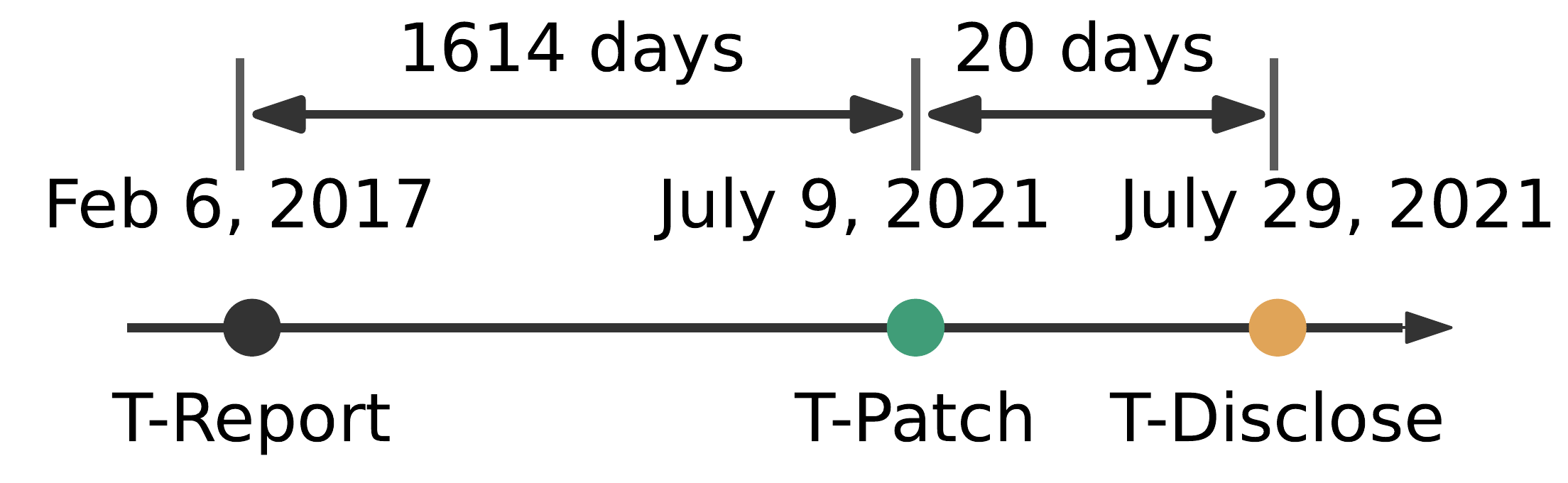} 
  \end{center}
\end{wrapfigure}
}
\newcommand\figureLowpriorityfirstcase[0]{
\begin{wrapfigure}{r}{0.32\textwidth}
  \begin{center}
    \includegraphics[trim=0cm 5cm 0cm 5cm, width=0.32\textwidth]
    {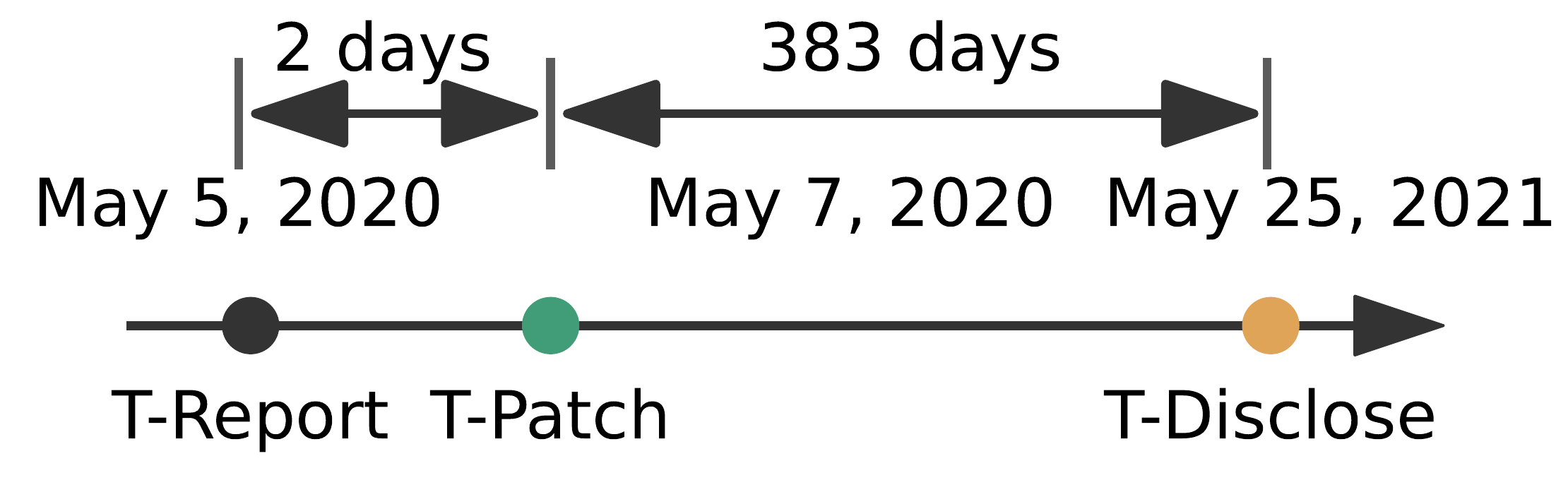} 
  \end{center}
\end{wrapfigure}
}
\newcommand\figureLowprioritythirdcase[0]{
\begin{wrapfigure}{r}{0.32\textwidth}
  \begin{center}
    \includegraphics[trim=0cm 5cm 0cm 5cm, width=0.32\textwidth]
    {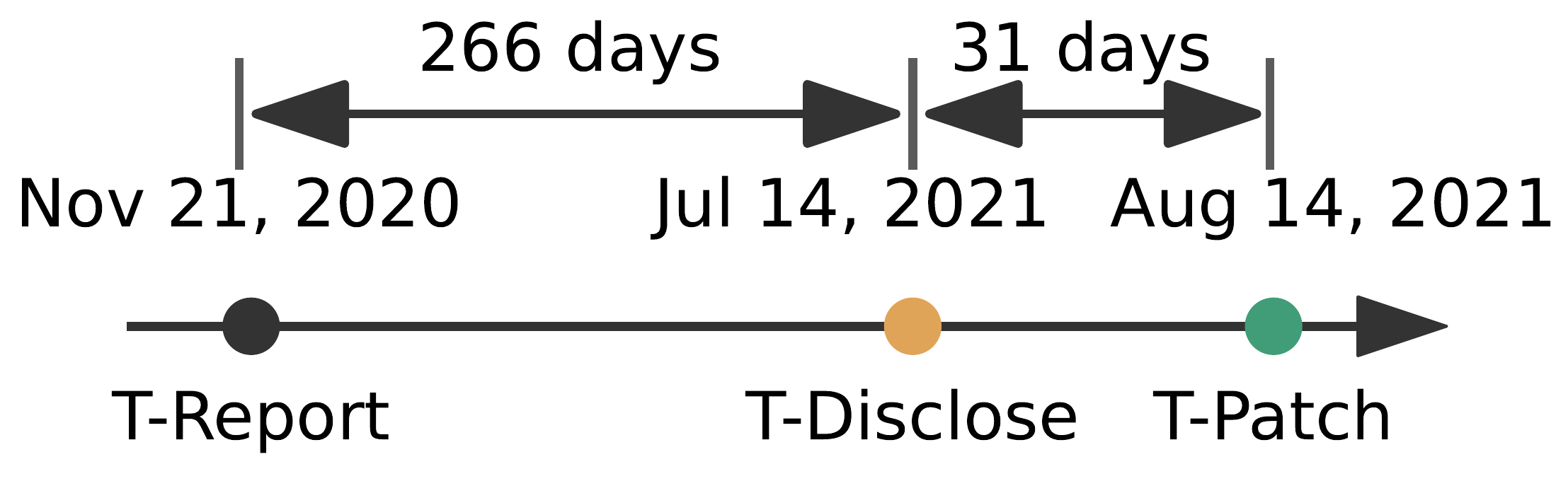} 
  \end{center}
\end{wrapfigure}
}
\usepackage{wrapfig}
\newcommand\figureLowResource[0]{
\begin{wrapfigure}{r}{0.32\textwidth}
  \begin{center}
    \includegraphics[trim=0cm 5cm 0cm 5cm, width=0.32\textwidth]{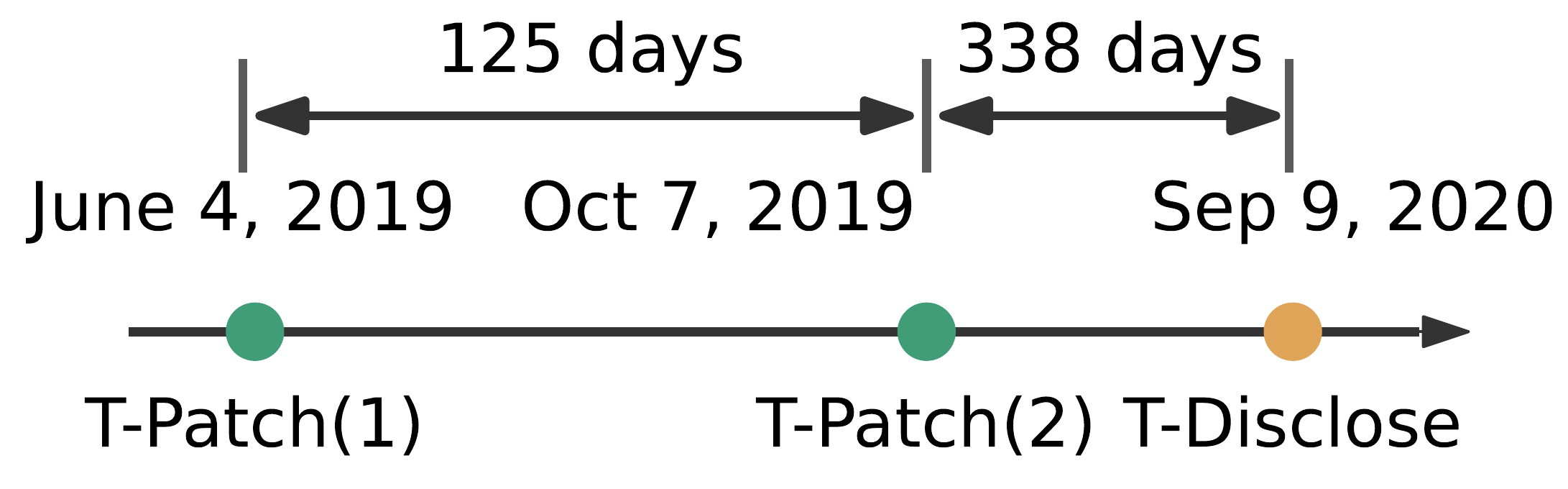}
  \end{center}
\end{wrapfigure}
}
\newcommand\figureCVEIneffectivePatch[0]{
\begin{wrapfigure}{r}{0.32\textwidth}
  \begin{center}
    \includegraphics[trim=0cm 5cm 0cm 5cm, width=0.32\textwidth]{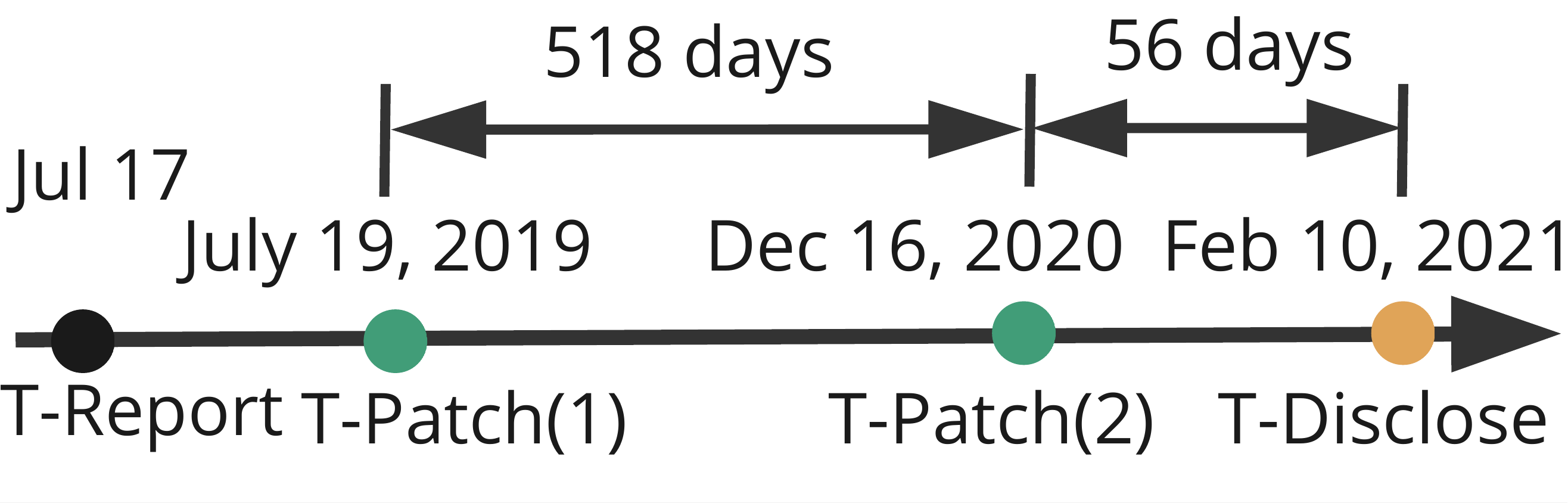}
  \end{center}
\end{wrapfigure}
}

\newcommand\figureLackofClarityregarding[0]{
\begin{wrapfigure}{r}{0.32\textwidth}
  \begin{center}
    \includegraphics[trim=0cm 5cm 0cm 5cm, width=0.32\textwidth]    {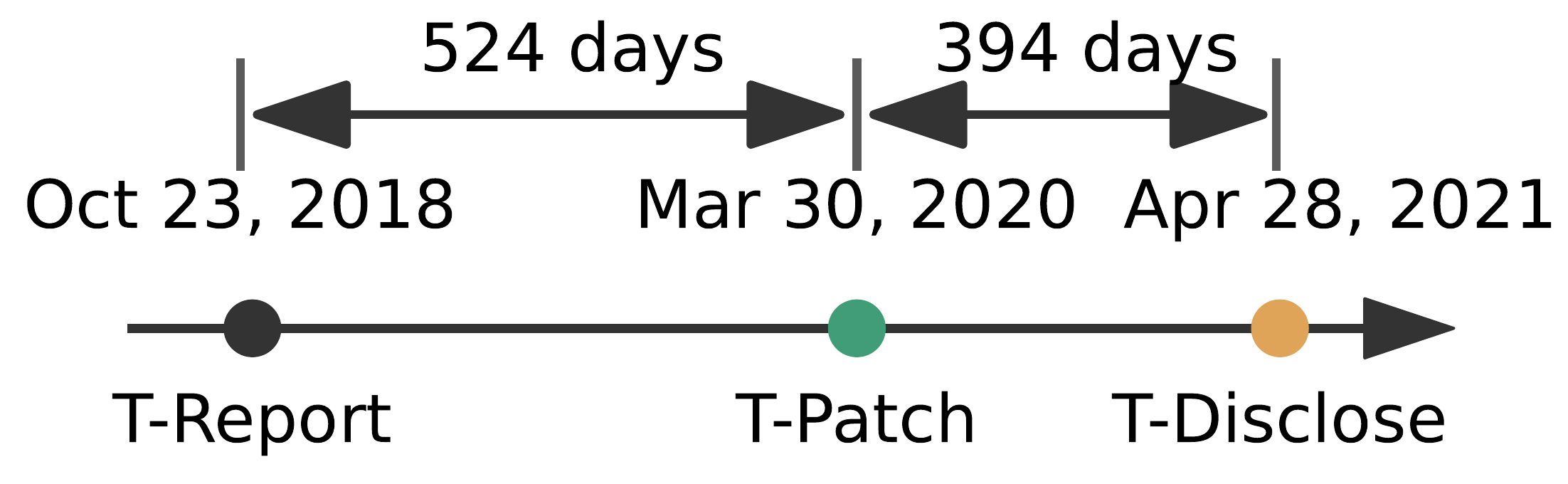}
  \end{center}
\end{wrapfigure}
}

\newcommand\figureUnknown[0]{
\begin{wrapfigure}{r}{0.32\textwidth}
  \begin{center}
    \includegraphics[trim=0cm 5cm 0cm 5cm, width=0.32\textwidth]    {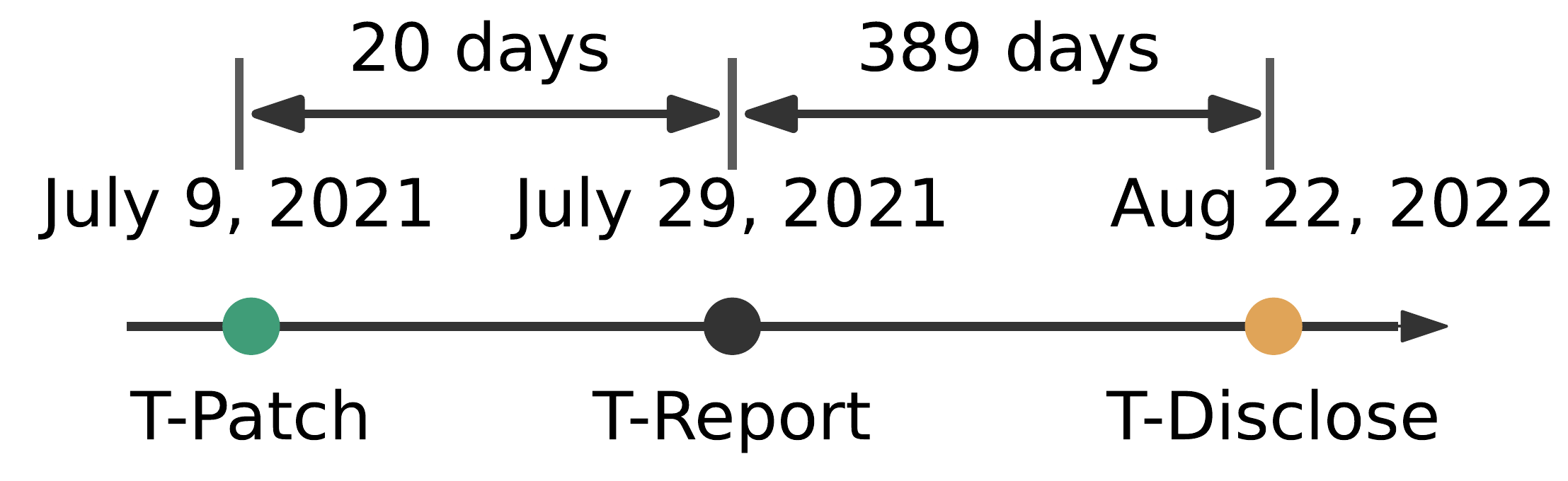}
  \end{center}
\end{wrapfigure}
} 
\newcommand\CVEDisc[0]{
\begin{wrapfigure}{r}{0.32\textwidth}
  \begin{center}
    \includegraphics[trim=0cm 5cm 0cm 5cm, width=0.32\textwidth]
     {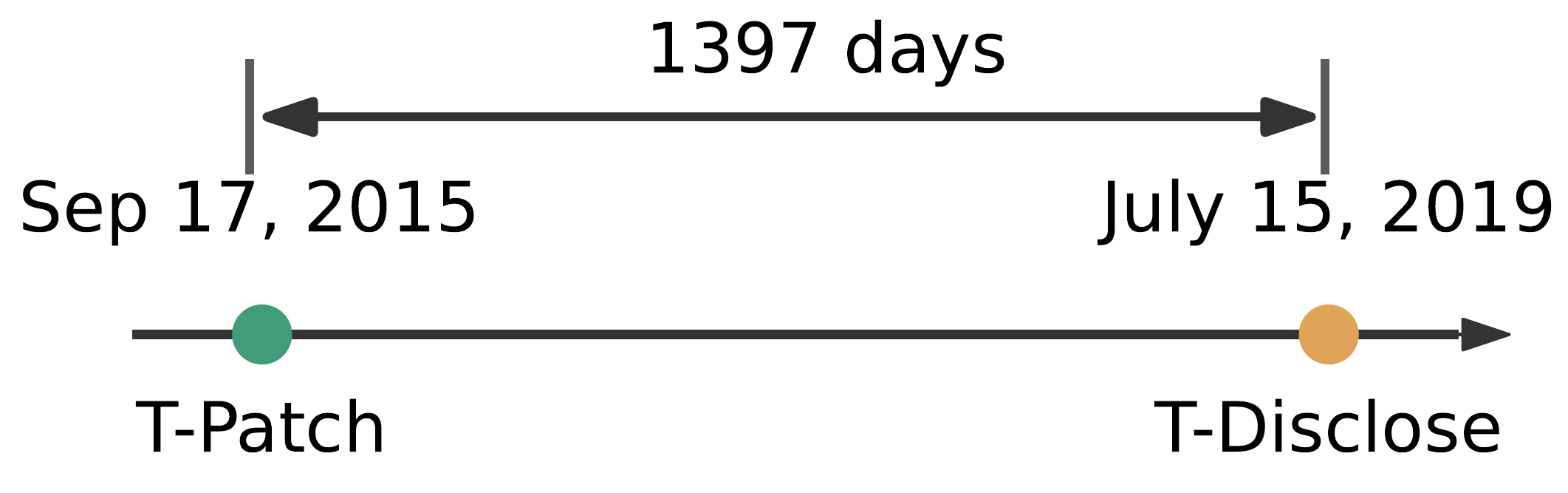}
  \end{center}
\end{wrapfigure}
} 

\newcommand\figureCVEONE[0]{
\begin{wrapfigure}{r}{0.32\textwidth}
  \begin{center}
    \includegraphics[trim=0cm 5cm 0cm 5cm, width=0.32\textwidth]
     {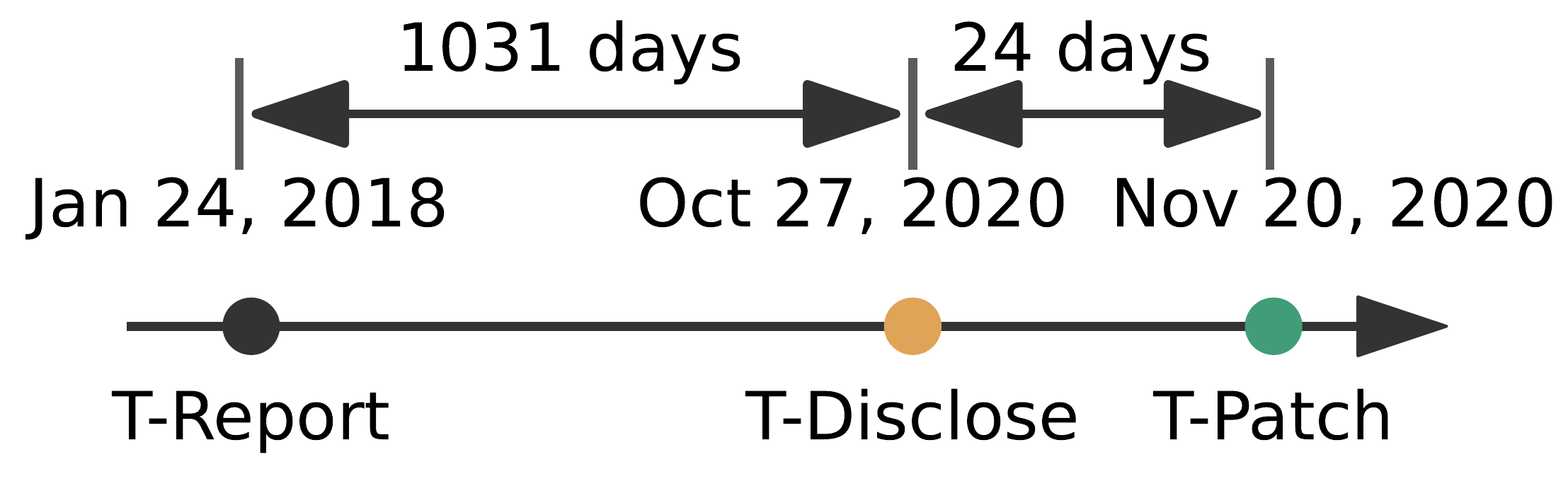}
  \end{center}
\end{wrapfigure}
}

\newcommand{\timelinetwothree}{\begingroup
\setbox0=\hbox{\includegraphics[width=12mm, height=2mm]{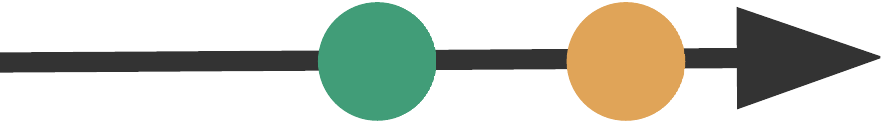}}\parbox{\wd0}{\box0}\endgroup} 

\newcommand{\timelineonethree}{\begingroup
\setbox0=\hbox{\includegraphics[width=12mm, height=2mm]{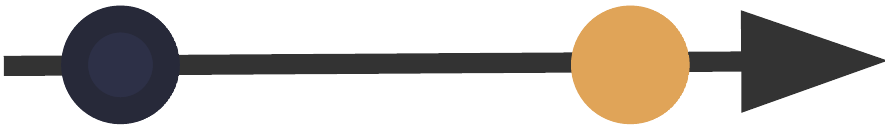}}\parbox{\wd0}{\box0}\endgroup} 

\newcommand{\timelinetwoonethree}{\begingroup
\setbox0=\hbox{\includegraphics[width=12mm, height=2mm]{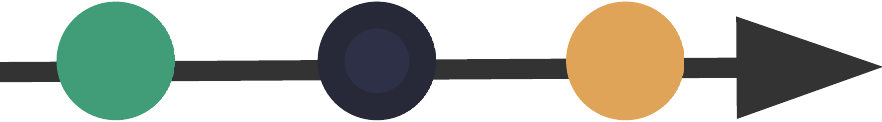}}\parbox{\wd0}{\box0}\endgroup} 

\newcommand{\timelineonethreetwo}{\begingroup
\setbox0=\hbox{\includegraphics[width=12mm, height=2mm]{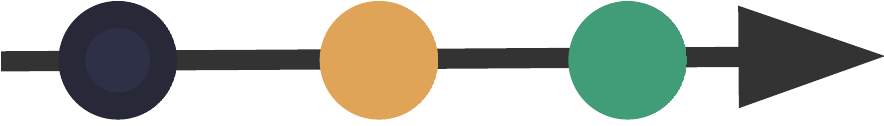}}\parbox{\wd0}{\box0}\endgroup} 

\newcommand{\timelineCVD}{\begingroup
\setbox0=\hbox{\includegraphics[width=12mm, height=2mm]{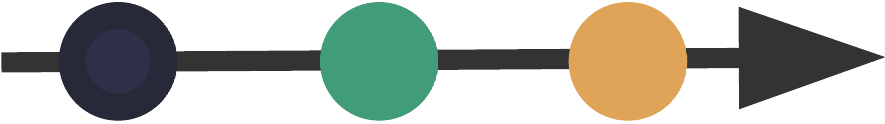}}\parbox{\wd0}{\box0}\endgroup} 

\newcommand{\patchdot}{\begingroup
\setbox0=\hbox{\includegraphics[width=2mm, height=2mm]{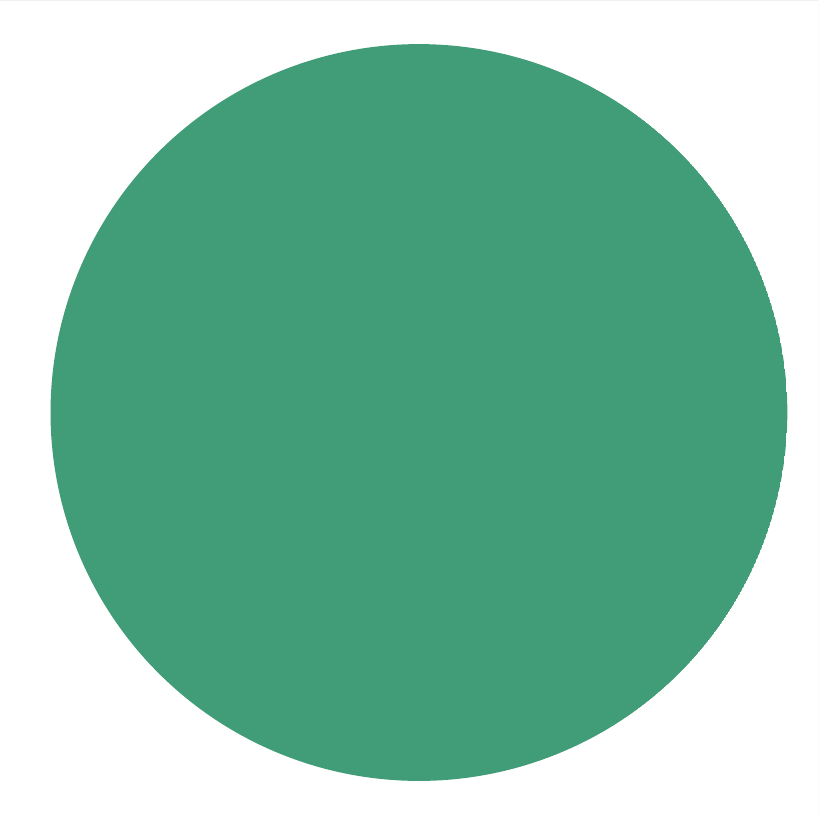}}\parbox{\wd0}{\box0}\endgroup} 

\newcommand{\reportdot}{\begingroup
\setbox0=\hbox{\includegraphics[width=2mm, height=2mm]{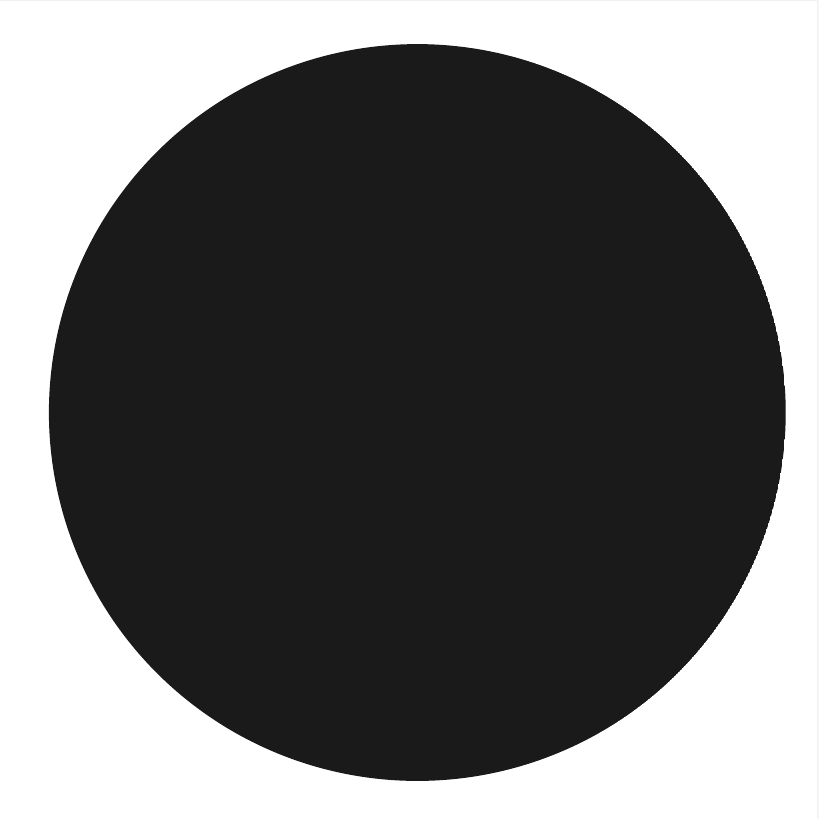}}\parbox{\wd0}{\box0}\endgroup} 

\newcommand{\disclosedot}{\begingroup
\setbox0=\hbox{\includegraphics[width=2mm, height=2mm]{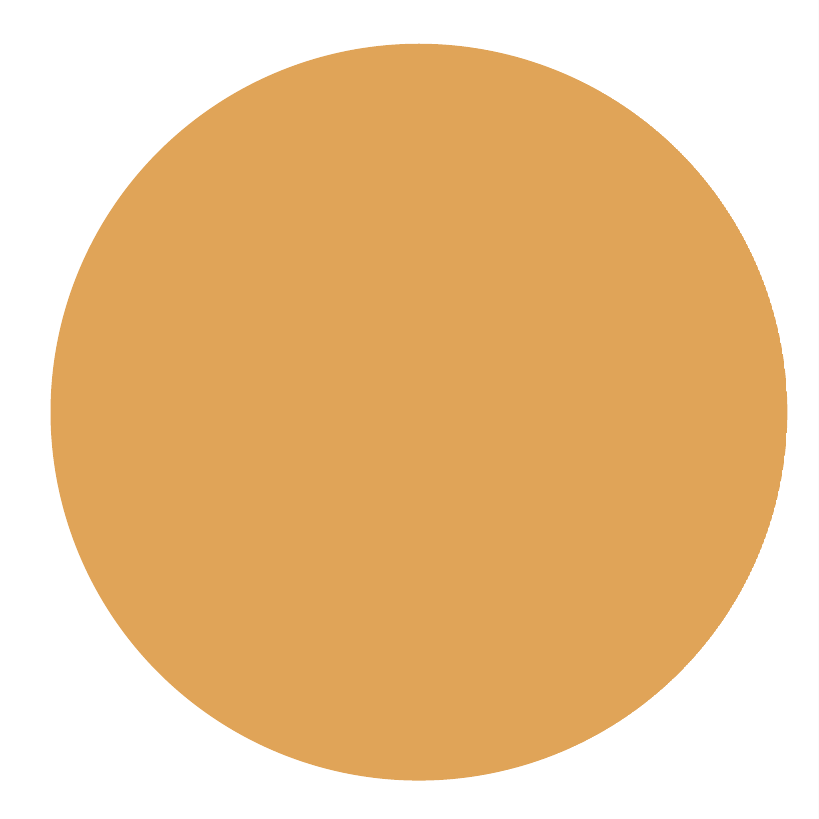}}\parbox{\wd0}{\box0}\endgroup} 

\setcopyright{none}
\renewcommand\footnotetextcopyrightpermission[1]{}
\begin{document}

\title{Detecting Protracted Vulnerabilities in Open Source Projects}

\author{Arjun Sridharkumar}
\authornote{These authors contributed equally to this work.}
\affiliation{ 
    \institution{University of Toronto}
    \country{Canada}
}
\email{arjunsridharkumar@mail.utoronto.ca}

\author{Sara Al Hajj Ibrahim}
\authornotemark[1]
\affiliation{
    \institution{University of Toronto}
    \country{Canada}
}
\email{sara.alhajjibrahim@mail.utoronto.ca}

\author{Jiayuan Zhou}
\affiliation{
    \institution{Queen’s University}
    \country{Canada}
}
\email{jiayuan.zhou@queensu.ca}

\author{Yuliang Wang}
\affiliation{
    \institution{University of Toronto}
    \country{Canada}
}
\email{stevenyuliang.wang@mail.utoronto.ca}

\author{Safwat Hassan}
\affiliation{
    \institution{University of Toronto}
    \country{Canada}
}
\email{safwat.hassan@utoronto.ca}

\author{Ahmed E. Hassan}
\affiliation{
    \institution{Queen’s University}
    \country{Canada}
}
\email{ahmed@cs.queensu.ca}

\author{Shurui Zhou}
\affiliation{
    \institution{University of Toronto}
    \country{Canada}
}
\email{shurui.zhou@utoronto.ca}

\renewcommand{\shortauthors}{Sridharkumar et al.}

\begin{abstract}

Timely resolution and disclosure of vulnerabilities are essential for maintaining the security of open-source software. However, many vulnerabilities remain unreported, unpatched, or undisclosed for extended periods, exposing users to prolonged security threats. While various vulnerability detection tools exist, they primarily focus on predicting or identifying known vulnerabilities, often failing to capture vulnerabilities that experience significant delays in resolution. In this study, we examine the vulnerability lifecycle by analyzing protracted vulnerabilities (PCVEs), which remain unresolved or undisclosed over long periods. We construct a dataset of PCVEs and conduct a qualitative analysis to uncover underlying causes of delay. To assess current automated solutions, we evaluate four state-of-the-art (SOTA) vulnerability detectors on our dataset. These tools detect only 1,059 out of 2,402 PCVEs, achieving approximately 44\% coverage. To address this limitation, we propose \projectname, an enhanced detection approach designed specifically for protracted cases. \projectname\ integrates multiple development artifacts and code signals, supported by a Large Language Model (LLM)-based summarization component. For comparison, we also evaluate a standalone LLM. Our results show that \projectname\ improves detection performance, achieving a 14\% increase in coverage across all PCVEs and reaching 90\% coverage on the \projectname\ PCVE subset, outperforming existing SOTA detectors and standalone LLM based inference. 

\end{abstract}



\maketitle
\section{Introduction} \label{introduction_paper}
Practitioners and researchers have developed protocols for the process of addressing vulnerabilities. Among these, the Coordinated Vulnerability Disclosure (CVD), or responsive disclosure model~\cite{ding2019ethical, vuldisclosure}, outlines how vulnerability reporters and open-source software (OSS) developers should coordinate to ensure that vulnerability details are not disclosed until the issues are resolved~\cite{DisclosureGitHubDocs, householder2017cert, GoogleOSSVuln, SecuritytechniquesVulnerabilitydisclosure2018, MicrosoftOSSguide, TheApacheSoftwareFoundation}. As part of this process, OSS maintainers are also advised to disclose the vulnerability as  
soon as a fix becomes available, since users depend on this information to stay informed and apply necessary updates~\cite{zhou2021finding, imtiaz2022open, li2017large}.

\figuretimeline 

In the figure on the right, we illustrate a simplified vulnerability lifecycle comprising four key events: 
(1) \tintroBOLD, when a vulnerability is introduced into the codebase;
(2) \treportBOLD, when the vulnerability is reported to the OSS developers; 
(3) \tpatchBOLD, when the fix for the vulnerability becomes available;
and (4) \tdiscloseBOLD, 
when security advisories make vulnerability information available to the public.
According to the previously described best practices, it is crucial to minimize the time difference  between each pair of the events, such as 
\tpatch -- \treport~and \tdisclose -- \tpatch.

However, these best practices are often neglected. 
Developers sometimes fix vulnerabilities silently without intending to publicly disclose them~\cite{zhou2021finding}. 
Furthermore, previous studies have highlighted delays at various stages of the vulnerability lifecycle.
For example, 
Decan et al. examined 369 vulnerability reports affecting 269 NPM packages and found a median delay of 11 days for vulnerabilities to be fixed after being reported~\cite{decan2018impact}.  
Additionally, certain vulnerabilities were only resolved following their public disclosure. 
In a study by Alfadel et al.~\cite{alfadel2023empirical}, 1,396 vulnerability reports concerning 686 Python packages were analyzed, revealing that 40.86\% of these vulnerabilities were fixed post-public disclosure, with a median delay of two months from the initial reporting of the vulnerability. 
Prior works have also found that the interval between patching and public disclosure can vary widely, ranging from days to years  ~\cite{zhou2021finding,li2017large,MandiantOSS}. 
\figurecveintro
In the figure on the right, 
we illustrate the timeline of \emph{\href{https://nvd.nist.gov/vuln/detail/CVE-2021-3538}{CVE-2021-3538}}, as analyzed in our study, showing delays of 203 days in \tpatch~and 904 days in \tdisclose. 
These delays expose vulnerabilities that malicious third parties can exploit, posing risks to end users who depend on public disclosure to integrate patches~\cite{alfadel2023empirical}. 

Although previous research has found that delays throughout the vulnerability timeline are common \cite{rodriguez2018analysis}, the underlying causes of the large delays and potential solutions to mitigate the associated risks remain unclear. 
Consequently, in this project, we conducted a comprehensive study to understand the factors that contribute to delays, particularly in patching (\tpatch) and public disclosure (\tdisclose). 
To enhance clarity in this paper, we define vulnerabilities with significant gaps between any of the timestamps (i.e., \treport-- \tpatch~or \tpatch-- \tdisclose) as \textbf{protracted vulnerabilities} (\textbf{\emph{PCVEs}}). 

In this work, we study whether a vulnerability is present during the prolonged period before it is recognized or disclosed. In such protracted cases, evidence of an underlying vulnerability may surface incrementally across different development artifacts and at different points in time, rather than appearing only in finalized source code or as a completed security patch. In practice, vulnerability detection is most commonly framed as a source-code analysis problem and generally excludes information beyond the code itself \cite{li2016vulpecker}. Our setting instead asks whether a vulnerability is already present based on the evidence available at a given point in time, regardless of which development artifacts provide that evidence.

Consequently, SOTA techniques are applied to the artifacts available at each point in time in the same technical manner as originally designed, but with a different evaluation objective. Rather than assuming the presence of finalized code or completed fixes, these techniques are used to assess whether intermediate artifacts exhibit signals consistent with an underlying vulnerability. This usage reflects how protracted vulnerabilities unfold in practice, where relevant signals surface incrementally across artifacts prior to formal recognition. Incorporating multiple artifact types therefore enables analysis of the extended pre-disclosure period during which the vulnerability exists but has not yet been identified as security relevant.

To the best of our knowledge, there is no publicly available dataset consisting of PCVEs and corresponding artifacts, including code changes and vulnerability-related discussions. 
Thus, we first parsed the National Vulnerability Database (NVD)~\cite{NVD}, one of the popular security repositories, to collect 20,951 CVEs that have development artifacts on GitHub, spanning the period from 1999 through July 2024; then we gathered a subset of CVEs with potential delays of more than one year, resulting in a total of 3,228 CVEs (15.4\%) classified as PCVEs.

\begin{figure}[b]
\centering
\includegraphics[width=0.9\textwidth]{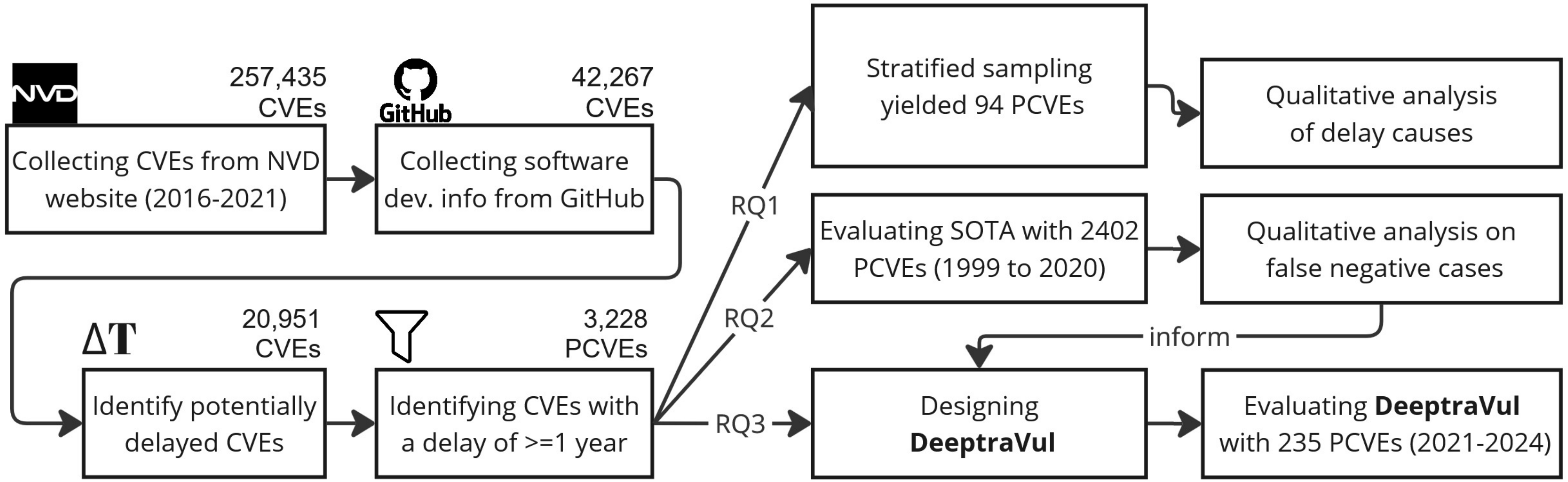}
\caption{Research method overview.}
\label{fig:overview}
\end{figure}

To achieve our research objectives, we formulate three research questions outlined below. 
Figure~\ref{fig:overview} provides an overview of our project along with the associated research methods.

\textbf{RQ1: \RQone}
\noindent To understand the potential causes of delays, we conducted a mixed-methods analysis on the lifecycle of 94 stratified sampled PCVEs. This stratified sample was chosen to ensure a 95\% confidence level with a 10\% margin of error, representing a portion of the total CVEs in the dataset. The analysis reveals eight key factors contributing to delays in PCVE resolution, with the most prevalent being (a) delayed reporting to the NVD, (b) misjudgment of vulnerability severity and relevance, and (c) lack of active project maintenance. Among the 94 CVEs analyzed, the median time to patch vulnerabilities was 9 days, while the median time to disclose vulnerabilities publicly was 619 days.
 
\textbf{RQ2: \RQtwo}
Based on the results from RQ1 concerning the complexities of addressing vulnerabilities, it is evident that previous SOTA vulnerability detection approaches have been limited to a narrow range of artifacts. 
Consequently, our objective is to evaluate the effectiveness of four SOTA methods 
(i.e., \emph{LineVul}~\cite{fu2022linevul}, \emph{DeepDFA}~\cite{steenhoek2024dataflow}, \emph{VulCurator}~\cite{nguyen2022vulcurator}, 
\emph{MemVul}~\cite{pan2022automated}, and \emph{PatchRNN}~\cite{wang2021patchrnn}).

The evaluation results show that out of 2,402 PCVEs, 1,059 (44\%) were detected by the selected SOTA methods. Also, each SOTA method is good at identifying a unique group of PCVEs, which raises the question naturally of if we could take advantage of the collective effort to detect more PCVEs. 
Subsequently, we conducted a qualitative analysis on a subset of the remaining 1,343 PCVEs, which represent 56\% of the dataset. This analysis yielded five critical insights aimed at enhancing PCVE detection. 
 These insights encompass potentially valuable textual and non-textual artifacts that could contribute to the creation of a more effective vulnerability detection method.

\textbf{RQ3: \RQthree}
Building on the findings from RQ1 and RQ2, that current SOTA performs poorly due to input artifact restrictions and can be improved by including external knowledge and additional artifacts, 
we performed a feasibility study to understand the potential of leveraging various input artifacts, including issues, commits, PRs, and  Common Weakness Enumeration (CWE) information, for the accurate identification of PCVEs. 
We designed \projectname, a novel approach that integrates a comprehensive set of textual artifacts to identify additional PCVEs that existing SOTA methods fail to detect. For context, our implementation incorporates an LLM-based summarization component and includes a direct LLM baseline for comparison, allowing us to evaluate how LLMs behave relative to our model.
Our approach achieves superior performance, detecting 14\% more PCVEs overall and 46\% more on the \projectname\ dataset compared to existing SOTA methods.
In addition, we conducted an analysis of the false negative instances in which \projectname~ did not succeed in detection, uncovering additional indicators that highlight potential directions for future research.

This study emphasizes the widespread nature of protracted vulnerabilities, underscores the necessities for continued systematic investigation, and calls for the development of tooling support to accurately identify PCVEs and reduce delays in future research endeavors.
Specifically, we make the following contributions:
\begin{itemize} 
\item To the best of our knowledge, we are the first to study the causes of the delay in patching and disclosure of CVEs;
\item We evaluate the SOTA vulnerability detection methods using the PCVE dataset; 
\item We conduct a feasibility study using \projectname,  an innovative method for detecting vulnerabilities leveraging a wide range of development artifacts and external vulnerability-related knowledge;
\item  We demonstrate that \projectname~surpasses  current SOTA methods, including an LLM-based approach, in identifying PCVEs.
\item We provide a  comprehensive dataset of 3,228 PCVEs and a replication package~\cite{replicationpkg} to support future research efforts.
\end{itemize}

\section{Related Work}\label{relatedwork}

\subsection{Vulnerability Timeline and Relevant Activities}
\begin{figure}[t]
\centering
\includegraphics[width=\columnwidth]{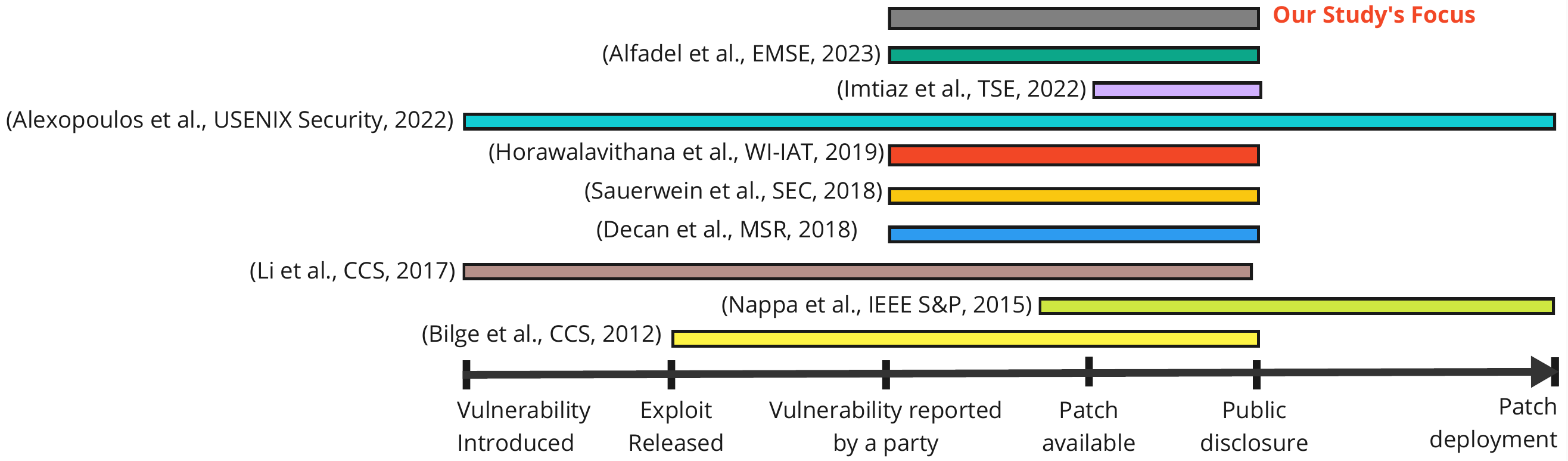}
\caption{Summary of the time frames within the vulnerability lifecycle examined by previous studies.}
\label{fig:timeline-summary}
\end{figure}

Previous studies have explored various stages of the vulnerability lifecycle, each focusing on different timestamps and specific objectives~\cite{alfadel2023empirical,imtiaz2022open,alexopoulos2022long,horawalavithana2019mentions,sauerwein2018tweet,decan2018impact,li2017large,nappa2015attack,bilge2012before,shahzad2012large}. 
Figure~\ref{fig:timeline-summary} depicts the specific time frames examined in these prior studies.  
For instance, Nappa et al. investigated the patch deployment process for 1,593 vulnerabilities, finding that the time between patch releases ranged up to 118 days, with a median duration of 11 days~\cite{nappa2015attack}. 
Similarly, Bilge et al. assessed the prevalence and duration of zero-day attacks by examining the interval between exploit availability and public disclosure, using multiple vulnerability databases to automatically identify instances of zero-day attacks~\cite{bilge2012before}. 
Li et al. conducted a large-scale analysis of 4,080 fixes related to 3,094 vulnerabilities, discovering that over a quarter (26.4\%) of these security issues remained unpatched 30 days after disclosure~\cite{li2017large}.

In this project, we focus on the period between the discovery of a vulnerability (\treport) and its public disclosure (\tdisclose). Our goal is to understand the reasons for delay and to develop heuristics for the early detection of vulnerabilities. 
Several related studies have also examined this time frame, though with varying research goals. 
For example, Decan et al. found that it took an average of 11 days to fix vulnerabilities reported in 369 instances affecting 269 NPM packages~\cite{decan2017empirical}.
Similarly, Alfadel et al. analyzed 1,396 vulnerability reports in the PyPi ecosystem  affecting 686 Python packages, and found that 40.86\% of these vulnerabilities were addressed after public disclosure, with a median delay of two months.  Other studies have  investigated the use of social media platforms (e.g., Twitter, Reddit) for disseminating information about vulnerabilities throughout their lifecycle~\cite{horawalavithana2019mentions, sauerwein2018tweet}.   
Unlike previous studies, our research examines delays in the vulnerability lifecycle across a diverse  set of open-source projects. Our goal is to identify the underlying causes of these delays and to develop more efficient methods for early vulnerability detection

\subsection{SOTA Vulnerability Detection Methods} 

Researchers have designed several deep learning (DL)-based detectors to identify vulnerabilities, including function-level DL-based detectors~\cite{zhou2019devign}, slice-based detectors~\cite{li2018vuldeepecker,li2021sysevr}, and line-level detectors~\cite{hin2022linevd}. These approaches differ in their code representation granularity and contextual modeling strategies. To systematically assess their effectiveness, Steenhoek et al. comprehensively reviewed and benchmarked the performance of nine SOTA DL-based detectors on C/C++ projects~\cite{steenhoek2023empirical}. Their findings revealed that \emph{LineVul}, a transformer-based line-level vulnerability prediction method, exhibited the best overall performance among the evaluated models~\cite{fu2022linevul}.

Building upon the success of transformer-based token models, subsequent research has explored enriching such models with additional program semantics. For example, \emph{DeepDFA} incorporates control-flow and data-flow graphs (CFGs and DFGs) to improve detection accuracy on several benchmark datasets~\cite{steenhoek2024dataflow}. Empirical results show that \emph{DeepDFA} outperforms prior token-based baselines by leveraging these additional structural representations. Consistent with prior work comparing vulnerability detectors~\cite{li2024effectiveness,sejfia2024toward, das2025we}, we therefore include both \emph{LineVul} and \emph{DeepDFA} as baselines for RQ2 and RQ3.

In addition, other studies have developed methods to detect vulnerabilities using artifacts beyond source code.  
For example, early work classified commit messages and corresponding code changes as either vulnerable or not using linear SVM models with Bag-of-Words representations~\cite{sabetta2018practical}. 
This direction was later extended by \emph{PatchRNN}, which uses recurrent neural networks to jointly model code diffs and commit messages for identifying security-relevant patches~\cite{wang2021patchrnn}. 
More recent approaches incorporate additional artifact information. 
\emph{VulCurator} replaces traditional SVM techniques with transformer-based models and integrates issue reports, demonstrating improved detection performance~\cite{nguyen2022vulcurator}.  
Likewise, \emph{MemVul} augments issue-report analysis with external CWE knowledge to improve classification accuracy~\cite{pan2022automated}. 
Given that our CVE dataset includes both source code and development-related artifacts, we selected \emph{PatchRNN}, \emph{VulCurator}, and \emph{MemVul} as additional baselines and evaluated their performance on our PCVE dataset.

\section{RQ1: \RQone} \label{rqone}
This section begins by describing the process used to construct the PCVE dataset, followed by a detailed quantitative and qualitative analysis that highlights the delays observed throughout the vulnerability lifecycle.

\subsection{Constructing the PCVE Dataset}\label{identificationofPCVEs}

\begin{figure}[b]
    \centering
    \footnotesize
    \subfloat[Statistics of $\Delta T$. \label{tab:delta}]{
    \begin{tabular}{lc}
      \hline
    \textbf{Statistic} & {\textbf{\# Days}} \\ 
    \hline
    Mean                           & 218.6                                     \\ 
    25th Quartile                  & 10                                          \\  
    Median                         & 35                                         \\  
    75th Quartile                  & 150                                        \\  
    90th Quartile                  & 597                                        \\  
    95th Quartile                   & 1,061   \\
        \hline
    \end{tabular}
    }
    \quad
    \subfloat[Log-scaled distribution. \label{fig:delta}]{
        \includegraphics[width=0.24\textwidth]{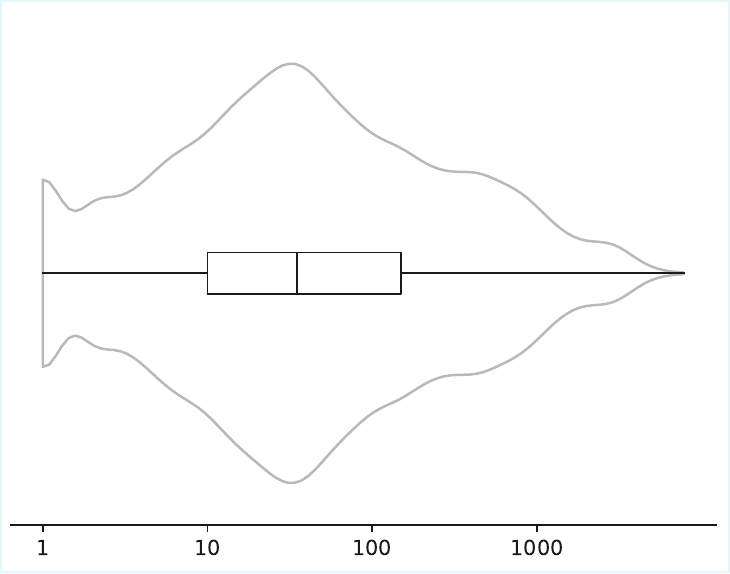}  
    }

    \caption{Distribution of $\Delta T$.}
     \vspace{-2em} 
\end{figure}

\begin{enumerate}[leftmargin=*]
\item
    \textbf{Collecting CVEs disclosed from 1999 to 2024.} We use the National Vulnerability Database (NVD) as our primary data source and retrieve CVEs through the NVD Data Feeds APIs~\cite{DATANVD}. For each CVE, we collect its identifier, disclosure date, and references to external sources containing additional vulnerability information. This step resulted in a total of 257,435 CVEs.
 \item
    \textbf{Collecting development information from GitHub.} To gather detailed information on the lifecycle of vulnerabilities, including discussions and code changes, we focus our study on projects hosted on GitHub. This selection criterion requires at least one GitHub URL in the reference list, resulting in 42,267 CVEs. For each GitHub URL associated with commits, issues, or PRs, we use the GitHub API~\cite{GitHubRestAPI} to gather the relevant information.
    
    \item
    \textbf{Identifying CVEs that have potential delays during their lifecycle.} To study these delays, we compared the timestamps of collected GitHub artifacts with the NVD disclosure dates, retaining only those CVEs where earlier artifacts indicated a delay. This process resulted in the identification of 20,951 CVEs.
\item
    \textbf{Sampling PCVEs with large time intervals between the artifacts creation and NVD disclosure.} We calculated the time difference ($\Delta T$) between the earliest collected development artifact timestamp  (\tearliest) and the NVD disclosure date (\tdisclose) for each CVE. The log-scaled distribution of $\Delta T$  for 20,951 CVEs is shown in Figure~\ref{fig:delta}, with a summary in Figure~\ref{tab:delta}. 
This study focuses on CVEs with a $\Delta T$ of at least 365 days, identifying 3,228 CVEs (15.4\%) as  \emph{PCVEs}.

\end{enumerate}
The 365-day threshold is motivated by both the empirical distribution of $\Delta T$ and the conceptual notion of protracted vulnerabilities. Empirically, $\Delta T$ follows a highly skewed distribution (Figure~\ref{fig:delta}). The 75th percentile is 150 days and the mean is 218.6 days, while 365 days lies in the long-tail region that contains only 15.4\% of CVEs. This indicates that CVEs with $\Delta T \geq 365$ days represent atypical cases rather than typical disclosure delays.

From a conceptual perspective, a one-year interval corresponds to a full development cycle for many software projects~\cite{kumar2018efficient}. A vulnerability with an earliest development artifact dated more than one year prior to public disclosure remains present across multiple cycles of development, testing, and release. This extended duration reflects the defining characteristic of PCVEs. Accordingly, the 365-day threshold serves as the criterion for identifying protracted vulnerabilities in this study.

 \subsection{\textbf{Distribution of the Delays during PCVEs' lifecycle}}
First, we aim to quantitatively analyze the delays occurring throughout the lifecycle of the PCVEs.

\textbf{Stratified sampling.}  
Manually reviewing each CVE's reference links, including understanding the vulnerability, its technical details, and related discussions, requires approximately three hours of focused effort per CVE. To keep the qualitative analysis manageable while preserving representativeness, we apply stratified sampling to the set of 3,228 CVEs~\cite{parsons2014stratified}.  
The PCVEs are grouped into nine stratified buckets based on their $\Delta T$ values, using three-month intervals. From these buckets, we select a total of 94 samples, achieving a 95 percent confidence level with a 10 percent margin of error for the $\Delta T$ distribution.

\textbf{Various types of vulnerability lifecycles.}  
As previously discussed, we identify artifacts corresponding to each timestamp (i.e., \treport, \tpatch, \tdisclose) for each PCVE and compute the intervals between them to examine delays.  
It is important to note that not all three timestamps are available for every PCVE, which leads to incomplete data in some cases.  
Among the 94 PCVEs selected, we successfully retrieved all three timestamps for 48 cases.  
Table~\ref{tab:timeline-summary-horizontal} presents a categorization of the 94 PCVEs based on the presence and sequence of the three activities. We organize them into three distinct lifecycle types.

\begin{enumerate}[leftmargin=*]
\item \textbf{PCVEs with only two recovered timestamps.\\}
 [\timelinetwothree] For 21 PCVEs, we could only obtain the \tpatch, with the \treport~ remaining unknown, such as in the cases of \href{https://nvd.nist.gov/vuln/detail/CVE-2018-11232}{\emph{CVE-2018-11232}} and \href{https://nvd.nist.gov/vuln/detail/CVE-2020-27601}{\emph{CVE-2020-27601}}. The mean interval between \tpatch~ and \tdisclose~ is 962 days.

[\timelineonethree] For 22 PCVEs, only the timestamp for \treport~ was determined, with no traceable patch artifacts, as seen with \href{https://nvd.nist.gov/vuln/detail/CVE-2020-19268}{\emph{CVE-2020-19268}} and \href{https://nvd.nist.gov/vuln/detail/CVE-2016-11014}{\emph{CVE-2016-11014}}. On average, the time gap between \treport~ and \tdisclose~ spans 714 days. 

\begin{table*}[b]

\centering
\footnotesize
     \begin{tabular}
     {ccccc}
     \toprule

      \textbf{Timeline} 
      &\textbf{\#} 
      &  \begin{tabular}{c}
      \tpatchBOLD   --  \treportBOLD 
      \end{tabular}
      & \begin{tabular}{c}
      \tdiscloseBOLD   --   \tpatchBOLD 
      \end{tabular}
      & \begin{tabular}{c}
      \tdiscloseBOLD   --  \treportBOLD 
      \end{tabular}
      \\ 
      \midrule
      \includegraphics[width=0.1\textwidth]{Figures/RQ_1/23.pdf}
      &     21
      & &  
      \raisebox{-0.5\totalheight}{\includegraphics[width=0.16\textwidth]{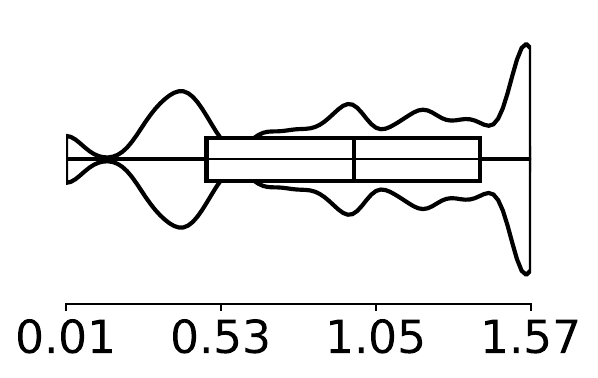}}
         &    
       
      \\
 
 \includegraphics[width=0.1\textwidth]{Figures/RQ_1/13.pdf}
 &
       22
      &    
      &   &   
      \raisebox{-0.5\totalheight}{\includegraphics[width=0.16\textwidth]{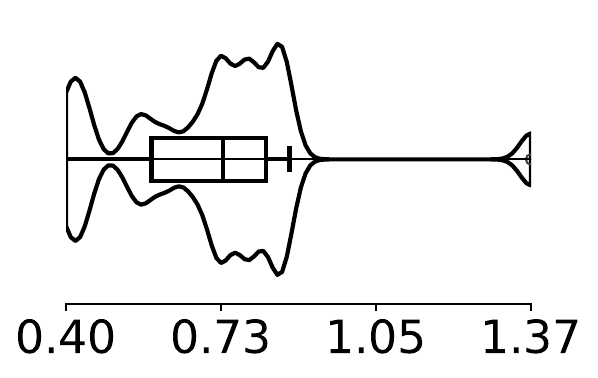}}
      
      \\
      \includegraphics[width=0.1\textwidth]{Figures/RQ_1/123.pdf}
   &      
   39
      & 
      \raisebox{-0.5\totalheight}{\includegraphics[width=0.16\textwidth]{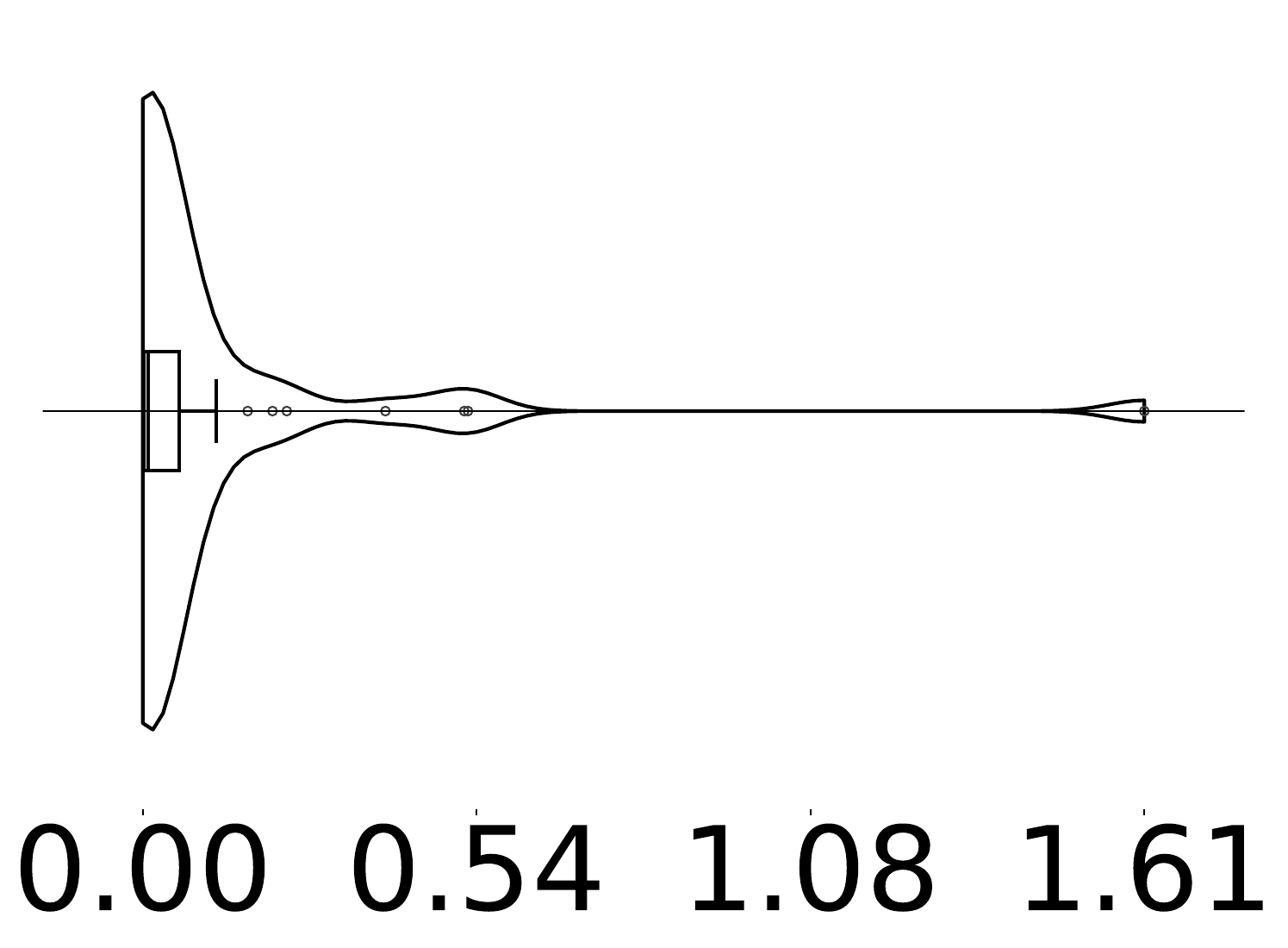}}
      &
      \raisebox{-0.5\totalheight}{\includegraphics[width=0.16\textwidth]{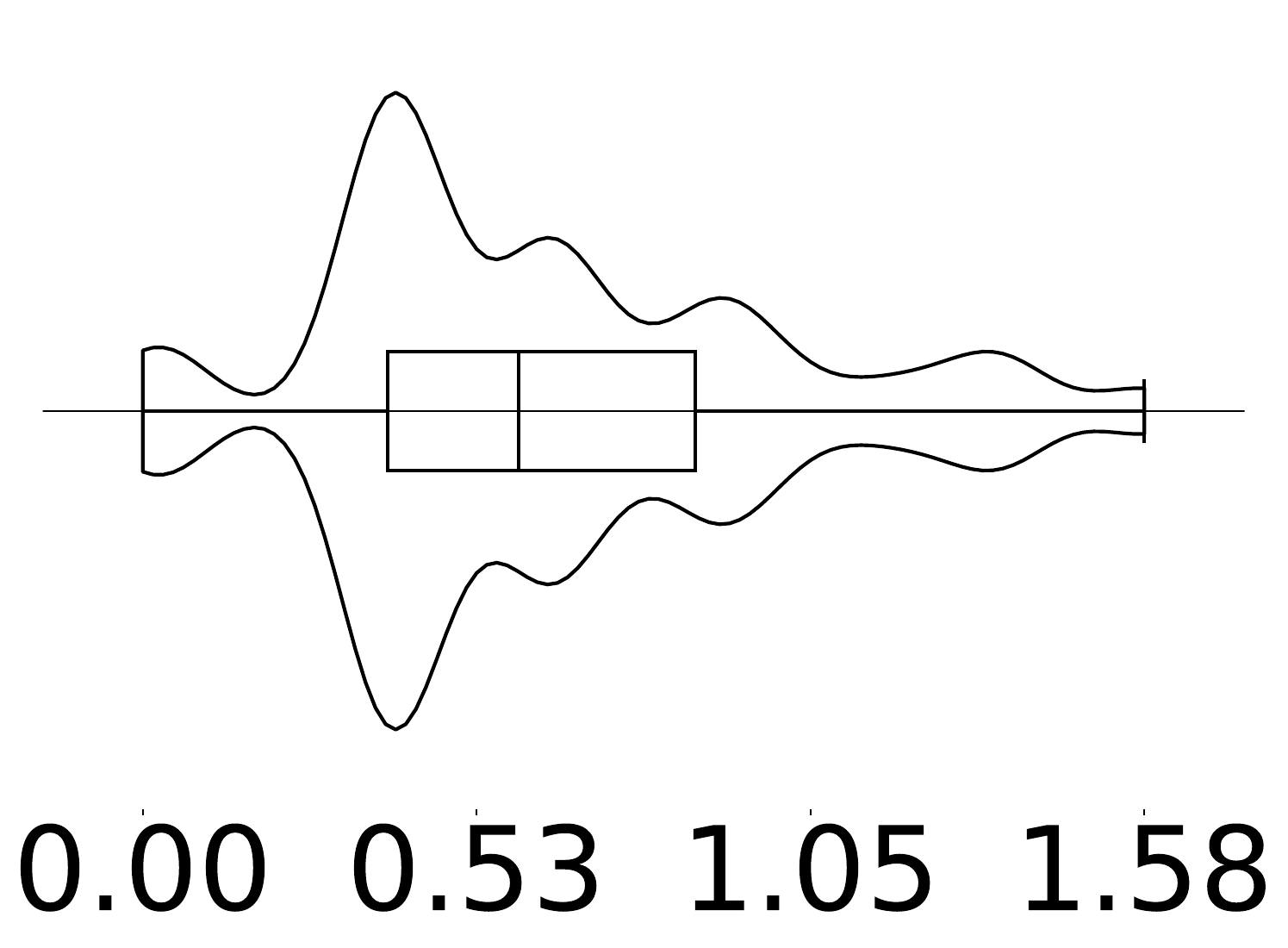}}
      &      \raisebox{-0.5\totalheight}{\includegraphics[width=0.16\textwidth]{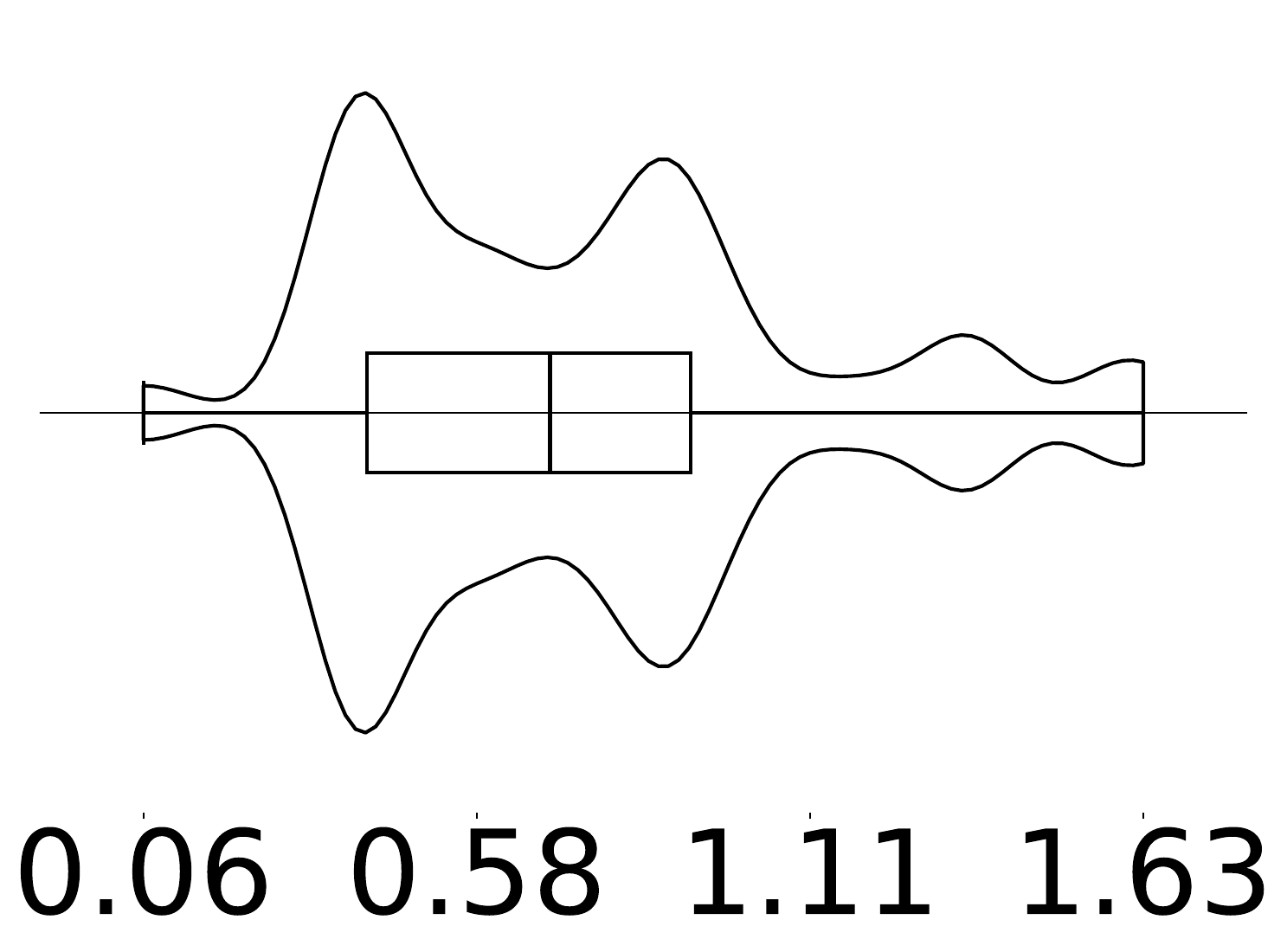}}  
      
      
      \\
 \includegraphics[width=0.1\textwidth]{Figures/RQ_1/132.pdf}
   & 6
      &

    \raisebox{-0.5\totalheight}{\includegraphics[width=0.16\textwidth]{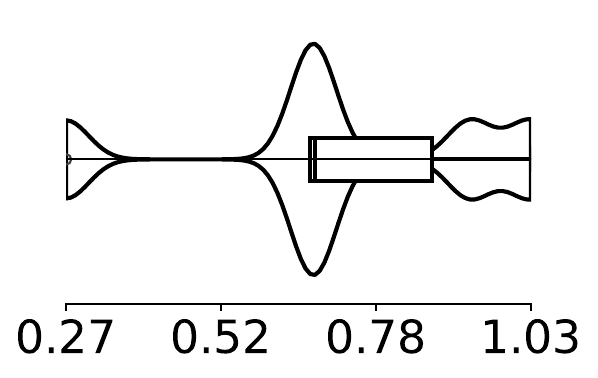}}
    &
    
    \raisebox{-0.5\totalheight}{\includegraphics[width=0.16\textwidth]{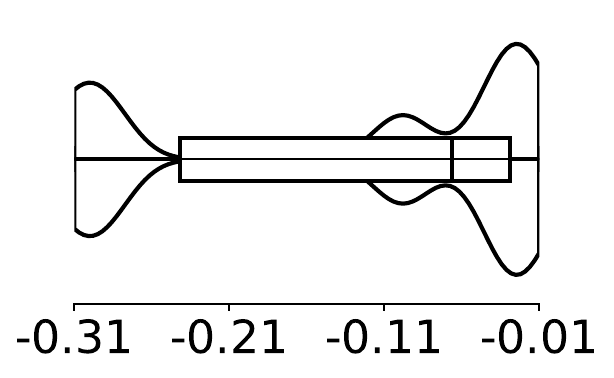}}
    &
    \raisebox{-0.5\totalheight}{\includegraphics[width=0.16\textwidth]{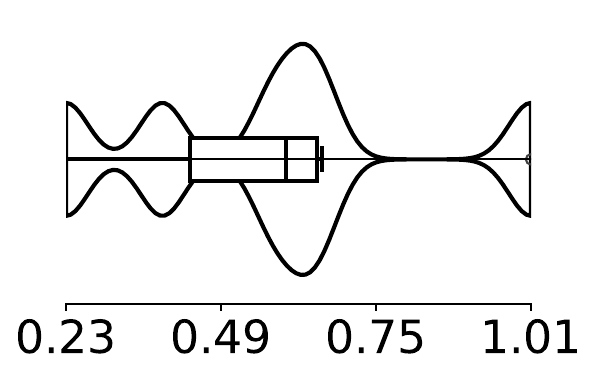}}
      
    \\
   \includegraphics[width=0.1\textwidth]{Figures/RQ_1/213.pdf}
 
      &
      6
      &
      \raisebox{-0.5\totalheight}{\includegraphics[width=0.16\textwidth]{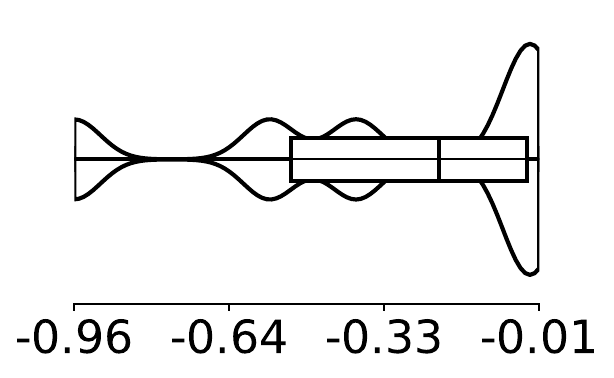}}
      &
      \raisebox{-0.5\totalheight}{\includegraphics[width=0.16\textwidth]{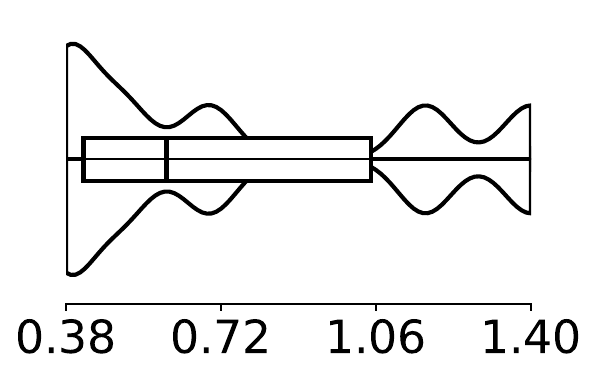}}
      &
         \raisebox{-0.5\totalheight}{\includegraphics[width=0.16\textwidth]{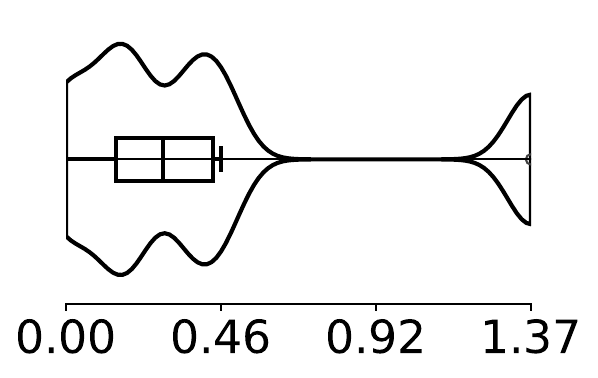}}

      \\ 
      
      \\ \bottomrule
      \end{tabular}
      \caption{Categorization of PCVE timeline types.   The unit of measurement on the x-axis is thousands of days. \reportdot -- \treportBOLD,  \patchdot -- \tpatchBOLD, \disclosedot -- \tdiscloseBOLD. }  
      \label{tab:timeline-summary-horizontal}
\end{table*}
\item \textbf{Following the CVD Model but with Delays.\\}
[\timelineCVD] 39 PCVEs adhered to a sequence of reporting, patching, and subsequent public disclosure. The average time to patch these CVEs was 108 days, with a median resolution period of 9 days. The process of public disclosure took an average of 634 days, with a median of 592 days. These results align with the research conducted by Decan et al. and Alfadel et al., who examined the timelines for vulnerability patching within the NPM and PyPI ecosystems respectively \cite{alfadel2023empirical,decan2018impact}. \href{https://nvd.nist.gov/vuln/detail/CVE-2021-23418}{\emph{CVE-2021-23418}} is an example where, despite early reporting, patch development and public disclosure were delayed.

\item \textbf{PCVEs that Violate the CVD Model.\\}
[\timelineonethreetwo] 6 PCVEs had their vulnerabilities patched only after public disclosure. 
On average, it took 708 days to resolve these vulnerabilities post-reporting, with a median of 676 days. These PCVEs were disclosed approximately 125 days on average before a patch was issued, with a median of 63 days. Such delays compromise the security of the OSS ecosystem, as hackers could exploit the disclosed vulnerabilities to target unaware end users. \href{https://nvd.nist.gov/vuln/detail/CVE-2020-7731}{\emph{CVE-2020-7731}} was publicly disclosed before a patch was available, resulting in a delay between disclosure and resolution.

[\timelinetwoonethree] 6 PCVEs were reported only after patches had already been made available. On average, these vulnerabilities were patched 331 days prior to reporting, with a median interval of 214 days. Public disclosure of these PCVEs occurred, on average, 756 days after patching, with a median interval of 599 days. This pattern is indicative of silent fixes. Such practices among OSS developers result in delays in vulnerability disclosure, providing malicious third parties the opportunity to analyze code changes and exploit vulnerabilities. \href{https://nvd.nist.gov/vuln/detail/CVE-2016-6786}{\emph{CVE-2016-6786}} involved delayed reporting and disclosure, resulting in limited timely visibility into the vulnerability.
\end{enumerate}

\subsection{\textbf{Reasons for delays in  PCVEs.}} \label{methodRQ1}
Next, we construct timelines for PCVEs to qualitatively examine the factors contributing to delays, followed by a summary of the root/observed causes underlying these delays.

\subsubsection{\textbf{Qualitative Data Analysis.}} \label{Qualitative_Analysis_research_method}
For each sampled PCVE, we reviewed discussion threads in referenced GitHub issues and PRs, if available, to determine potential reasons for the delays in either \tpatch~and \tdisclose~respectively.
Specifically, we used an open coding procedure~\cite{wicks2017coding, service2009book} to categorize these reasons.  
Note that a PCVE can have multiple causes throughout its lifecycle, contributing to delays in key activities such as reporting, fixing, and public disclosure. Thus, each PCVE might be assigned multiple codes. 

One author conducted an initial analysis of 25 CVEs, developing a preliminary codebook to establish an initial categorization framework. To ensure consistency in labeling, two additional authors independently reviewed the coding decisions, engaging in collaborative discussions to refine the categorization process. Discrepancies were resolved through iterative dialogue, leading to the finalization of a consistent coding scheme. Following this initial phase, the team expanded the analysis to a stratified sample of 94 CVEs. This sample size was determined to achieve a 95\% confidence level with a 10\% margin of error, ensuring a representative subset of the dataset. As the analysis progressed, the emergence of new themes diminished, and by the time coding was completed for these 94 CVEs, no significant novel patterns were identified. To confirm this, an additional 25 CVEs were analyzed, contributing only marginal new insights and further supporting the conclusion that thematic saturation had been reached \cite{francis2010adequate}. 

Throughout the study, an iterative and dynamic coding approach was used, with ongoing discussions to refine and validate the categorization framework. Any remaining discrepancies were resolved collaboratively, ensuring the robustness of the classification methodology.

\subsubsection{\textbf{Reason for delays in PCVEs lifecycle.}}\label{codebooksection}

\begin{table}[t]
    \centering
    \footnotesize
    \caption{Codebook showcasing the underlying reasons behind the delays.}

\begin{tabular}{@{}p{0.02\textwidth}p{0.15\textwidth}
p{0.6\textwidth}p{0.1\textwidth}@{}}
\toprule
\textbf{\#} & \textbf{Reason for delay} & \textbf{Description} & \textbf{Incidence} \\ \midrule

1 & \makecell[l]{\textbf{Delayed}\\\textbf{NVD}\\\textbf{Disclosure} }&   
\makecell[l]{The vulnerability was fixed, but there was a delay in reporting it \\to the NVD or publishing the CVE, resulting in a gap between the fix \\ and public disclosure.  This may have been due to a silent fix or a slow \\ reporting process.} &
65 (46.76\%) \\ \midrule

2 &\makecell[l]{\textbf{Misjudgment}\\\textbf{of Relevance}\\\textbf{and Severity}} &   
\begin{tabular}[t]{@{}l@{}}
\makecell[l]{
The issue was reported, but the maintainers either: \\
1. Explicitly stated it was a lower priority \\
2. Focused on other contributions during the same period \\
3. Debated whether it qualified as a vulnerability or reported it as a CVE.}
\end{tabular} &
27 (19.42\%) \\ \midrule 

3 & \makecell[l]{\textbf{Lack of}\\\textbf{Active}\\\textbf{Maintenance}} &   
 \makecell[l]{The issue has remained open, with the repository experiencing months\\ of inactivity; lacking contributions, updates, or maintenance, resulting\\ in problems staying unresolved for significant periods.} &
16 (11.51\%) \\ \midrule

4 & \makecell[l]{\textbf{Incomplete or}\\\textbf{Insufficient Fix}} &  
The initial fix did not fully address the issue, requiring further updates. &
10 (7.19\%) \\ \midrule

5 & \makecell[l]{\textbf{Disagreement}\\\textbf{on Resolution}} & 
The team had conflicting views on how to resolve the issue. &
8 (5.76\%) \\ \midrule
 
6 & \makecell[l]{\textbf{Other or}\\\textbf{Unknown}} & 
\makecell[l]{The delay occurred between the reported issue and the fix, but there is no \\ clear evidence explaining the cause, making it difficult to determine what \\was delayed.} &
7 (5.04\%) \\ \midrule

7 & \makecell[l]{\textbf{Resource}\\\textbf{Constraints}} &
\makecell[l]{The team lacked the necessary resources (e.g., time, developers,\\ or testing facilities) to address the issue quickly. }&
4 (2.88\%) \\ \midrule

8 & \makecell[l]{\textbf{Lack of}\\\textbf{Expertise}} &
\makecell[l]{The maintainers lacked the necessary expertise and \\explicitly sought external help.} &
2 (1.44\%) \\  

\bottomrule
\end{tabular}
\label{tab:codebook-simple}
\end{table}

Among the 94 sampled PCVEs, we identified eight potential factors contributing to the observed delays.  
Table~\ref{tab:codebook-simple} summarizes the codebook, providing a description of each factor with the distribution of PCVEs that exhibit delays in \tpatch~or \tdisclose.  
We describe each identified factor in detail and present illustrative examples to support the analysis.

\textbf{Category 1: Delayed Reporting to NVD.}  This category includes cases where a vulnerability was patched in a public codebase, but its corresponding entry in the NVD was published only after a significant delay. The delay we refer to is the time gap between the commit that fixed the issue (\tpatch) and the official disclosure date in the NVD (\tdisclose). We identified \textbf{65} such instances where vulnerabilities were patched in open-source projects, yet the corresponding NVD entries appeared days, weeks, months, or even years later. In reviewing these cases, we found no observable artifacts,   such as GitHub issues, PRs, advisory discussions, or maintainer comments,  addressing the delay between \tpatch~and \tdisclose. For example, in \emph{\href{https://nvd.nist.gov/vuln/detail/CVE-2019-1010308}{CVE-2019-1010308}},\CVEDisc the GitHub repository contained a silent \href{https://github.com/aquaverde/aquarius-core/commit/e1af89aa9df07ea265d879518ede9eb98aa494e0}{\emph{commit\#e1af89a}} on September 17, 2015, addressing the vulnerability, while the disclosure happened on July 15, 2019. When the NVD eventually published the CVE, it referenced the commit directly but provided no further insights. This absence of intermediate documentation made it challenging to understand the reason for the delayed disclosure.

\textbf{Category 2: Misjudgment of Relevance and Severity.}  
This category includes cases where the vulnerability was reported, but maintainers either prioritized other contributions, considered the issue to be of lower severity, or questioned whether it warranted a CVE assignment, resulting in delayed action. These cases generally fall into three types.  
  
\textbf{Case 1: Issues are explicitly stated as lower priority.}  
This type captures situations where maintainers clearly assigned the report a lower priority relative to other ongoing work.  A representative example is \href{https://nvd.nist.gov/vuln/detail/CVE-2020-10064}{\emph{CVE-2020-10064}}, reported in 
\href{https://github.com/zephyrproject-rtos/zephyr/issues/24970}{\emph{zephyrproject-rtos/zephyr}} on May~5,~2020.  The issue was labeled as medium priority within the project's workflow, indicating that it was treated as a routine \figureLowpriorityfirstcase matter at the time. Maintainers closed the report on May~7,~2020, and added \href{https://github.com/zephyrproject-rtos/zephyr/commit/38970c07abfcddcfc6a5958189f096a55c49594a}{commit\#38970c0} referencing the issue on the same day.  The vulnerability was later assigned a CVE and publicly disclosed on May~25,~2021.  
This example illustrates cases where an issue appeared ordinary initially but later proved to have security implications.

\textbf{Case 2: Issues are deprioritized while maintainers focused on other contributions.}  
This type describes reports that were acknowledged but repeatedly postponed, while maintainers focused on other development tasks.  
For this case, in \emph{\href{https://nvd.nist.gov/vuln/detail/CVE-2021-23418}{CVE-2021-23418}}, the vulnerability was reported in the 
\href{https://github.com/nicolargo/glances/}{\emph{nicolargo/glances}} repository as 
\href{https://github.com/nicolargo/glances/issues/1025}{\emph{issue\#1025}} on February~6,~2017.  
Although acknowledged and scheduled for Glances~2.9, it was repeatedly deferred across milestones—2.9, 2.9.1, 2.11, 3.0, 3.1.3, 3.1.7, and later. During this period, maintainers continued regular development activity, \figureLowprioritysecondcase including several commits shortly after the report (February~12 and February~19,~2017) that did not address the issue.  
A patch was applied on July~9,~2021, and public disclosure followed on July~29,~2021.  
This example illustrates how delays can stem from competing development priorities rather than from lack of awareness.

\textbf{Case 3: Issues where maintainers were initially uncertain whether the report qualified as a vulnerability.}
This category captures situations in which the security relevance of a report was unclear and required additional assessment before being treated as a vulnerability.
A representative example is \href{https://nvd.nist.gov/vuln/detail/CVE-2020-36420}{\emph{CVE-2020-36420}}
 in the Gentoo ecosystem, documented in \href{https://bugs.gentoo.org/755896}{issue\#755896}, which was reported on November~21,~2020.
In this case, the affected package was no longer maintained.
During the discussion, the maintainers described it as a \emph{“trivial issue with a minor impact”} and noted that they were \emph{“not sure if there is a need to stabilize this package”}. Routine development continued while its potential security implications were evaluated. The vulnerability was publicly disclosed on July~14,~2021, and a patch was committed on August~14,~2021 to \figureLowprioritythirdcase remove the unmaintained package. This case shows how delays may arise when the severity or security impact of a report is not immediately evident. 

Overall, we identified \textbf{27} similar cases where delays resulted from misjudged severity or competing development priorities.

\textbf{Category 3: Lack of Active Maintenance.} 
This category includes vulnerabilities that remain unpatched due to the absence of sustained development activity. In many open-source projects, maintenance relies on voluntary contributions. When these contributions stop for a significant period, reported vulnerabilities may remain unresolved regardless of their severity. A clear example is the project \href{https://github.com/taosir/wtcms/}{\emph{taosir/wtcms}}, which has shown minimal activity since its creation. The commit history shows only a few short periods of development, including September 2017, April 2018, and June and December 2019. 

For the remainder of the time between 2017 and 2025, the repository was inactive \figureRepoisnotActive and no commits were made. On September 4, 2019, a vulnerability in this project was reported as \href{https://nvd.nist.gov/vuln/detail/CVE-2020-20343}{\emph{CVE-2020-20343}}. No patch was provided, no discussion took place, and the issue remains unresolved at the time of writing.  Although no corrective action was taken, the vulnerability was eventually disclosed publicly on September 1, 2021. This case shows how the lack of active maintenance can result in serious vulnerabilities remaining unpatched for extended periods. Of the 94 PCVEs analyzed in our study, \textbf{16} cases involved inactive projects.

\textbf{Category 4: Incomplete or Insufficient Fix.}  
This category includes cases where open-source developers release patches that do not fully address the vulnerability, requiring multiple revisions. This delays both the release of a secure version and public disclosure, increasing user exposure. A clear example is a path traversal vulnerability in \href{https://github.com/QuorumDMS/ftp-srv/}{\emph{QuorumDMS/ftp-srv}}, which was later assigned \href{https://nvd.nist.gov/vuln/detail/CVE-2020-26299}{\emph{CVE-2020-26299}}. It was first reported on July 17, 2019, in \href{https://github.com/QuorumDMS/ftp-srv/issues/167}{\emph{issue\#167}}, with a \href{https://drive.google.com/file/d/1Ywi5YPzJkOWZqTTcou_15bKbj29voHcn/view?pli=1}{\emph{proof-of-concept video}} shared in the discussion thread demonstrating the exploit. An initial fix was submitted two days later in \href{https://github.com/QuorumDMS/ftp-srv/pull/168}{\emph{PR\#168}}. \figureCVEIneffectivePatch The patch was incomplete. One contributor wrote, \emph{\textquotedblleft I'd like to see if we can confirm this is still the case\textquotedblright}, and later testing confirmed that the issue was still present. On December 9, 2020, another contributor reported, \emph{\textquotedblleft I was able to reproduce it on 4cd88b1\textquotedblright}. A second fix was committed on December 16 via \href{https://github.com/QuorumDMS/ftp-srv/commit/457b859450a37cba10ff3c431eb4aa67771122e3}{\emph{commit\#457b859}}, but concerns remained. A user commented, \emph{\textquotedblleft Still not fixed or I am doing something wrong\textquotedblright}. The vulnerability was eventually disclosed publicly on February 10, 2021, after multiple incomplete fixes and continued concerns. This case shows how incomplete patches can delay remediation and increase the risk of exposing technical details that could be exploited. Of the 94 PCVEs analyzed, \textbf{10} cases fall into this category.

\textbf{Category 5: Disagreement on Resolution.}  
This category includes cases where fixing the vulnerability was delayed due to disagreements among developers about how to handle the issue. These discussions often extended over time, slowing down decision-making and implementation. \figureCVEONE A notable example is \emph{\href{https://nvd.nist.gov/vuln/detail/CVE-2018-21269}{CVE-2018-21269}}, reported on January 24, 2018, in \href{https://github.com/OpenRC/openrc/issues/201}{\emph{issue\#201}}. Developers disagreed on whether certain usage patterns should be restricted. One contributor questioned the need for enforcement, saying, \emph{“All we are doing is scanning the path and warning if there are any symbolic links”}. Another pushed for stricter handling and a deprecation period, suggesting, \emph{“My vote is to ban them... and take that value from the upstream build system”}. Compatibility concerns also came up and a developer noted, \emph{“What if, for example, \texttt{/var} is a symlink... Either \texttt{/opt} or \texttt{/opt/package} might legitimately be a symlink”}. The ongoing back-and-forth delayed progress, with public disclosure occurring on October 27, 2020, and a complete fix not applied until November 20, 2020, nearly three years after the initial report. We identified \textbf{8} cases where similar disagreements led to delays in fixing vulnerabilities.

\textbf{Category 6: Other or Unknown Cases.} In certain cases, delays in vulnerability resolution cannot be attributed to any specific cause due to missing or unclear information. These situations arise when there are no recorded discussions, commits, or documentation explaining the time gap between the initial report (T\textsubscript{Report}) and the patch (T\textsubscript{Patch}), making it difficult to determine  \figureUnknown why the fix or disclosure was delayed. For instance, \href{https://nvd.nist.gov/vuln/detail/CVE-2021-3639}{\emph{CVE-2021-3639}} was reported on July 9, 2021, and a fix was implemented on July 29, 2021. There were no recorded commits or discussions during this period to explain the delay.  Public disclosure took place later, on August 22, 2022, more than a year after the patch was applied. Without clear evidence, it remains uncertain whether the delay was due to internal decisions, security concerns, or other unknown factors. Of the 94 PCVEs analyzed, \textbf{7} cases fall into this category.

\textbf{Category 7: Resource Constraints.}
This category refers to delays in vulnerability resolution caused by limited development resources, such as time and contributor availability. Such constraints are common in open-source projects, which often depend on volunteer contributions and informal coordination. A clear example is \href{https://nvd.nist.gov/vuln/detail/CVE-2020-15163}{\emph{CVE-2020-15163}} from \href{https://github.com/theupdateframework/python-tuf}{\emph{theupdateframework/python-tuf}}, where progress was slowed due to a lack of available contributors. An initial patch was submitted on June 4, 2019, through \href{https://github.com/theupdateframework/python-tuf/pull/885}{\emph{PR\#885}}, but the work remained incomplete for several months. On June 13, 2019, the project manager requested assistance with writing test cases, commenting, \emph{“Maybe an intern”}.\figureLowResource In a later comment on the same PR, they reiterated the request more explicitly: \emph{“Do we have an intern who can take over, that is, fix the things I noted below and add some tests?”}. The tests were eventually added on October 3, 2019, and the final fix was merged on October 7, 2019, approximately four months after the initial call for help. Public disclosure occurred later, on September 9, 2020, nearly a year after the patch was completed. This example illustrates how even acknowledged vulnerabilities can face prolonged resolution times when contributors are unavailable. We identified \textbf{4} cases in which patch completion was delayed due to resource-related challenges, highlighting the need for sustained contributor engagement and structured collaboration in open-source security work.
  
\textbf{Category 8: Lack of Expertise.} 
In some cases, vulnerability resolution is delayed due to limited security expertise among project maintainers. This can result in uncertainty, extended discussions, and hesitation in applying a fix. Without a clear understanding of the issue or its implications, maintainers may postpone action, delaying both patching and disclosure. For example, \href{https://nvd.nist.gov/vuln/detail/CVE-2020-22781}{\emph{CVE-2020-22781}} from \href{https://github.com/ether/etherpad-lite}{\emph{ether/etherpad-lite}} was reported on October 23, 2018, but the patch was not implemented until March 30, 2020, and public disclosure followed over a year later on April 28, 2021. The delay was partly due to the absence of a formal reporting process and the maintainers’ unfamiliarity with handling security issues. One maintainer openly stated, \figureLackofClarityregarding  \emph{“I'm not clever enough to resolve this. I need someone with expertise to help!”}. This highlights a broader challenge in some open-source projects, where maintainers may lack experience in secure coding, vulnerability triage, or disclosure coordination. In such cases, progress often depends on support from external contributors or security researchers. \textbf{2} cases out of the 94 PCVEs involved delays linked to limited expertise, underscoring the importance of domain knowledge for timely vulnerability resolution.

Our analysis of PCVE resolution timelines shows that the longest delays often occur between implementing a fix and reporting it, with 65 cases involving delayed publication to the NVD. These gaps are difficult to explain due to the absence of documentation or recorded discussions between the fix (T\textsubscript{Patch}) and the disclosure (T\textsubscript{Disclose}) times. The second most common cause of delay is the misjudgment of a vulnerability’s relevance or severity by maintainers, which can lead to deprioritization or extended debates over classification. Additional factors contributing to delays include organizational and technical challenges such as inactive maintenance, disagreements on appropriate fixes, and incomplete or ineffective patches. Limited resources and expertise also affect patch deployment, as many projects face contributor shortages, lack testing infrastructure, or require external support for security-related tasks. Collectively, these challenges point to broader issues in open-source vulnerability management, particularly inconsistent reporting practices and communication gaps that hinder timely and effective resolution.

\begin{tcolorbox}[boxsep=2pt,left=2pt,right=2pt,top=2pt,bottom=2pt]
\textbf{Finding:}  We identify eight key factors that contribute to delays in PCVE resolution, with the most prevalent being (a) delayed reporting to the NVD, (b) misjudgment of vulnerability severity and relevance, and (c) lack of active project maintenance. To minimize these delays, developers should prioritize security vulnerabilities appropriately and ensure timely disclosure of fixes.
\end{tcolorbox}

\section{RQ2: \RQtwo}
\subsection{Data Preparation}

As described in Sec.~\ref{relatedwork}, we selected five SOTA methods as the baseline for evaluation,  \emph{LineVul}~\cite{fu2022linevul}, \emph{DeepDFA}~\cite{steenhoek2024dataflow}, \emph{VulCurator}~\cite{nguyen2022vulcurator}, \emph{MemVul}~\cite{pan2022automated}, and \emph{PatchRNN}~\cite{wang2021patchrnn}. For RQ2, we selected a subset of 2,402 PCVEs released for 21 years (from 1999 to 2020).\footnote{PCVEs released from 2021 to 2024 are utilized for the evaluation of RQ3} 
To accommodate the diverse input data requirements of various SOTA methods (as summarized in Table~\ref{tab:rq_2_artifact_level_performance}, column `Artifact type'), we organized the PCVEs accordingly to ensure a fair evaluation setup. 
For methods that require code-level inputs (e.g., diffs, functions, or code slices), we included only PCVEs whose patches or affected files are written in a supported programming language, specifically C, C++, or Java.
Overall, we successfully collected 1,213 out of 2,402 PCVEs, each containing at least one of the necessary artifacts that can be used to evaluate one or more SOTA methods in a compatible programming language. In addition, we included non-vulnerability data points for each method to facilitate accurate comparisons. The detailed data collection procedure is described below. 
 
\subsubsection{\textbf{MemVul}.} \label{memvuln_data_collection}
Utilizing a DL-based approach that incorporates language models and external vulnerability knowledge from the CWE, \emph{MemVul} takes GitHub issues as input for vulnerability identification, focusing on the titles and bodies, as they were first created. 
Among the 2,402 PCVEs from our dataset, 1,059 PCVEs have at least one referenced issue, totalling 1,109 unique issues, all created before the NVD disclosure date of their corresponding CVEs. 
We removed the 399 issues that have been used for \emph{MemVul}'s training purpose~\cite{pan2022automated}, resulting in a total of 683 PCVEs and corresponding 710 issues as summarized in Table~\ref{tab:rq_2_artifact_level_performance}, column  `\#Vuln'. 
Finally, we collected the titles and bodies of all the 710 issues at the time they were created from the GitHub Archive~\cite{GHArchive}. 
  
\subsubsection{\textbf{VulCurator}.} \label{VulCurator_experiment_setting_rq_2}

\emph{VulCurator} leverages DL techniques to identify fixing commits for vulnerabilities through the analysis of issue–commit pairs. In particular, the issue title, body, and comments, along with the commit message and patch, are used. To extract issue–commit pairs in our dataset, we first extract issue IDs from commit messages using the same regular-expression matching method described in the \emph{VulCurator} paper~\cite{nguyen2022hermes}. Further, to recover additional pairs, we leverage the issue timeline events to identify linked commits using the GitHub API~\cite{GitHubRESTIssuesTimeline}.

\emph{Data filtering.}
Given our emphasis on gathering information prior to the NVD disclosure date to deduce the heuristics that might result in an earlier disclosure, we verify that both the issues and commits are created before the NVD disclosure date in all instances. To make a fair comparison with \emph{MemVul}, which only accounts for information available at the time the issue was created, we also refine the issue–commit pairing by solely considering the information available when the issue–commit link was established. For example, if the issue is created before the commit, we extract its title, body, and comments as of the commit creation date. If the commit is created before the issue, we extract the issue title and body as of the issue creation date.

After applying these rules and restricting code-related artifacts to supported programming languages (C, C++, and Java), our dataset contains 586 commits, 198 issues, and 631 issue–commit pairs related to 210 PCVEs to evaluate \emph{VulCurator}.

\subsubsection{\textbf{LineVul}. } \label{linevuln_experiment_setting_rq_2}   
The original \emph{LineVul}, targeting C/C++, employs a transformer-based, fine-grained approach for predicting vulnerabilities and analyzing source code at the line level to identify potential security weaknesses \cite{fu2022linevul}. Originally designed for C/C++, we extended \emph{LineVul}'s functionality to support Java. The fine-tuning process involved constructing the training dataset and designing experiments, resulting in an F1 score of 0.65 across Java. 

To obtain the modified C, C++, and Java source code files from our dataset, we initially pinpointed 486 PCVEs referenced with at least one commit. Subsequently, we extract a total of 534 commits, from which we gather 793 files, encompassing source files written in C, C++, and Java as detected using their file extensions (i.e., \texttt{.c, .cpp, .cxx, .java}). 
Further, as \emph{LineVul} requires the source code files to be split into functions, we extract a total of 1,656 functions from the files by utilizing \texttt{SrcML} \cite{collard2013srcml}. 
From these 1,656 functions, we identify the functions that are modified by the commits. 
We identified 1,656 modified functions linked to 486 unique PCVEs.

 \subsubsection{\textbf{DeepDFA}.} \label{deepdfa_experiment_setting_rq_2}

\emph{DeepDFA} is a deep learning–based vulnerability detection model that enhances token-level representations such as  CFGs and DFGs, enabling richer structural and semantic program analysis~\cite{steenhoek2024dataflow}. DeepDFA is designed for detecting vulnerabilities in C/C++ projects by integrating data-flow information directly into the learning process.

In our study, we reuse the dataset constructed for \emph{LineVul} and restrict it to C/C++ source files to align with the original scope of \emph{DeepDFA}. After filtering, the resulting dataset contains 407 unique PCVEs associated with 453 commits and 590 modified C/C++ files. Since \emph{DeepDFA} operates at the function level, we further evaluate it on 1,212 modified C/C++ functions, for which we generate the required CFG and DFG representations.

\subsubsection{\textbf{PatchRNN}.} \label{patchrnn_experiment_setting_rq_2}
\emph{PatchRNN} is an RNN-based model designed to identify vulnerability-fixing commits by jointly encoding commit messages and patch diffs \cite{wang2021patchrnn}. Unlike \emph{LineVul}, which requires function-level representations, \emph{PatchRNN} operates directly on commit-level artifacts, including the commit message and code.

During the data collection process, we retained only commits that (1) modify source files written in supported programming languages (C, C++, or Java), and (2) are created prior to the NVD disclosure date. Consequently, the \emph{PatchRNN} dataset inherits these constraints and contains only commits for which valid diffs and corresponding file modifications were available. Following this filtering process, we obtained a total of 660 PCVEs associated with 1,199 commits. For each commit, we collect the commit message and the full patch diff, which constitute the required input to \emph{PatchRNN}.

\begin{table*}[t]
\footnotesize
\centering
\caption{(RQ2) Data preparation results and the performance of SOTA methods on the PCVE dataset.}
\vspace{-1em}

\begin{tabular}{@{}p{1.2cm} p{1.4cm} cc p{.5cm}p{.5cm}p{.5cm}p{.5cm}ccll@{}}
\toprule

\multirow{2}{*}{\textbf{SOTA}} &
\multicolumn{3}{c}{\textbf{Input Data}} &
\multicolumn{8}{c}{\textbf{Evaluation Results}} \\

\cmidrule(lr){2-4} \cmidrule(lr){5-12}

& \multicolumn{1}{l}{\textbf{\begin{tabular}[c]{@{}l@{}}Artifact\\Type\end{tabular}}}
& \multicolumn{1}{l}{\textbf{\#Vuln}}
& \multicolumn{1}{l}{\textbf{\#Non-Vul}}
& \multicolumn{1}{c}{\textbf{TP}}
& \multicolumn{1}{c}{\textbf{FP}}
& \multicolumn{1}{c}{\textbf{FN}}
& \multicolumn{1}{c}{\textbf{TN}}
& \multicolumn{1}{c}{\textbf{Prec.}}
& \multicolumn{1}{c}{\textbf{F1}}
& \multicolumn{1}{c}{\textbf{\begin{tabular}[c]{@{}l@{}}Applicable\\_Recall\end{tabular}}}
& \multicolumn{1}{c}{\textbf{\begin{tabular}[c]{@{}l@{}}All\\_Recall\end{tabular}}}
\\
\midrule

\textbf{MemVul} &
Issue &
\multicolumn{1}{c}{683} &
\multicolumn{1}{c}{549} &
\multicolumn{1}{c}{\textbf{525}} &
\multicolumn{1}{c}{273} &
\multicolumn{1}{c}{276} &
\multicolumn{1}{c}{\textbf{158}} &
\multicolumn{1}{c}{0.66} &
\multicolumn{1}{c}{0.71} &
\multicolumn{1}{c}{0.77} &
\multicolumn{1}{c}{\textbf{0.22}}
\\
\midrule

\textbf{VulCurator} &
\begin{tabular}[c]{@{}l@{}}Issue–\\Commit\\Pair\end{tabular} &
\multicolumn{1}{c}{210} &
\multicolumn{1}{c}{156} &
\multicolumn{1}{c}{109} &
\multicolumn{1}{c}{\textbf{84}} &
\multicolumn{1}{c}{72} &
\multicolumn{1}{c}{101} &
\multicolumn{1}{c}{0.60} &
\multicolumn{1}{c}{0.56} &
\multicolumn{1}{c}{0.52} &
\multicolumn{1}{c}{0.05}
\\
\midrule

\textbf{LineVul} &
\begin{tabular}[c]{@{}l@{}}Source Code\\Function\end{tabular} &
\multicolumn{1}{c}{486} &
\multicolumn{1}{c}{476} &
\multicolumn{1}{c}{219} &
\multicolumn{1}{c}{473} &
\multicolumn{1}{c}{267} &
\multicolumn{1}{c}{3} &
\multicolumn{1}{c}{0.32} &
\multicolumn{1}{c}{0.37} &
\multicolumn{1}{c}{0.45} &
\multicolumn{1}{c}{0.09}
\\
\midrule
\textbf{DeepDFA} &
\begin{tabular}[c]{@{}l@{}}Source Code\\Function\end{tabular} &
\multicolumn{1}{c}{393} &
\multicolumn{1}{c}{157} &
\multicolumn{1}{c}{323} &
\multicolumn{1}{c}{151} &
\multicolumn{1}{c}{\textbf{70}} &
\multicolumn{1}{c}{6} &
\multicolumn{1}{c}{\textbf{0.68}} &
\multicolumn{1}{c}{\textbf{0.75}} &
\multicolumn{1}{c}{\textbf{0.82}} &
\multicolumn{1}{c}{0.13}
\\
\midrule

\textbf{PatchRNN} &
Commit &
\multicolumn{1}{c}{660} &
\multicolumn{1}{c}{616} &
\multicolumn{1}{c}{444} &
\multicolumn{1}{c}{496} &
\multicolumn{1}{c}{216} &
\multicolumn{1}{c}{120} &
\multicolumn{1}{c}{0.47} &
\multicolumn{1}{c}{0.56} &
\multicolumn{1}{c}{0.67} &
\multicolumn{1}{c}{0.18}
\\

\bottomrule
\end{tabular}
\label{tab:rq_2_artifact_level_performance}
\end{table*}

\subsubsection{\textbf{Collection of Non-vulnerable artifacts}.} \label{non_vuln_data_rq2} 
We constructed a dataset comprising non-vulnerable artifacts, maintaining a 1:1 ratio at the CVE level and a 1:5 ratio at the artifact level between vulnerable and non-vulnerable data, using the augmentation process described in previous studies~ \cite{sabetta2018practical,nguyen2022vulcurator,nguyen2022hermes}.
For example, if a CVE is associated with one issue and two commits, we randomly sample five non-vulnerable issues and ten non-vulnerable commits to create a corresponding non-vulnerable data point.
However, some repositories may not have sufficient artifacts due to limited activity, resulting in a discrepancy between the number of vulnerable and non-vulnerable data points. 
In particular, we employ a four-step method to gather non-vulnerable artifacts, including issues and commits. The following description uses commits as an example:
\begin{itemize}
    \item \emph{Step 1}: For each CVE referencing a GitHub commit, we collected all the commits from the corresponding repository.
    \item \emph{Step 2}: We removed all the commits  referenced by any CVE published between 1999 (the first vulnerability logged in the NVD) and 2020 by comparing them against the NVD database.
    \item \emph{Step 3}: To ensure the non-vulnerable commit is created within the same time frame as the vulnerable artifact, maintaining relevance and timeliness, we only retained commits  made within a six-month window surrounding the vulnerable artifact's creation date~\cite{sabetta2018practical}.
    \item \emph{Step 4}: We randomly selected  five commits from the retained list.
\end{itemize}

\noindent
As such, for each SOTA method, we follow the same data preparation process to combine the collected non-vulnerable artifacts based on the corresponding CVEs, programming languages, and the specific model under evaluation.
We obtained 2,109 non-vulnerable issues used for \emph{MemVul}, producing 549 non-vulnerable data points.
For \emph{VulCurator}, we collected 429 commits and 803 issues, with 4,792 non-vulnerable issue–commit pairs, yielding 156 non-vulnerable data points.
For \emph{LineVul}, we collected 4,786 non-vulnerable source code functions and 383 commits, resulting in 476 non-vulnerable data points.
For \emph{DeepDFA}, after restricting to C/C++ projects, we collected 3,151 non-vulnerable C/C++ functions associated with 237 non-vulnerable commits, corresponding to 397 non-vulnerable data points.
For \emph{PatchRNN}, we collected 1,167 non-vulnerable commits, resulting in 616 non-vulnerable data points.
The summary of the input data is presented in Table~\ref{tab:rq_2_artifact_level_performance}, column `\#Non-Vuln'.
The replication package includes our script for these steps~\cite{replicationpkg}.

\subsection{Results} \label{rq_2_results}
 
\subsubsection{\textbf{Overview of CVE Detection Rates.}} \label{OverviewofCVEDetectionRate}

In Table~\ref{tab:rq_2_artifact_level_performance}, the column titled `Evaluation Results' displays the outcomes of the evaluation. 
Given the diverse input data requirements of various SOTA methods, each   applies to a subset of PCVEs for experimentation. 
As mentioned before, 1,213 out of 2,402 PCVEs are applicable to at least one of the SOTA methods.
Therefore, we report the performance of the SOTA methods on our dataset using two recall metrics:
\begin{itemize}
    \item Recall in \textbf{applicable} PCVEs per each model:   {\small  \[ Applicable\_Recall = \frac{\#Detected~PCVEs}{\#Applicable~PCVEs}\]}
 
    \item Recall in \textbf{all} PCVEs, where $\#All~PCVEs = 2,204$: 

{\small \[All\_Recall= \frac{\#Detected~PCVEs}{\#All~PCVEs~}\]}
\end{itemize}

\begin{figure}[b]
\centering
 
\begin{minipage}[t]{0.43\textwidth}
  \centering 
  \vspace{0.3em}
  \includegraphics[width=0.97\linewidth]{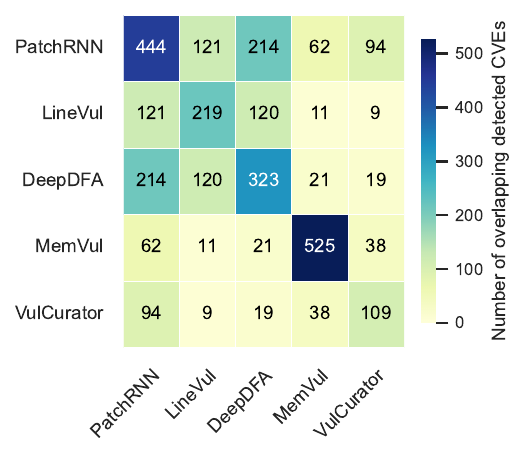}
  \caption{Overlap in detection results among different SOTA methods.}
  \label{fig:matrix}
\end{minipage}
\hfill 
\begin{minipage}[t]{0.52\textwidth}
  \vspace{0pt} 
  \begin{table}[H] 
\centering
  \setlength{\tabcolsep}{2.5pt} 
  \renewcommand{\arraystretch}{1.3}  
\caption{Additional iInformation sources for improving SOTA.}
\footnotesize
\vspace{-0.8em}
\begin{tabular}{@{}c p{0.68\columnwidth} r@{}}
\toprule
\textbf{Source \#} & \textbf{Information sources for vulnerability detection} & \textbf{\#PCVEs} \\
\midrule
1 & Keywords corresponding to CWE information & 71 (63.96) \\
2 & Manually linked artifacts offer additional information & 44 (39.64) \\
3 & Non-textual information offers additional information & 36 (32.43) \\
4 & GitHub PRs offer additional information  & 25 (22.52) \\
5 & Manually assigned labels to issues or PRs provide crucial information & 16 (14.41) \\
\bottomrule
\end{tabular}
\label{tab:rq_3_codebook}  
\end{table}
\end{minipage} 
\end{figure}
Our findings reveal that \emph{DeepDFA} attained the highest applicable recall, achieving a recall value of 0.82 on the applicable dataset. In contrast, \emph{MemVul} achieved the highest all recall, with a value of 0.22 across all PCVEs. Additionally, \emph{DeepDFA} achieved the highest precision of 0.68 among the SOTA methods. \emph{DeepDFA} also achieved the highest F1 score, indicating a strong balance between precision and recall within its supported language scope. Notably, compared to \emph{LineVul}, which achieves an applicable recall of 0.45 and an F1 score of 0.37, \emph{DeepDFA} demonstrates substantially stronger performance within the same source-code setting, suggesting that the integration of control-flow and data-flow representations contributes to improved detection effectiveness. These results highlight the distinction between applicable recall, which measures performance within a model's supported subset, and all recall, which measures coverage across the full PCVE set.

In total, 1,059 unique PCVEs are identified collectively by the SOTA methods after accounting for overlaps, leading to an overall all recall of approximately 0.44 over the 2,402 collected PCVEs. This collective recall exceeds the performance of any individual method, demonstrating that the models capture complementary subsets of vulnerabilities rather than detecting identical PCVEs.

Figure~\ref{fig:matrix} illustrates the overlap in detection results among different SOTA methods. Specifically, \emph{MemVul} detected 525 PCVEs out of a total of 2,402, including 450 unique PCVEs not identified by any other SOTA methods. \emph{DeepDFA} uniquely detected 54 PCVEs, indicating additional but comparatively narrower unique coverage. \emph{PatchRNN} identified 85 unique PCVEs, while \emph{LineVul} detected 46 unique PCVEs. \emph{VulCurator} identified 4 unique PCVEs. These results further confirm that each approach specializes in identifying distinct vulnerability patterns and that measurable complementarity exists across methods.

\begin{tcolorbox}[boxsep=2pt,left=2pt,right=2pt,top=2pt,bottom=2pt]
\textbf{Finding}: SOTA methods exhibit different performance under two recall definitions across 1,213 applicable PCVEs. \emph{DeepDFA} achieves the highest applicable recall and precision, while \emph{MemVul} achieves the highest all recall across the full PCVE set. Collectively, the methods detect 1,059 unique PCVEs, achieving an overall all recall of approximately 0.44. The overlap analysis further indicates that each method specializes in identifying partially distinct subsets of PCVEs, suggesting that integrating these approaches could further improve overall PCVE detection rates.
\end{tcolorbox}

\subsubsection{\textbf{Understanding the reason behind undetected PCVEs.}} \label{rqtwotwo}
Next, we qualitatively analyze the reasons behind the undetected PCVEs and derive insights for improving the performance of SOTA methods. 
Among the 1,213 applicable PCVEs, SOTA methods failed to detect 154. 
Similar to the qualitative analysis described in Sec.~\ref{Qualitative_Analysis_research_method}, for each PCVE, we analyze the lifecycle focusing on the information present in the GitHub artifacts. 
We performed stratified sampling by creating a bucket for each SOTA method using a confidence level of 95\% and 5\% margin of error, containing the set of PCVEs they failed to detect, resulting in 111 samples. We identified five types of information that could be useful for vulnerability detection, as summarized in Table~\ref{tab:rq_3_codebook}.

\paragraph{\textbf{
Source 1: (63.96\%) Keywords that correspond to CWE information.}} 
Our analysis revealed that out of 111 PCVEs, \textbf{71} PCVEs contain feature artifacts with keywords closely associated with CWE keywords. 
For example, when \emph{\href{https://nvd.nist.gov/vuln/detail/CVE-2020-28471}{CVE-2020-28471}} from the \emph{steveukx/properties} project was reported in \emph{\href{https://github.com/steveukx/properties/issues/40}{issue\#40}}, the description highlights a scenario involving ``\emph{prototype attributes being modified via a crafted properties file such that global object state becomes polluted}''.
In this case, the corresponding CWE-1321 (Improperly Controlled Modification of Object Prototype Attributes -- Prototype Pollution) is semantically similar to the vulnerability report, which describes the attack mechanism. 
If utilized, the CVE could potentially be fixed and disclosed earlier.
Similar instances can be found in \emph{\href{https://nvd.nist.gov/vuln/detail/CVE-2020-22203}{CVE-2020-22203}}, where \emph{\href{https://github.com/blindkey/cve_like/issues/6}{issue\#6}}, titled ``\textit{phpcms2008 /yp/job.php genre parameter SQL inject}'', is semantically similar to \emph{\href{https://cwe.mitre.org/data/definitions/89.html}{CWE-89}}, which describes improper neutralization of special elements in SQL commands leading to SQL injection vulnerabilities.
\textbf{Insight: }
One of the SOTA methods, \emph{MemVul} has already leveraged the CWE information in conjunction with GitHub issues to detect vulnerabilities. 
 Our results indicate that the incorporation of various other artifacts along with the CWE information can potentially improve the efficiency of vulnerability classification. 

\begin{figure}[b]
    \centering
    \includegraphics[width=0.7\textwidth]{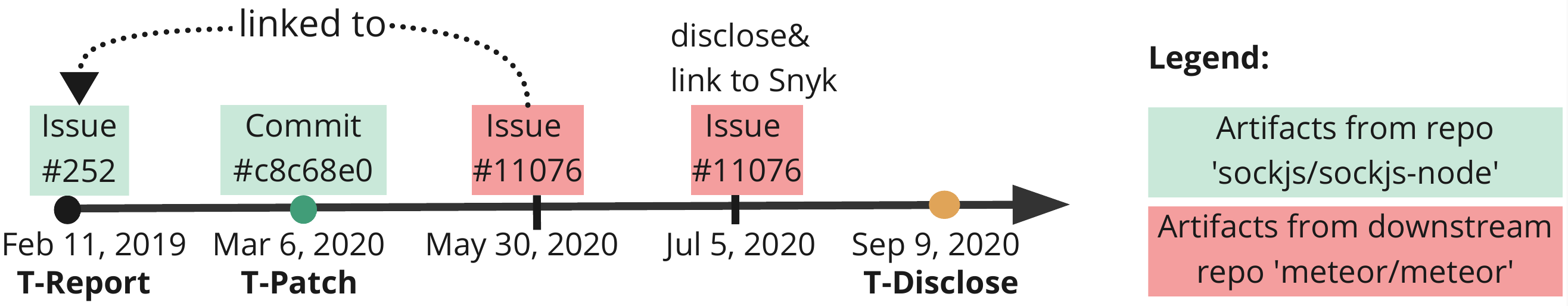}
    \caption{Timeline of CVE-2020-7693.}
    \label{fig:cve-link}
    \vspace{-1em}
\end{figure}

\paragraph{\textbf{Source 2: (39.64\%) Manually linked artifacts offer additional information.}}
During the discussion of resolving vulnerabilities, OSS developers often manually link relevant GitHub artifacts for efficient communication.
Such activity is supported by GitHub's cross-reference feature, including Autolinked references, URLs~\cite{autolink-github}, and cross-referenced issue events~\cite{cross-ref-issue}.
Currently, amongst the SOTA only \emph{VulCurator} utilizes linked artifacts by evaluating only the links between issues and commits. However, linkages between other types of artifacts not restricted to issue-commit links are useful for the identification of vulnerabilities.
For example, we present the timeline of \href{https://nvd.nist.gov/vuln/detail/CVE-2020-7693}{\emph{CVE-2020-7693}} in Figure \ref{fig:cve-link}.
On February 11, 2019, it was reported in \emph{\href{https://github.com/sockjs/sockjs-node/issues/252}{issue\#252}} and later fixed on March 6, 2020 via \emph{\href{https://github.com/sockjs/sockjs-node/commit/c8c68e0}{commit\#c8c68e0}}.
On May 30, 2020, a developer created \emph{\href{https://github.com/meteor/meteor/issues/11076}{issue\#11076}} to discuss the vulnerability's impact, as it causes a denial of service. This issue was cross-referenced with the original \emph{\href{https://github.com/sockjs/sockjs-node/issues/252}{issue\#252}} in the description, providing additional context about the vulnerability.
On July 5, 2020, in the discussion thread, a developer added a link to Snyk, a security advisory, further indicating that the issue is a vulnerability.
Similar example can be identified in \href{https://nvd.nist.gov/vuln/detail/CVE-2016-9581}{\emph{CVE-2016-9581}}. In total, we identified \textbf{44} PCVEs in this category.

\textbf{Insight: } Prior studies have explored the cross-referenced links between code reviews and issues to interpret the intention behind changes~\cite{hirao2019review,li2018issue}.
 However, this data has not been employed for vulnerability detection. The linked artifacts provide a richer source of information for identifying vulnerabilities.
 
\paragraph{\textbf{Source 3: (32.43\%) Non-textual information. } }
We discovered that developers use non-textual artifacts, such as images and proof-of-concept (POC) files, to communicate about vulnerabilities. Specifically, \textbf{36} PCVEs in this category included POC files or snapshots to replicate the vulnerability.  These snapshots contained log traces and configuration details. 
For example, \emph{\href{https://nvd.nist.gov/vuln/detail/CVE-2020-19720}{CVE-2020-19720}} from project \emph{axiomatic-systems/Bento4} was reported in  \emph{\href{https://github.com/axiomatic-systems/Bento4/issues/413}{issue\#413}}, in which the reporter provided a \texttt{poc_input4.zip} file and two annotated snapshots of the source code. 

\textbf{Insight: }The investigation of identifying source code from non-textual sources, including images and videos, has been previously explored~\cite{ott2018deep}. We posit that applying these techniques to enhance information for vulnerability detection holds significant promise.

\paragraph{\textbf{Source 4: (22.52\%) PRs offer additional information.}}
The current SOTA methods do not include GitHub PRs for identifying vulnerabilities, even though PRs are similar to issues and contain discussions and commits, making their inclusion both intuitive and feasible. 
For example, \emph{\href{https://nvd.nist.gov/vuln/detail/CVE-2017-17718}{CVE-2017-17718}} could be identified 
earlier if the corresponding \emph{\href{https://github.com/ruby-ldap/ruby-net-ldap/pull/259}{PR\#259}}, created for discussing and patching the vulnerability, were detected.
We identified \textbf{25} out of 111 PCVEs that could have been detected earlier if their PR information had been included.

\paragraph{\textbf{Source 5: (14.41\%) Manually assigned labels to issues and PRs provide crucial information.}}
GitHub supports the addition of labels to issues and PRs ~\cite{izquierdo2015gila}, which can indicate security-related information and facilitate early vulnerability detection.
For example, \emph{\href{https://nvd.nist.gov/vuln/detail/CVE-2019-16140}{CVE-2019-16140}} was reported on Jan 1, 2018 through \emph{\href{https://github.com/sagebind/isahc/issues/2}{issue\#2}}.
On Feb 2, 2018, the developer fixed the vulnerability in \emph{\href{https://github.com/sagebind/isahc/commit/9e9f1fb44114078c000c78c72e691eeb9e7ac260}{commit\#9e9f1fb}} without any discussion. Later, a “bug” label was assigned on Jul 6, 2019. On Aug 31, 2019, a contributor pointed out that the issue appeared to be an exploitable use-after-free vulnerability and recommended filing a security advisory. Consequently, a “security” label was added on Sep 1, 2019. This sequence illustrates how manually applied labels later revealed the issue’s security relevance. We observed similar patterns in our dataset. For example, security-related labels such as “cat:security” were added in \emph{\href{https://nvd.nist.gov/vuln/detail/CVE-2018-25031}{CVE-2018-25031}} within \emph{\href{https://github.com/swagger-api/swagger-ui/issues/4872}{issue\#4872}}
. In total, we identified \textbf{16} PCVEs exhibiting this label evolution pattern.

  \textbf{Insight:} Prior work has shown that issue and PR labels are meaningful signals
(Izquierdo et al.~\cite{izquierdo2015gila}; Kallis et al.~\cite{kallis2021predicting}),
including security-specific labels
(Buhlmann et al.~\cite{buhlmann2022developers}; Palacio et al.~\cite{palacio2019learning}).
These labels provide strong indicators of security relevance and are valuable input features,
potentially with higher weighting.

\begin{tcolorbox}[boxsep=2pt,left=2pt,right=2pt,top=2pt,bottom=2pt]
\textbf{Finding}: Various types of resources can aid in detecting more vulnerabilities. These include the similarity between issue discussions and CWE keywords, cross-referenced artifacts slightly removed from directly related information, PR-related information, developer-assigned labels, and non-textual information. Future research could enhance vulnerability detection performance by integrating these diverse artifacts.
\end{tcolorbox}
\begin{figure*}[h]
\centering
\includegraphics[width=\textwidth]{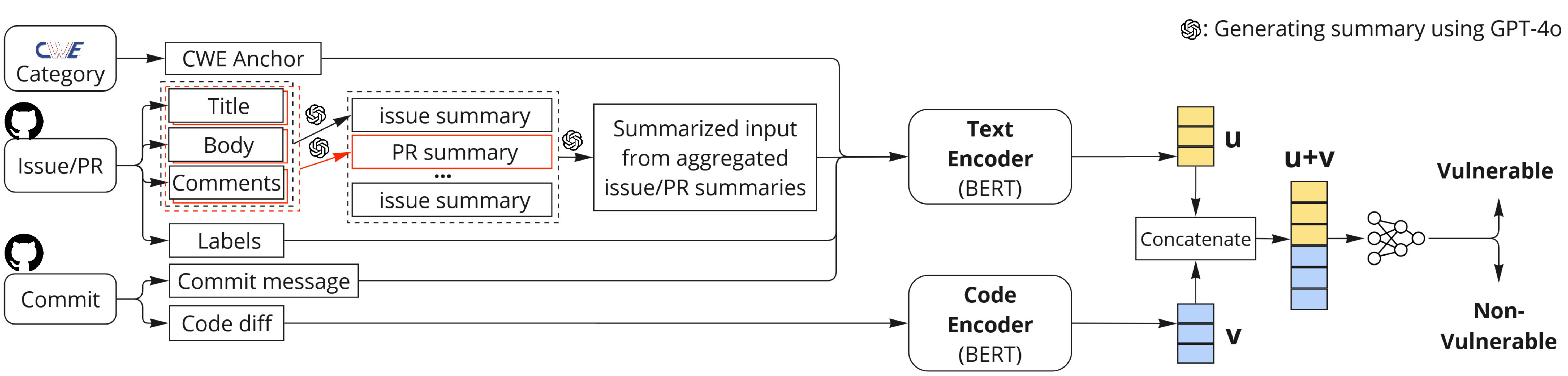}
\caption{Architecture of \projectname.}
\label{fig:model_arch}
\vspace{-1em}
\end{figure*} 

\section{RQ3: \RQthree}

\subsection{Research Method}
This section utilizes the findings from RQ2 to create an improved method for detecting a greater number of PCVEs. We restrict our analysis to text-based artifacts, intentionally excluding non-textual elements like screenshots and attached files, considering their complexity and encouraging future exploration.
We define our method as \projectname, which is short for \textbf{De}t\textbf{e}cting \textbf{P}ro\textbf{tra}cted \textbf{Vul}nerabilities. 
We first introduce the overall architecture of \projectname, followed by the model training and inference details.
We present the architecture in Figure~\ref{fig:model_arch}. 

\subsubsection{\textbf{Input Representation.}}

There are two types of artifacts used in~\projectname, including \emph{text-based} input and \emph{code-based} input.

\textbf{Text-based input.}
In addition to the artifacts that are covered by the SOTA methods,
as presented in Table~\ref{tab:rq_2_artifact_level_performance} column `Artifact type',
we added four new types of artifacts as input, including (1) PRs, (2) labels of both issues and PRs,
(3) linked artifacts in the issue/PR timeline events~\cite{GitHubRESTIssuesTimeline}, and
(4) CWE Anchor, a representation of the CWE category-related information reused from prior work~\cite{pan2022automated}. A PCVE is included in our dataset when it includes at least one commit and one issue or PR, ensuring that every case contains both code evidence and its accompanying development context.

We used the Bidirectional Encoder Representations from Transformers (BERT) architecture to convert text-based input, however, it only supports encoding up to 512 tokens. 
To meet the token constraint and keep as much information as possible, we designed a two-step summarization using \emph{GPT-4o}~\cite{openai2024gpt4o}. 
Earlier research has shown that GPT models perform well in summarization tasks, including the summarization of both generic and vulnerability focused reports~\cite{chintagunta2021medically,prodan2023prompt,reddy2023sumren,yoshimura2023semi,mcclanahan2024chatgpt,althebeiti2023enriching}. More recently, models in the GPT-4 class have been adopted in empirical studies because they offer stronger performance than GPT-3 models and have been evaluated for a variety of security analysis tasks such as vulnerability detection, classification, repair, and broader security assessment~\cite{chen2025chatgpt,fu2023chatgpt}. They have also been used to summarize vulnerability reports and other security relevant technical content~\cite{chopra2024chatnvd}. These developments support our choice of \emph{GPT-4o} for preprocessing long textual artifacts.

In our pipeline, we (1) summarize each GitHub issue and PR individually, including their titles, bodies, and comments, then (2) aggregate these summaries to produce a consolidated overall summary. The prompts used for both steps are presented in Figure~\ref{fig:sum_prompt}.
\begin{figure}[t]
\centering

\begin{subfigure}{0.95\linewidth}
\begin{lstlisting}[style=prompt, basicstyle=\scriptsize\ttfamily]
Objective (O): We are building a deep learning system for vulnerability
classification. You are tasked with summarizing GitHub Issues and Pull
Requests. The summary should be informative and technical.

Style (S): Capture key discussion points.
Tone (T): Reflect sentiment present in the discussion.
Audience (A): A classification model.
Constraint: Ensure no source code appears in the summary.

Input:
<TITLE>
<BODY>
<COMMENTS>
\end{lstlisting}
\caption{Step 1 prompt used to summarize each Issue or PR, including title, body, and comments.}
\end{subfigure}

\vspace{0.6em}

\begin{subfigure}{0.95\linewidth}
\begin{lstlisting}[style=prompt, basicstyle=\scriptsize\ttfamily]
Summarize the text concisely and ensure the summary is brief and strictly
to the point, using as few characters as possible.
Ensure there are no source code components in the summary.

Input:
<STEP_1_SUMMARY_COLLECTION>
\end{lstlisting}
\caption{Step 2 prompt used to generate the aggregated summary.}
\end{subfigure}

\caption{Two-step summarization prompts used prior to \projectname\ encoding.}
\label{fig:sum_prompt}
\end{figure}

Finally, we leveraged BERT, trained with the Robustly Optimized BERT Pretraining Approach (ROBERTA)~\cite{liu2019roberta,devlin2018bert}, incorporating training weights from \emph{MemVul}. In this context, the ROBERTA encoder was fine-tuned for vulnerability prediction using textual data sourced from GitHub issues~\cite{pan2022automated}.

\textbf{Code-based input} 
is the code \texttt{diff} of each commit, including both the referenced commit in NVD, commits included in PRs, and the linked commits in issue/PR discussion threads. 
We leveraged CodeBERT (a variant of BERT) to create the feature vectors~\cite{feng2020codebert}.
Regarding the neural network classifier, we use a one-layer neural network classifier for the classification task.



\begin{table*}[t]
\footnotesize
\centering 
\caption{Summary of Training and Evaluation Data used for \projectname\ by Programming Language.} 
\renewcommand{\arraystretch}{1.2}
\begin{tabular}{ll
                rr
                rr
                rr
                rr}
\toprule
\multirow{4}{*}{\textbf{Artifact}} &
\multirow{4}{*}{\textbf{\begin{tabular}[c]{@{}l@{}}Vulnerability\\Status\end{tabular}}} &
\multicolumn{6}{c}{\textbf{Programming Language}} &
\multirow{4}{*}{\textbf{\begin{tabular}[c]{@{}l@{}}Total\\Train\end{tabular}}} &
\multirow{4}{*}{\textbf{\begin{tabular}[c]{@{}l@{}}Total\\Eval\end{tabular}}} \\
\cmidrule(lr){3-8}
& &
\multicolumn{2}{c}{\textbf{C}} &
\multicolumn{2}{c}{\textbf{C++}} &
\multicolumn{2}{c}{\textbf{Java}} &
& \\ 
\cmidrule(lr){3-4}
\cmidrule(lr){5-6}
\cmidrule(lr){7-8}
& & \textbf{Train} & \textbf{Eval}
  & \textbf{Train} & \textbf{Eval}
  & \textbf{Train} & \textbf{Eval}
  &  &  \\
\midrule

\textbf{Issue} & Vuln
  & 194 & 64
  & 47  & 30
  & 33  & 6
  & 274 & 100 \\
& Non-Vuln
  & 611 & 249
  & 148 & 88
  & 104 & 17
  & 863 & 354 \\

\midrule

\textbf{Commit} & Vuln
  & 605 & 250
  & 175 & 93
  & 187 & 57
  & 967 & 400 \\
& Non-Vuln
  & 1847 & 747
  & 410  & 284
  & 558  & 143
  & 2815 & 1174 \\

\midrule

\textbf{PR} & Vuln
  & 98 & 48
  & 39 & 17
  & 55 & 23
  & 192 & 88 \\
& Non-Vuln
  & 325 & 178
  & 95  & 56
  & 173 & 66
  & 593 & 300 \\

\bottomrule
\end{tabular}
\label{tab:merged_training_eval_data}
\end{table*}


\subsubsection{\textbf{Data collection \& preprocessing.}}\label{rq_3_data_collection_process}
We gathered relevant artifacts from each CVE from three sources:
(a) GitHub artifacts listed in the NVD References section, such as issues, PRs, and commits created prior to disclosure;
(b) cross-referenced artifacts derived from timeline events of issues and PRs; and
(c) artifacts connected through hyperlinks found in the text of (a) using the same pattern matching as prior work~\cite{nguyen2022hermes}.
Similarly, we construct a collection of non-vulnerability datasets for training and evaluation following the procedure in Sec.~\ref{non_vuln_data_rq2}.  
In Table~\ref{tab:merged_training_eval_data}, we present the dataset statistics per programming language.
We further split the full dataset into 80\% training, 10\% validation (covering years 1999–2020), and 10\% testing (covering years 2020–2024). \projectname~achieved an average F1 score of 0.84, precision of 0.84, and an applicable recall of 0.85 across the training and validation sets.

\subsubsection{\textbf{LLM-Based Detector (Chat\emph{GPT-4o}).}}
To complement neural encoders, we investigate whether modern reasoning-capable
LLMs can detect vulnerability signals directly from heterogeneous development
artifacts rather than relying on engineered representations. Prior studies report
that LLMs can infer latent security patterns, reason over causal relationships
in developer discussions, and surface vulnerability indicators that traditional
token-based models may fail to capture~\cite{cheshkov2023evaluation,zhang2024prompt,chen2025chatgpt}.
Thus, evaluating an LLM alongside conventional baselines enables us to assess
whether such reasoning benefits extend to PCVE detection.

To ensure comparability with \projectname, we adopt an artifact preparation process
consistent with our pipeline, with the exception that CWE anchor information
is omitted because the LLM processes raw linguistic cues directly. All associated
evidence, including issue text, PR descriptions, commit messages, and pre-disclosure
patch diffs, are combined into a single consolidated instance per CVE. As in
\projectname, artifacts are limited to supported source languages (C, C++, and Java)
when code content is present.

Chat\emph{GPT-4o} is used in a zero-shot classification setting~\cite{openai2024gpt4o}: for
each consolidated artifact bundle, the model is asked to determine whether it
indicates a security vulnerability. A fixed prompt, as shown in Figure~\ref{fig:chatgpt_prompt}, produces a binary decision
(“Yes’’ / “No’’) similar to other baselines, enabling direct comparison.

 \begin{figure}[t]
\centering
\begin{minipage}{0.90\linewidth}
\begin{lstlisting}[style=prompt, basicstyle=\scriptsize\ttfamily]
You are a security analysis assistant.
Identify whether the provided artifacts indicate a security vulnerability.

Respond strictly in this format:
Answer: Yes / No
Justification: (one concise sentence explaining why)

=== PULL REQUEST DESCRIPTION ===
<PR text, if available>

=== ISSUE DESCRIPTION ===
<Issue text, if available>

=== COMMIT MESSAGE ===
<Commit message, if available>

=== COMMIT DIFF ===
<Commit diff, if available>
\end{lstlisting}
\end{minipage}
\caption{Prompt used to evaluate \emph{GPT-4o} on consolidated PCVE artifacts.}
\label{fig:chatgpt_prompt}
\end{figure}

\subsubsection{\textbf{Evaluation Setup.}}

The evaluation dataset for \projectname\ comprised 235 PCVEs from 2021 until 2024.
For comparative analysis with the SOTA methods, we included additional relevant artifacts as input for other models.
In particular, \emph{MemVul} was evaluated using 3,672 issue/PR records yielding 524 vulnerable and 449 non-vulnerable data points.
Similarly, \emph{VulCurator} was evaluated using 7,325 issue-commit and PR-commit pairs,
\emph{LineVul} processed 430 commits containing valid function data,
\emph{DeepDFA} processed 305 commits containing valid function data,
and \emph{PatchRNN} processed 1,340 commits with required filtering and language constraints.
Table~\ref{tab:rq3_results} provides the vulnerable and non-vulnerable counts for each baseline.
Our experiments were executed on a CentOS 7 environment using an NVIDIA A100-40GB GPU.

\subsection{Results}

In Table~\ref{tab:rq3_results}, we summarize the performance of the SOTA methods, \projectname, and the GPT model. As in RQ2, the baselines remain constrained to the artifacts they support.

\begin{table*}[t]
\footnotesize
\centering
\caption{(RQ3) Data preparation results and the performance of SOTA methods, \emph{GPT-4o} and \projectname\ on the PCVE dataset.}
\label{tab:rq3_results}
\vspace{-1em}

\begin{tabular}{@{}p{1.2cm} p{1.6cm} cc p{.5cm}p{.5cm}p{.5cm}p{.5cm}ccll@{}}
\toprule

\multirow{2}{*}{\textbf{SOTA}} &
\multicolumn{3}{c}{\textbf{Input Data}} &
\multicolumn{8}{c}{\textbf{Evaluation Results}} \\
\cmidrule(lr){2-4} \cmidrule(lr){5-12}

& \multicolumn{1}{l}{\textbf{\begin{tabular}[c]{@{}l@{}}Artifact\\Type\end{tabular}}}
& \multicolumn{1}{l}{\textbf{\#Vuln}}
& \multicolumn{1}{l}{\textbf{\#Non-Vul}}
& \multicolumn{1}{c}{\textbf{TP}}
& \multicolumn{1}{c}{\textbf{FP}}
& \multicolumn{1}{c}{\textbf{FN}}
& \multicolumn{1}{c}{\textbf{TN}}
& \multicolumn{1}{c}{\textbf{Prec.}}
& \multicolumn{1}{c}{\textbf{F1}}
& \multicolumn{1}{c}{\textbf{\begin{tabular}[c]{@{}l@{}}Applicable\\_Recall\end{tabular}}}
& \multicolumn{1}{c}{\textbf{\begin{tabular}[c]{@{}l@{}}All\\_Recall\end{tabular}}}
\\
\midrule

\textbf{MemVul} &
Issue + PR &
\multicolumn{1}{c}{524} &
\multicolumn{1}{c}{449} &
\multicolumn{1}{c}{\textbf{285}} &
\multicolumn{1}{c}{138} &
\multicolumn{1}{c}{239} &
\multicolumn{1}{c}{\textbf{311}} &
\multicolumn{1}{c}{0.67} &
\multicolumn{1}{c}{0.60} &
\multicolumn{1}{c}{0.54} &
\multicolumn{1}{c}{\textbf{0.35}}
\\
\midrule

\textbf{VulCurator} &
\begin{tabular}[c]{@{}l@{}}Issue-Commit\\ + PR-Commit\end{tabular} &
\multicolumn{1}{c}{150} &
\multicolumn{1}{c}{121} &
\multicolumn{1}{c}{75} &
\multicolumn{1}{c}{70} &
\multicolumn{1}{c}{75} &
\multicolumn{1}{c}{51} &
\multicolumn{1}{c}{0.52} &
\multicolumn{1}{c}{0.51} &
\multicolumn{1}{c}{0.50} &
\multicolumn{1}{c}{0.09}
\\
\midrule

\textbf{LineVul} &
\begin{tabular}[c]{@{}l@{}}Source Code\\Function\end{tabular} &
\multicolumn{1}{c}{157} &
\multicolumn{1}{c}{154} &
\multicolumn{1}{c}{75} &
\multicolumn{1}{c}{148} &
\multicolumn{1}{c}{82} &
\multicolumn{1}{c}{6} &
\multicolumn{1}{c}{0.34} &
\multicolumn{1}{c}{0.40} &
\multicolumn{1}{c}{0.48} &
\multicolumn{1}{c}{0.09}
\\
\midrule
\textbf{DeepDFA} &
\begin{tabular}[c]{@{}l@{}}Source Code\\Function\end{tabular} &
\multicolumn{1}{c}{125} &
\multicolumn{1}{c}{71} &
\multicolumn{1}{c}{109} &
\multicolumn{1}{c}{68} &
\multicolumn{1}{c}{16} &
\multicolumn{1}{c}{3} &
\multicolumn{1}{c}{0.62} &
\multicolumn{1}{c}{0.72} &
\multicolumn{1}{c}{0.87} &
\multicolumn{1}{c}{0.13}
\\
\midrule

\textbf{PatchRNN} &
Commit &
\multicolumn{1}{c}{253} &
\multicolumn{1}{c}{277} &
\multicolumn{1}{c}{181} &
\multicolumn{1}{c}{191} &
\multicolumn{1}{c}{72} &
\multicolumn{1}{c}{36} &
\multicolumn{1}{c}{0.49} &
\multicolumn{1}{c}{0.58} &
\multicolumn{1}{c}{0.72} &
\multicolumn{1}{c}{0.22}
\\
\midrule

\textbf{DeeptraVul} &
\begin{tabular}[c]{@{}l@{}}Issue + PR\\ + Commit\end{tabular} &
\multicolumn{1}{c}{134} &
\multicolumn{1}{c}{101} &
\multicolumn{1}{c}{121} &
\multicolumn{1}{c}{\textbf{28}} &
\multicolumn{1}{c}{\textbf{13}} &
\multicolumn{1}{c}{73} &
\multicolumn{1}{c}{\textbf{0.81}} &
\multicolumn{1}{c}{\textbf{0.86}} &
\multicolumn{1}{c}{\textbf{0.90}} &
\multicolumn{1}{c}{0.15}
\\
\midrule

\textbf{LLM (GPT-4o)} &
\begin{tabular}[c]{@{}l@{}}Issue + PR\\ + Commit\end{tabular} &
\multicolumn{1}{c}{134} &
\multicolumn{1}{c}{101} &
\multicolumn{1}{c}{112} &
\multicolumn{1}{c}{63} &
\multicolumn{1}{c}{22} &
\multicolumn{1}{c}{38} &
\multicolumn{1}{c}{0.64} &
\multicolumn{1}{c}{0.72} &
\multicolumn{1}{c}{0.84} &
\multicolumn{1}{c}{0.14}
\\

\bottomrule
\end{tabular}
\vspace{-1em}
\end{table*}

Across 826 PCVEs, the baseline detectors are applicable on different PCVE subsets, as vulnerability evidence appears unevenly across commits, PRs, issues, and related artifacts. After applying artifact and language constraints, \emph{PatchRNN}, \emph{LineVul}, \emph{DeepDFA}, \emph{MemVul}, and \emph{VulCurator} are applicable to 253, 157, 125, 524, and 150 CVEs respectively. \projectname\ and \emph{GPT-4o} apply to 134 CVEs and contribute complementary coverage. Collectively, the models detect 486 of the 826 PCVEs, corresponding to an overall coverage of 0.58 within the constrained RQ3 setting.

Within this applicable subset, detection outcomes vary substantially across models. \emph{MemVul} detects 285 PCVEs, \emph{PatchRNN} detects 181, and both \emph{LineVul} and \emph{VulCurator} detect 75 PCVEs. \emph{DeepDFA} detects 109 PCVEs with an applicable recall of 0.87, demonstrating strong performance within its constrained source-code setting. Compared to \emph{LineVul}, \emph{DeepDFA} identifies a larger portion of vulnerabilities under the same language restrictions, reflecting the benefit of incorporating structural program information when analysis is limited to source code artifacts. \projectname\ detects 121 cases, while \emph{GPT-4o} detects 112. These differences in detection volume are also reflected in precision and F1 performance. \projectname\ achieves a precision of 0.81 and an F1 score of 0.86, followed by \emph{GPT-4o} at 0.64 precision and 0.72 F1, while \emph{DeepDFA} attains 0.62 precision and 0.72 F1. Overall, the results underscore meaningful differences in selectivity and coverage across detectors, with \projectname\ providing the most balanced trade-off between detection breadth and accuracy within its applicable scope.

\subsubsection{\textbf{Generalization Across Programming Languages.}}

\begin{figure}[b]
\centering

\begin{minipage}[t]{0.47\linewidth}
\centering
\vspace{1.2em}
\includegraphics[width=1.1\linewidth]{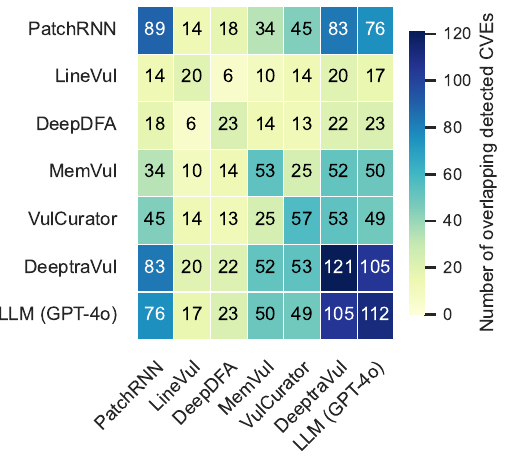}
\captionof{figure}{Overlap in detection results on \projectname{} dataset.}
\label{fig:matrix2}
\end{minipage}
\hfill
\begin{minipage}[t]{0.44\linewidth}
\centering

\begin{table}[H]
\caption{Language-level PCVE detection effectiveness across vulnerability detectors.}
 
\label{tab:lang_effectiveness} 
\vspace{-0.7em}  
  \setlength{\tabcolsep}{6.5pt}
  \renewcommand{\arraystretch}{1.1}\footnotesize
\centering
\begin{tabular}{lcccc}
\hline
\textbf{Model} & \textbf{C} & \textbf{C++} & \textbf{Java} & \textbf{Overall} \\
\hline
VulCurator  & 0.42 & 0.49 & 0.72 & 0.50 \\
LineVul     & 0.33 & 0.78 & 0.82 & 0.48 \\
DeepDFA     & 0.38 & 0.92 & - & 0.87 \\
PatchRNN    & 0.77 & 0.80 & 0.33 & 0.72 \\
DeeptraVul  & 0.91 & 0.95 & 0.54 & 0.90 \\
GPT-4o  & 0.89 & 0.82 & 0.63 & 0.84 \\
\hline
\end{tabular}
\end{table}

\vspace{-1em}

\begin{table}[H] 
  \caption{\projectname\ AUC performance under different artifact configurations.}
  
\label{tab:auc_delta_addback}
\vspace{-0.7em}   
  \setlength{\tabcolsep}{8.3pt}
  \renewcommand{\arraystretch}{1.1} 
  \centering
\footnotesize
\begin{tabular}{lcc}
\hline
\textbf{Features} & \textbf{AUC} & \textbf{$\Delta$AUC} \\
\hline
Code                     & 0.49 & +0.32 \\
Issue + PR               & 0.64 & +0.17 \\
Commit (Msg + Diff)      & 0.66 & +0.15 \\
Issue + PR + Commit Msg  & 0.78 & +0.04 \\
All Features     & 0.81 & ---   \\
\hline
\end{tabular}

\end{table}

\end{minipage}

\end{figure}

Table~\ref{tab:lang_effectiveness} reports language-level detection effectiveness for the five evaluated models that expose language metadata. Overall, the results show substantial variation in performance across programming languages.

\projectname~achieves the strongest detection capability for C with a recall detection score of 0.91 and for C++ with a score of 0.95, outperforming all other models on native-code languages, while maintaining moderate effectiveness for Java at 0.54. \emph{GPT-4o} ranks second in terms of balanced performance, with high detection effectiveness for C at 0.89 and C++ at 0.82, and comparatively stronger performance on Java at 0.63. \emph{PatchRNN} exhibits a similar trend, achieving strong effectiveness for C at 0.77 and C++ at 0.80, but showing a pronounced decline for Java at 0.33, indicating that its learned representations generalize more effectively to vulnerability patterns common in low-level languages, consistent with its predominantly C and C++ training data. \emph{DeepDFA} achieves strong effectiveness on C++ at 0.92 and lower performance on C at 0.31. Since \emph{DeepDFA} was originally designed and evaluated for C and C++ code, it is applied here within its supported language scope and therefore does not report results for Java.

In contrast, \emph{LineVul} and \emph{VulCurator} demonstrate a different pattern. \emph{LineVul} achieves its strongest effectiveness on Java at 0.82, followed by C++ at 0.78, while its performance on C at 0.33 is substantially lower. Similarly, \emph{VulCurator} attains its highest effectiveness on Java at 0.72, followed by C++ at 0.49 and C at 0.42.

These findings highlight systematic differences in cross-language generalization, with \emph{LineVul} and \emph{VulCurator} performing better on Java, and \emph{GPT-4o}, \emph{PatchRNN}, \emph{DeepDFA}, and \projectname~exhibiting stronger performance on C and C++. \textbf{Language sensitivity emerges as a consistent factor in model performance, shaped by architectural decisions, training composition, and language-specific vulnerability patterns. No detector exhibits universally strong behavior across languages.}

\subsubsection{\textbf{Review of Misclassification Cases.}}
The observed variability in detector performance across languages extends beyond aggregate language-level metrics and is reflected in the models' decision behavior. Specifically, the same architectural and data-driven factors that shape language sensitivity also influence how models balance coverage and reliability. Although the baseline detectors, such as \emph{MemVul} and \emph{PatchRNN}, identify more CVEs in absolute terms, this is largely because they rely on a single artifact type. \emph{PatchRNN} uses commits and \emph{MemVul} uses issues, whereas \projectname\ requires the presence of at least one commit linked to an issue or a PR, which naturally reduces the number of detectable CVEs. Consequently, approaches that rely on a single artifact type exhibit higher false positive and false negative rates. By comparison, both \projectname\ and \emph{GPT-4o} produce markedly fewer false positives, indicating that they learn more selective and accurate decision boundaries. This behavior suggests that the reasoning processes of \projectname\ and \emph{GPT-4o} are more effective at distinguishing true vulnerability evidence from non-security signals, even when their overall detection volumes differ.

Therefore, to enable a fair comparison, we restrict the evaluation to the subset of 134 PCVEs for which \projectname\ was applicable and for which all models had access to the necessary artifacts. Within this controlled setting, the performance distinction becomes clearer: \projectname\ detects 121 PCVEs and \emph{GPT-4o} detects 112 PCVEs, while the baseline methods detect fewer cases within this PCVE subset.

Figure \ref{fig:matrix2} visualizes these overlaps, illustrating both shared coverage and remaining gaps across detectors. This contrast shows that \projectname’s advantage is most apparent when all models are required to reason over the same evidence rather than over their preferred artifact types. 
To understand the remaining gaps, we analyse the cases \projectname~failed to detect. Eight of these involve non-textual artifacts, such as compressed test suites, proof of concept exploit files, or screenshots highlighting affected code regions, that are not currently supported by the 
pipeline. \textbf{This observation suggests that extending the model to handle multi-modal inputs could recover a notable portion of these missed detections.}
 
Four of the remaining cases point to limitations in evidence availability rather than model capacity alone. Two of these were also missed by \emph{GPT-4o}, indicating that the underlying artifacts provide limited or ambiguous security cues. To probe this ambiguity, we previously asked \emph{GPT-4o} to justify its predictions, as illustrated in Figure~\ref{fig:chatgpt_prompt}. The LLM response shows a recurring pattern in which corrective changes are presented as benign stability or feature improvements, masking the fact that their existence reflects a prior vulnerability. For example, \href{https://nvd.nist.gov/vuln/detail/CVE-2021-4289}{\emph{CVE-2021-4289}}
 in \href{https://github.com/openmrs/openmrs-module-referenceapplication/pull/89}{\emph{PR\#89}}
 introduces encoding of the \texttt{AppId} field to prevent XSS, which the model summarized as \textit{“addressing an XSS vulnerability”}. The change was interpreted as a positive remediation action rather than as vulnerability-related evidence. Likewise, \href{https://nvd.nist.gov/vuln/detail/CVE-2022-4963}{\emph{CVE-2022-4963}}
 in \href{https://github.com/folio-org/spring-module-core/pull/39}{\emph{PR\#39}}
 replaces unsafe SQL construction with prepared statements and was summarized as a maintenance improvement, and treated as not vulnerability-related. In both cases, positive framing in the model summaries led \projectname~and \emph{GPT-4o} to infer quality improvement rather than security relevance. \textbf{Security-motivated code changes may not identified as vulnerability-related when described using positive, maintenance-oriented language.}

The final \projectname~miss illustrates a different issue. A representative example is \href{https://nvd.nist.gov/vuln/detail/CVE-2022-21657}{\emph{CVE-2022-21657}}, in \href{https://github.com/envoyproxy/envoy/pull/630/commits}{\emph{commit\#630}}, where routing and configuration logic is restructured across multiple files with little explanatory text, as in commit messages. Although the vulnerability corresponds to CWE 295 (Improper Certificate Validation), vulnerability evidence exists but is embedded in many code edits, demonstrating that the presence of changed code does not necessarily translate into accessible narrative signals for automated reasoning. This motivates a need to investigate which artifact types and representations most influence \projectname’s behaviour. Accordingly, we analyze how different artifact types and representation choices affect \projectname’s ability to surface actionable vulnerability evidence. 
\subsubsection{\textbf{Evaluation of \projectname~Under Different Settings.}}

To examine how artifact availability influences detection performance, we evaluated \projectname~under several feature configurations. These included: (1) code-only input, following the setting of \emph{LineVul} and \emph{DeepDFA}; (2) issue and PR descriptions, as used by \emph{MemVul}; (3) commit messages paired with code diffs, consistent with \emph{PatchRNN}; (4) issue, PR, and commit message texts combined; and (5) a full configuration integrating issue descriptions, PR descriptions, commit messages, and code changes.

\textbf{Insight 1: Multi-artifact evidence is critical for performance.}  
On average, removing artifact types reduces AUC by approximately 27\% relative to the full-evidence configuration, indicating the importance of integrating information from multiple sources. As shown in Table~\ref{tab:auc_delta_addback}, the full configuration achieves the highest performance, with an AUC of 0.81.

\textbf{Insight 2: Code-only evidence is insufficient.}  
The code-only configuration yields an AUC of 0.49, representing a decline of roughly 40\%, indicating that diffs alone provide insufficient context for distinguishing vulnerable from benign updates.

\textbf{Insight 3: Textual context substantially improves detection.}  
Issue and PR descriptions yield an AUC of 0.64, a reduction of 21\% relative to full evidence, while commit messages paired with diffs reach 0.66, a 19\% reduction. Combining issue, PR, and commit text achieves an AUC of 0.78, only 4\% below the full configuration, capturing most contextual signal even without code changes.

Overall, these results show that \projectname~achieves its strongest performance when multiple artifact types are available, and that removing any single artifact degrades detection performance in predictable ways.

\begin{tcolorbox}[boxsep=2pt,left=2pt,right=2pt,top=2pt,bottom=2pt]
\textbf{Finding}: \projectname~achieves the strongest overall detection performance among evaluated models, with 0.81 precision, 0.90 recall, and an F1 score of 0.86. Its effectiveness remains stable across C and C++, and its feature evaluation indicates that performance improves when multiple artifact sources are provided. Our qualitative analysis also suggests that supporting non-textual artifacts could further reduce the remaining missed PCVEs.
\end{tcolorbox}
\section{Discussion}

\subsection{For SE Researchers}
This section highlights key areas where software engineering researchers can investigate and mitigate delays in the vulnerability lifecycle. The findings are also relevant to SE practitioners aiming to improve practices related to vulnerability reporting, detection, prioritization, and resolution. 

\begin{enumerate} [leftmargin=*]
    \item  \textbf{Delays in CVE Reporting.} Timely disclosure of vulnerabilities is essential to minimize exposure and allow users to apply patches immediately. Our analysis shows that \textbf{46.27\%} of PCVEs experienced delays during the NVD disclosure process, leaving systems vulnerable even after the fixes were implemented. 
    
    Many relevant artifacts, such as reasons for delays or related discussions, were missing from the NVD references. As a result, it becomes difficult to trace the full vulnerability timeline and understand the decisions behind these delays. For example, \href{https://nvd.nist.gov/vuln/detail/CVE-2018-10757}{\emph{CVE-2018-10757}} was patched in \href{https://github.com/dukereborn/cmum/commit/c89158ec646c4e8e95587b650f6fd86b502ff8b5}{\emph{commit\#c89158e}} on April 26, 2015, but the CVE was not published until May 5, 2018, over three years later. The NVD lists the patch reference, but provides no explanation or metadata to clarify the delay.
    
    In other cases, the delays were more transparent. \href{https://nvd.nist.gov/vuln/detail/CVE-2020-35132}{\emph{CVE-2020-35132}} was delayed due to confusion around the reporting process. In the related \href{https://github.com/leenooks/phpLDAPadmin/issues/130}{\emph{issue\#130}}, a contributor asked if a CVE had been assigned. The reporter replied: \emph{“I don't think so — do you know how I can go about this? I apologize for the lack of knowledge / context here''}. The reporter also mentioned submitting a private ticket on Launchpad but was unsure how to request a CVE. Another contributor noted the time it takes to gather the necessary information and the risk of duplicate assignments. This shows how limited familiarity with the CVE process can unintentionally delay disclosure, even when issues are reported responsibly. This highlights the need for automated tools that monitor repository activity, reconstruct vulnerability histories, and flag disclosure delays. 
    
    To better understand these delays, it is useful to identify the common characteristics of CVE (i.e., patterns of delayed CVEs) that are associated with delays.
    For example, certain categories of CVEs may be more prone to delays due to their complexity, reporting practices, or lack of automation support. Understanding these patterns can inform the design of tools for earlier detection and intervention. Below are examples of delayed CVEs that indicate common patterns of delays.

    \begin{enumerate} [leftmargin=3em]
    \item \textbf{Memory corruption issues}, such as \href{https://nvd.nist.gov/vuln/detail/CVE-2016-9581}{\emph{CVE-2016-9581}}, may involve complex debugging and verification, which can slow down disclosure. \item \textbf{XSS vulnerabilities}, like \href{https://nvd.nist.gov/vuln/detail/CVE-2020-18475}{\emph{CVE-2020-18475}}, may be underreported due to lack of proper documentation or awareness. \item \textbf{Authentication flaws}, such as \href{https://nvd.nist.gov/vuln/detail/CVE-2020-19268}{\emph{CVE-2020-19268}}, often suffer from insufficient automation and visibility, contributing to delays. \end{enumerate}
    
    Beyond technical classifications, the reporting ecosystem itself plays a critical role in shaping disclosure timelines. Informal sources such as mailing lists and issue trackers often provide essential context missing from official databases. For instance, \href{https://nvd.nist.gov/vuln/detail/CVE-2017-20005}{\emph{CVE-2017-20005}} was discussed in \href{https://trac.nginx.org/nginx/ticket/1368}{\emph{Nginx issue \#1368}} and later referenced in the \href{https://lists.debian.org/debian-lts-announce/2021/06/msg00009.html}{\emph{Debian LTS mailing list}}. Tools such as MITRE’s CVE Services API~\cite{cveservices} and GitHub Security Advisories~\cite{githubAdvisories} support structured submission workflows, but broader adoption and better integration are needed to improve the timeliness and completeness of disclosures.

    \item \textbf{Finer-Grained Analysis of Early Vulnerability Detection.} In line with the need for transparency in disclosure timelines, our study did not investigate the exact timing of artifact creation relative to vulnerability resolution. This limits our understanding of when vulnerabilities are first detected. Future work should focus on reconstructing CVE timelines at a finer granularity to assess how early existing tools and techniques can detect vulnerabilities. Some vulnerabilities in our dataset remained unresolved across multiple release cycles, suggesting earlier detection was possible but missed. Leveraging historical repository data with predictive modeling may help identify vulnerabilities earlier. Time-series analyses across ecosystems could uncover patterns in how vulnerabilities emerge, enabling the development of early-warning systems that support more proactive detection and response strategies.

    \item  \textbf{Vulnerability Prioritization.} Delays are not solely the result of external reporting gaps. Internal project decisions also play a role. Our data shows that \textbf{19.4\%} of delayed PCVEs were due to \emph{misjudgment of severity}, where vulnerabilities were deprioritized or overlooked. For example, \href{https://nvd.nist.gov/vuln/detail/CVE-2018-10727}{\emph{CVE-2018-10727}} remained unresolved despite active development activity, while \href{https://nvd.nist.gov/vuln/detail/CVE-2020-36420}{\emph{CVE-2020-36420}} involved prolonged use of a known vulnerable package. A developer\textquotesingle s comment in a discussion \href{ https://bugs.gentoo.org/755896}{\emph{thread}}, \textit{\textquotedblleft It would've been appreciated if you could've given us a heads up / let us in security@ and treecleaner@ know about your intentions\textquotedblright}, underscores the lack of internal visibility. To address this issue, development workflows should incorporate automated risk scoring systems based on objective security metrics~\cite{parente2025frape}. Such tools can help teams identify and prioritize critical vulnerabilities more effectively, reducing delays caused by oversight or misclassification.

    \item \textbf{Investing in Tools for Patch Quality.} Even when vulnerabilities are detected and prioritized, the patching process itself can contribute to delays. In \textbf{7.64\%} of PCVEs, initial fixes were incomplete and required multiple revisions. For example, \href{https://nvd.nist.gov/vuln/detail/CVE-2020-18705}{CVE-2020-18705} required follow-up fixes less than two months after the initial patch in  \href{https://github.com/quokkaproject/quokka/issues/676}{\emph{issue\#676}}. These cases often result from time pressure, limited testing, or insufficient peer review. To mitigate such delays, future research should explore automated patch validation tools that help identify incomplete fixes before deployment. Techniques that improve first-attempt patch success rates can streamline the remediation process and reduce the window of exposure caused by ineffective patches.

 \begin{figure}[t]
\centering
\begin{tikzpicture}[scale=0.65, every node/.style={transform shape},
  node distance=1.2cm,
  function/.style={draw, rounded corners, fill=gray!5, minimum width=4.4cm, minimum height=0.9cm, font=\small\ttfamily, align=center},
  vuln/.style={draw, rounded corners, fill=red!20, minimum width=4.4cm, minimum height=0.9cm, font=\small\ttfamily\bfseries, align=center},
  label/.style={font=\bfseries, align=right}
]

\node[function] (A) at (0,0) {handleUpload()\\\textit{UserInputHandler.java}};
\node[function, below=of A] (B) {compress()\\\textit{Compressor.java}};
\node[function, below=of B] (C) {compress()\\\textit{Snappy.java}};
\node[function, below=of C] (D) {encodeBlock()\\\textit{SnappyEncoder.java}};
\node[vuln, below=of D] (E) {shuffle() (vul)\\\textit{Shuffle.java}};

\node[label, left=1.8cm of A] {Application Code};
\node[label, left=1.8cm of C] {Library API};
\node[label, left=1.8cm of E] {Internal Logic};

\draw[-{Latex}] (A) -- (B);
\draw[-{Latex}] (B) -- (C);
\draw[-{Latex}] (C) -- (D);
\draw[-{Latex}] (D) -- (E);

\end{tikzpicture}
\caption{Function-level dependency chain showing indirect reachability to the vulnerable \texttt{shuffle()} method (CVE-2023-34453). Layer labels on the left indicate architectural separation, while boxes show the function path.}
\label{fig:vuln-call-graph}
\end{figure}
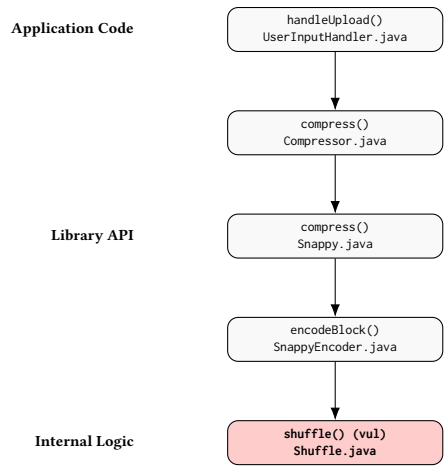

    \item \textbf{Leveraging Image-Based Code Analysis for Vulnerability Detection.} Most vulnerability detection techniques rely on static, text-based code analysis. However, these approaches can miss important structural patterns that are easier to detect visually. Image-based code analysis provides a promising alternative by converting source code into visual formats, such as control-flow graphs or architectural diagrams, and applying computer vision techniques like convolutional neural networks (CNNs). When combined with traditional static analysis, these methods can improve detection accuracy and reduce false positives. For example, visualizing data flow or function-level dependencies may reveal vulnerabilities spread across multiple files that text-based tools might miss. Consider \href{https://nvd.nist.gov/vuln/detail/CVE-2020-18705}{CVE-2023-34453}, which involved an integer overflow in the \texttt{shuffle(int[] input)} method within the snappy-java library. Although this method might not be directly invoked in application code, its reachability can still pose a risk. For example, function-level dependency analysis could reveal that this method is indirectly reachable via internal library calls triggered by standard compression routines. While text-based scans may overlook such indirect paths, a visual dependency graph can help uncover how user-controlled input could propagate through these internal layers and eventually reach the vulnerable method. As illustrated in Figure~\ref{fig:vuln-call-graph}, such a graph can make hidden call relationships explicit, highlighting both direct and indirect paths to vulnerable functions.
    
    Our review of CVE-related GitHub repositories shows that visual artifacts are already used in practice. Many vulnerability reports include screenshots to demonstrate how the issue occurs or to point out the exact vulnerable functions. For instance, \href{https://nvd.nist.gov/vuln/detail/CVE-2017-17480}{\emph{CVE-2017-17480}}, discussed in \href{https://github.com/uclouvain/openjpeg/issues/1044}{\emph{issue\#1044}} of the \emph{uclouvain/openjpeg} repository, contains images that help clarify the vulnerability. These findings highlight the value of incorporating visual information into automated vulnerability analysis. Using such visual context can make vulnerability reports more understandable and support the development of more accurate and practical detection tools.

    \item \textbf{Process Standardization and Ecosystem Comparisons.} Finally, delays often stem from inconsistencies in reporting workflows across software ecosystems. Manual triage by CVE authorities contributes to reporting bottlenecks, particularly in resource-constrained projects. Future research should investigate semi-automated classification methods to support faster CVE processing.
    A comparative study of proprietary and open-source ecosystems could help identify process models that promote timely reporting. Practices from structured proprietary pipelines may be adapted to open-source environments. Additionally, standardizing reporting frameworks to account for team structure, project scale, and communication patterns can improve consistency in disclosure practices. Improving coordination between development and security teams and integrating timelines into issue tracking can further reduce delays and enhance vulnerability response efforts.

\end{enumerate}

\subsection{For OSS Practitioners And Developers}
This section provides actionable strategies for open-source practitioners and maintainers to improve vulnerability management and reduce delays in patching and disclosure. The recommendations are informed by recurring patterns observed in delayed CVEs.

\begin{enumerate}[leftmargin=*]
\item \textbf{Improving Vulnerability Identification Practices.} Prior studies label components without CVE assignments as non-vulnerable, although this does not confirm that the code is safe. Croft et al.\ show that real-world datasets lack ground-truth labels for non-vulnerable code and therefore cannot validate their correctness \cite{croft2023data}. Similarly, \emph{CVEfixes} collects CVE-linked vulnerable code but does not verify non-vulnerable samples \cite{bhandari2021cvefixes}. This limitation also appears in VCMATCH, where negative samples are randomly selected commits treated as non-vulnerable~\cite{wang2022vcmatch}. Likewise, \emph{DiverseVul} labels unchanged and post-patch versions of modified functions as non-vulnerable based on fixing commits, without validating these functions \cite{chen2023diversevul}. These practices show that non-vulnerable code identification relies on heuristics that may not hold.
    
    For OSS practitioners and developers, these findings highlight the importance of avoiding implicit assumptions that unreported or unpatched code is non-vulnerable. Practitioners are encouraged to treat vulnerability status as evolving and incomplete, particularly for widely reused components and dependencies. Maintaining clear documentation of security-relevant changes, adopting structured vulnerability reporting practices, and proactively auditing critical or high-impact code paths can help reduce reliance on informal heuristics. Furthermore, OSS projects may benefit from explicitly distinguishing between code that has been reviewed for security and code whose vulnerability status remains unknown, thereby improving transparency for downstream users and integrators.

    \item \textbf{Enhancing Security Expertise Through Training and Reviews.} Delays in addressing vulnerabilities often stem from misjudgments in severity. Developers may fail to prioritize or properly address security issues due to limited expertise. For example, \href{https://nvd.nist.gov/vuln/detail/CVE-2021-23418}{\emph{CVE-2021-23418}} remained unresolved in \href{https://github.com/nicolargo/glances/}{\emph{Glances}} despite active development, and \href{https://nvd.nist.gov/vuln/detail/CVE-2020-27794}{\emph{CVE-2020-27794}} experienced delays due to prolonged discussion without resolution. To mitigate such cases, developers should participate in regular security training, use AI-assisted auditing tools, and engage in periodic project-wide security reviews. Peer-reviewed security assessments can also help improve classification accuracy and reduce delays caused by misjudgment.

    \item \textbf{Incentivizing Security Contributions to Address Resource Constraints.} A lack of dedicated contributors is another common barrier to timely patching. For instance, \href{https://nvd.nist.gov/vuln/detail/CVE-2020-15163}{\emph{CVE-2020-15163}} in \href{https://github.com/theupdateframework/python-tuf/}{\emph{python-tuf}} faced delays due to contributor shortages. Approximately \textbf{3\%} of PCVEs were affected by such workforce limitations. Open-source communities should introduce incentives such as security-focused bug bounties, mentorship programs, and recognition systems to attract contributors. Automated security checks in PRs can also streamline the review process.

    \item \textbf{Strengthening Security Policy Enforcement and Public Disclosure.} Some vulnerabilities are fixed silently without proper public disclosure, as in the case of \href{https://nvd.nist.gov/vuln/detail/CVE-2017-7495}{\emph{CVE-2017-7495}}. Others, like \href{https://nvd.nist.gov/vuln/detail/CVE-2021-32265}{\emph{CVE-2021-32265}}, were disclosed long after being patched. Projects should enforce structured security policies that mandate public disclosure of all security-related fixes. This includes integrating compliance checks into CI/CD pipelines and defining clear timelines and communication procedures for disclosure.

    \item \textbf{Ensuring Timely Patch Deployment Through Rigorous Validation.} Patches that undergo multiple revisions delay vulnerability resolution. An example is \href{https://nvd.nist.gov/vuln/detail/CVE-2020-26299}{\emph{CVE-2020-26299}}, which took over 500 days to fully resolve. Projects should implement rigorous validation practices, including mandatory peer reviews, automated test coverage for patches, and post-deployment monitoring of patch effectiveness.

    \item \textbf{Mitigating Risks from Under-Maintained and Abandoned Projects.} Inactive or abandoned projects often leave vulnerabilities unpatched, as seen in \href{https://nvd.nist.gov/vuln/detail/CVE-2016-11014}{\emph{CVE-2016-11014}} in \href{https://github.com/cybersecurityworks/Disclosed}{\emph{Disclosed}}. Projects should designate backup maintainers and integrate tooling that flags unmaintained repositories with known vulnerabilities. Package managers and dependency platforms can assist by warning users when using outdated or insecure libraries.
\end{enumerate} 
\subsection{For CVE Authorities (e.g., NVD)}
\begin{enumerate}
    \item \textbf{Streamlining Vulnerability Reporting.} Delays in CVE reporting increase exposure risk. Although GitHub recommends security policy setup, adoption remains inconsistent. \href{https://nvd.nist.gov/vuln/detail/CVE-2020-22781}{\emph{CVE-2020-22781}} offers a positive example, where \href{https://github.com/ether/etherpad-lite}{\emph{ether/etherpad-lite}} encouraged private reporting through a dedicated channel~\cite{etherpadSecurity2024}. CVE authorities should enforce standardized reporting workflows, clearly define submission responsibilities, and provide tools to help projects report vulnerabilities efficiently.

    \item \textbf{Enhancing Tooling Support.} High-recall detection techniques often result in developer fatigue due to false positives. Detection algorithms should be refined to balance precision and recall. Integrating triage automation into CVE pipelines would expedite classification and reduce manual workload.

    By improving workflows, promoting transparency, and expanding tooling support, CVE authorities can significantly enhance the overall responsiveness of the vulnerability disclosure ecosystem. 

\end{enumerate} 
\begin{tcolorbox}[boxsep=2pt,left=2pt,right=2pt,top=2pt,bottom=2pt] 
\textbf{Finding:} Delays in vulnerability disclosure often arise due to unclear
communication, missing context, and limited automation. To address this:

(1) Researchers should investigate these causes and develop improved detection techniques, automated severity scoring, and early
patch validation tools.

(2) OSS developers require stronger security training, AI-assisted reviews,
clearer disclosure practices, and better dependency and maintenance management.

(3) CVE authorities should standardize reporting
workflows, provide secure submission channels, and automate triage to
improve consistency and response times.
\end{tcolorbox}
\section{Conclusion} 
In this study, we investigated the issue of PCVEs and analyzed the reasons behind delayed patching and disclosure. Through a qualitative analysis, we identified \textbf{eight} key causes contributing to these delays: lack of active maintenance, misjudgment of relevance and severity, disagreement on resolution, incomplete or insufficient fixes, resource constraints, lack of expertise, delayed NVD disclosure, and unknown causes.

To assess the effectiveness of existing vulnerability detection techniques, we evaluated SOTA methods on a curated PCVE dataset. Our results indicate that these methods successfully detect only \textbf{44\%} of the instances, demonstrating their limitations in identifying protracted vulnerabilities. This suggests that conventional models, which typically rely on a single artifact, fail to capture the complexities of PCVEs. The results emphasize the need for more comprehensive approaches that incorporate multiple artifacts to improve detection accuracy.

To address these limitations, we introduce \projectname, a model designed to enhance vulnerability detection. Unlike traditional approaches, \projectname~ integrates multiple software artifacts, to provide a richer context for analysis. Our experimental results show that \projectname\ significantly outperforms existing SOTA methods, including recent LLM-based approaches, achieving a higher detection rate on a subset of PCVEs. These findings highlight the importance of leveraging multiple artifacts for better vulnerability detection, ultimately aiding in the timely identification and mitigation of security risks.

\begin{acks} 
We acknowledge the support of the Natural Sciences and Engineering Research Council of Canada (NSERC): [RGPIN-2021-03538] and [RGPIN-2021-03969].
\end{acks}
\bibliographystyle{ACM-Reference-Format}

\begin{thebibliography}{82}


\ifx \showCODEN    \undefined \def \showCODEN     #1{\unskip}     \fi
\ifx \showDOI      \undefined \def \showDOI       #1{#1}\fi
\ifx \showISBNx    \undefined \def \showISBNx     #1{\unskip}     \fi
\ifx \showISBNxiii \undefined \def \showISBNxiii  #1{\unskip}     \fi
\ifx \showISSN     \undefined \def \showISSN      #1{\unskip}     \fi
\ifx \showLCCN     \undefined \def \showLCCN      #1{\unskip}     \fi
\ifx \shownote     \undefined \def \shownote      #1{#1}          \fi
\ifx \showarticletitle \undefined \def \showarticletitle #1{#1}   \fi
\ifx \showURL      \undefined \def \showURL       {\relax}        \fi
\providecommand\bibfield[2]{#2}
\providecommand\bibinfo[2]{#2}
\providecommand\natexlab[1]{#1}
\providecommand\showeprint[2][]{arXiv:#2}

\bibitem[vul(2018)]%
        {vuldisclosure}
 \bibinfo{year}{2018}\natexlab{}.
\newblock \bibinfo{title}{ISO/IEC 29147:2018: Security techniques -
  Vulnerability disclosure}.
\newblock
\newblock
\urldef\tempurl%
\url{https://www.iso.org/standard/72311.html}
\showURL{%
\tempurl}


\bibitem[aut(2024)]%
        {autolink-github}
 \bibinfo{year}{2024}\natexlab{}.
\newblock \bibinfo{title}{Autolinked references and URLs}.
\newblock
  \bibinfo{howpublished}{\url{https://docs.github.com/en/rest/using-the-rest-api/issue-event-types?apiVersion=2022-11-28\#cross-referenced}}.
\newblock
\newblock
\shownote{Accessed: 2024-06-03}.


\bibitem[cro(2024)]%
        {cross-ref-issue}
 \bibinfo{year}{2024}\natexlab{}.
\newblock \bibinfo{title}{cross-referenced issue event type}.
\newblock
  \bibinfo{howpublished}{\url{https://docs.github.com/en/get-started/writing-on-github/working-with-advanced-formatting/autolinked-references-and-urls}}.
\newblock
\newblock
\shownote{Accessed: 2024-06-03}.


\bibitem[GHA(2024)]%
        {GHArchive}
 \bibinfo{year}{2024}\natexlab{}.
\newblock \bibinfo{title}{GH Archive: A Public Dataset of GitHub Activity}.
\newblock \bibinfo{howpublished}{\url{https://www.gharchive.org/}}.
\newblock
\newblock
\shownote{Accessed: 2024-04-01}.


\bibitem[NVD(2024)]%
        {NVD}
 \bibinfo{year}{2024}\natexlab{}.
\newblock \bibinfo{title}{{National Vulnerability Database (NVD)}}.
\newblock
\newblock
\urldef\tempurl%
\url{https://nvd.nist.gov/}
\showURL{%
\tempurl}
\newblock
\shownote{Accessed: 2024-04-01}.


\bibitem[DAT(2024)]%
        {DATANVD}
 \bibinfo{year}{2024}\natexlab{}.
\newblock \bibinfo{title}{NVD Data Feed}.
\newblock \bibinfo{howpublished}{\url{https://nvd.nist.gov/vuln/data-feeds}}.
\newblock
\newblock
\shownote{Accessed: 2024-04-01}.


\bibitem[rep(2026)]%
        {replicationpkg}
 \bibinfo{year}{2026}\natexlab{}.
\newblock \bibinfo{booktitle}{\emph{Replication package}}.
\newblock
\urldef\tempurl%
\url{https://zenodo.org/records/17970073}
\showURL{%
\tempurl}


\bibitem[Alexopoulos et~al\mbox{.}(2022)]%
        {alexopoulos2022long}
\bibfield{author}{\bibinfo{person}{Nikolaos Alexopoulos},
  \bibinfo{person}{Manuel Brack}, \bibinfo{person}{Jan~Philipp Wagner},
  \bibinfo{person}{Tim Grube}, {and} \bibinfo{person}{Max M{\"u}hlh{\"a}user}.}
  \bibinfo{year}{2022}\natexlab{}.
\newblock \showarticletitle{How Long Do Vulnerabilities Live in the Code? A
  $\{$Large-Scale$\}$ Empirical Measurement Study on $\{$FOSS$\}$ Vulnerability
  Lifetimes}. In \bibinfo{booktitle}{\emph{31st USENIX Security Symposium
  (USENIX Security 22)}}. \bibinfo{pages}{359--376}.
\newblock


\bibitem[Alfadel et~al\mbox{.}(2023)]%
        {alfadel2023empirical}
\bibfield{author}{\bibinfo{person}{Mahmoud Alfadel},
  \bibinfo{person}{Diego~Elias Costa}, {and} \bibinfo{person}{Emad Shihab}.}
  \bibinfo{year}{2023}\natexlab{}.
\newblock \showarticletitle{Empirical analysis of security vulnerabilities in
  python packages}.
\newblock \bibinfo{journal}{\emph{Empirical Software Engineering}}
  \bibinfo{volume}{28}, \bibinfo{number}{3} (\bibinfo{year}{2023}),
  \bibinfo{pages}{59}.
\newblock


\bibitem[Althebeiti and Mohaisen(2023)]%
        {althebeiti2023enriching}
\bibfield{author}{\bibinfo{person}{Hattan Althebeiti} {and}
  \bibinfo{person}{David Mohaisen}.} \bibinfo{year}{2023}\natexlab{}.
\newblock \showarticletitle{Enriching vulnerability reports through automated
  and augmented description summarization}. In
  \bibinfo{booktitle}{\emph{International Conference on Information Security
  Applications}}. Springer, \bibinfo{pages}{213--227}.
\newblock


\bibitem[Bhandari et~al\mbox{.}(2021)]%
        {bhandari2021cvefixes}
\bibfield{author}{\bibinfo{person}{Guru Bhandari}, \bibinfo{person}{Amara
  Naseer}, {and} \bibinfo{person}{Leon Moonen}.}
  \bibinfo{year}{2021}\natexlab{}.
\newblock \showarticletitle{CVEfixes: automated collection of vulnerabilities
  and their fixes from open-source software}. In
  \bibinfo{booktitle}{\emph{Proceedings of the 17th International Conference on
  Predictive Models and Data Analytics in Software Engineering}}.
  \bibinfo{pages}{30--39}.
\newblock


\bibitem[Bilge and Dumitra{\c{s}}(2012)]%
        {bilge2012before}
\bibfield{author}{\bibinfo{person}{Leyla Bilge} {and} \bibinfo{person}{Tudor
  Dumitra{\c{s}}}.} \bibinfo{year}{2012}\natexlab{}.
\newblock \showarticletitle{Before we knew it: an empirical study of zero-day
  attacks in the real world}. In \bibinfo{booktitle}{\emph{Proceedings of the
  2012 ACM conference on Computer and communications security}}.
  \bibinfo{pages}{833--844}.
\newblock


\bibitem[B{\"u}hlmann and Ghafari(2022)]%
        {buhlmann2022developers}
\bibfield{author}{\bibinfo{person}{Noah B{\"u}hlmann} {and}
  \bibinfo{person}{Mohammad Ghafari}.} \bibinfo{year}{2022}\natexlab{}.
\newblock \showarticletitle{How do developers deal with security issue reports
  on github?}. In \bibinfo{booktitle}{\emph{Proceedings of the 37th ACM/SIGAPP
  Symposium on Applied Computing}}. \bibinfo{pages}{1580--1589}.
\newblock


\bibitem[Chen et~al\mbox{.}(2025)]%
        {chen2025chatgpt}
\bibfield{author}{\bibinfo{person}{Chong Chen}, \bibinfo{person}{Jianzhong Su},
  \bibinfo{person}{Jiachi Chen}, \bibinfo{person}{Yanlin Wang},
  \bibinfo{person}{Tingting Bi}, \bibinfo{person}{Jianxing Yu},
  \bibinfo{person}{Yanli Wang}, \bibinfo{person}{Xingwei Lin},
  \bibinfo{person}{Ting Chen}, {and} \bibinfo{person}{Zibin Zheng}.}
  \bibinfo{year}{2025}\natexlab{}.
\newblock \showarticletitle{When chatgpt meets smart contract vulnerability
  detection: How far are we?}
\newblock \bibinfo{journal}{\emph{ACM Transactions on Software Engineering and
  Methodology}} \bibinfo{volume}{34}, \bibinfo{number}{4}
  (\bibinfo{year}{2025}), \bibinfo{pages}{1--30}.
\newblock


\bibitem[Chen et~al\mbox{.}(2023)]%
        {chen2023diversevul}
\bibfield{author}{\bibinfo{person}{Yizheng Chen}, \bibinfo{person}{Zhoujie
  Ding}, \bibinfo{person}{Lamya Alowain}, \bibinfo{person}{Xinyun Chen}, {and}
  \bibinfo{person}{David Wagner}.} \bibinfo{year}{2023}\natexlab{}.
\newblock \showarticletitle{Diversevul: A new vulnerable source code dataset
  for deep learning based vulnerability detection}. In
  \bibinfo{booktitle}{\emph{Proceedings of the 26th International Symposium on
  Research in Attacks, Intrusions and Defenses}}. \bibinfo{pages}{654--668}.
\newblock


\bibitem[Cheshkov et~al\mbox{.}(2023)]%
        {cheshkov2023evaluation}
\bibfield{author}{\bibinfo{person}{Anton Cheshkov}, \bibinfo{person}{Pavel
  Zadorozhny}, {and} \bibinfo{person}{Rodion Levichev}.}
  \bibinfo{year}{2023}\natexlab{}.
\newblock \showarticletitle{Evaluation of chatgpt model for vulnerability
  detection}.
\newblock \bibinfo{journal}{\emph{arXiv preprint arXiv:2304.07232}}
  (\bibinfo{year}{2023}).
\newblock


\bibitem[Chintagunta et~al\mbox{.}(2021)]%
        {chintagunta2021medically}
\bibfield{author}{\bibinfo{person}{Bharath Chintagunta}, \bibinfo{person}{Namit
  Katariya}, \bibinfo{person}{Xavier Amatriain}, {and} \bibinfo{person}{Anitha
  Kannan}.} \bibinfo{year}{2021}\natexlab{}.
\newblock \showarticletitle{Medically aware GPT-3 as a data generator for
  medical dialogue summarization}. In \bibinfo{booktitle}{\emph{Machine
  Learning for Healthcare Conference}}. PMLR, \bibinfo{pages}{354--372}.
\newblock


\bibitem[Chopra et~al\mbox{.}(2024)]%
        {chopra2024chatnvd}
\bibfield{author}{\bibinfo{person}{Shivansh Chopra}, \bibinfo{person}{Hussain
  Ahmad}, \bibinfo{person}{Diksha Goel}, {and} \bibinfo{person}{Claudia
  Szabo}.} \bibinfo{year}{2024}\natexlab{}.
\newblock \showarticletitle{Chatnvd: Advancing cybersecurity vulnerability
  assessment with large language models}.
\newblock \bibinfo{journal}{\emph{arXiv preprint arXiv:2412.04756}}
  (\bibinfo{year}{2024}).
\newblock


\bibitem[Collard et~al\mbox{.}(2013)]%
        {collard2013srcml}
\bibfield{author}{\bibinfo{person}{Michael~L Collard},
  \bibinfo{person}{Michael~John Decker}, {and} \bibinfo{person}{Jonathan~I
  Maletic}.} \bibinfo{year}{2013}\natexlab{}.
\newblock \showarticletitle{srcml: An infrastructure for the exploration,
  analysis, and manipulation of source code: A tool demonstration}. In
  \bibinfo{booktitle}{\emph{2013 IEEE International conference on software
  maintenance}}. IEEE, \bibinfo{pages}{516--519}.
\newblock


\bibitem[Croft et~al\mbox{.}(2023)]%
        {croft2023data}
\bibfield{author}{\bibinfo{person}{Roland Croft}, \bibinfo{person}{M~Ali
  Babar}, {and} \bibinfo{person}{M~Mehdi Kholoosi}.}
  \bibinfo{year}{2023}\natexlab{}.
\newblock \showarticletitle{Data quality for software vulnerability datasets}.
  In \bibinfo{booktitle}{\emph{2023 IEEE/ACM 45th International Conference on
  Software Engineering (ICSE)}}. IEEE, \bibinfo{pages}{121--133}.
\newblock


\bibitem[Das et~al\mbox{.}(2025)]%
        {das2025we}
\bibfield{author}{\bibinfo{person}{Satyaki Das}, \bibinfo{person}{Syeda~Tasnim
  Fabiha}, \bibinfo{person}{Saad Shafiq}, {and} \bibinfo{person}{Nenad
  Medvidovic}.} \bibinfo{year}{2025}\natexlab{}.
\newblock \showarticletitle{Are we learning the right features? a framework for
  evaluating dl-based software vulnerability detection solutions}.
\newblock \bibinfo{journal}{\emph{arXiv preprint arXiv:2501.13291}}
  (\bibinfo{year}{2025}).
\newblock


\bibitem[Decan et~al\mbox{.}(2017)]%
        {decan2017empirical}
\bibfield{author}{\bibinfo{person}{Alexandre Decan}, \bibinfo{person}{Tom
  Mens}, {and} \bibinfo{person}{Ma{\"e}lick Claes}.}
  \bibinfo{year}{2017}\natexlab{}.
\newblock \showarticletitle{An empirical comparison of dependency issues in OSS
  packaging ecosystems}. In \bibinfo{booktitle}{\emph{2017 IEEE 24th
  international conference on software analysis, evolution and reengineering
  (SANER)}}. IEEE, \bibinfo{pages}{2--12}.
\newblock


\bibitem[Decan et~al\mbox{.}(2018)]%
        {decan2018impact}
\bibfield{author}{\bibinfo{person}{Alexandre Decan}, \bibinfo{person}{Tom
  Mens}, {and} \bibinfo{person}{Eleni Constantinou}.}
  \bibinfo{year}{2018}\natexlab{}.
\newblock \showarticletitle{On the impact of security vulnerabilities in the
  npm package dependency network}. In \bibinfo{booktitle}{\emph{Proceedings of
  the 15th international conference on mining software repositories}}.
  \bibinfo{pages}{181--191}.
\newblock


\bibitem[Devlin et~al\mbox{.}(2018)]%
        {devlin2018bert}
\bibfield{author}{\bibinfo{person}{Jacob Devlin}, \bibinfo{person}{Ming-Wei
  Chang}, \bibinfo{person}{Kenton Lee}, {and} \bibinfo{person}{Kristina
  Toutanova}.} \bibinfo{year}{2018}\natexlab{}.
\newblock \showarticletitle{Bert: Pre-training of deep bidirectional
  transformers for language understanding}.
\newblock \bibinfo{journal}{\emph{arXiv preprint arXiv:1810.04805}}
  (\bibinfo{year}{2018}).
\newblock


\bibitem[Ding et~al\mbox{.}(2019)]%
        {ding2019ethical}
\bibfield{author}{\bibinfo{person}{Aaron~Yi Ding},
  \bibinfo{person}{Gianluca~Limon De~Jesus}, {and} \bibinfo{person}{Marijn
  Janssen}.} \bibinfo{year}{2019}\natexlab{}.
\newblock \showarticletitle{Ethical hacking for boosting IoT vulnerability
  management: A first look into bug bounty programs and responsible
  disclosure}. In \bibinfo{booktitle}{\emph{Proceedings of the Eighth
  International Conference on Telecommunications and Remote Sensing}}.
  \bibinfo{pages}{49--55}.
\newblock


\bibitem[Documentation(2023)]%
        {GitHubRestAPI}
\bibfield{author}{\bibinfo{person}{GitHub Documentation}.}
  \bibinfo{year}{2023}\natexlab{}.
\newblock \bibinfo{booktitle}{\emph{GitHub REST API}}.
\newblock
\urldef\tempurl%
\url{https://docs.github.com/en/rest}
\showURL{%
\tempurl}
\newblock
\shownote{Accessed on September 14, 2023}.


\bibitem[{Etherpad Lite Community}(2024)]%
        {etherpadSecurity2024}
\bibfield{author}{\bibinfo{person}{{Etherpad Lite Community}}.}
  \bibinfo{year}{2024}\natexlab{}.
\newblock \bibinfo{title}{Security Policy for the Etherpad Lite Repository}.
\newblock \bibinfo{howpublished}{Website}.
\newblock
\urldef\tempurl%
\url{https://github.com/ether/etherpad-lite/security}
\showURL{%
\tempurl}
\newblock
\shownote{Accessed: May 16, 2024}.


\bibitem[Feng et~al\mbox{.}(2020)]%
        {feng2020codebert}
\bibfield{author}{\bibinfo{person}{Zhangyin Feng}, \bibinfo{person}{Daya Guo},
  \bibinfo{person}{Duyu Tang}, \bibinfo{person}{Nan Duan},
  \bibinfo{person}{Xiaocheng Feng}, \bibinfo{person}{Ming Gong},
  \bibinfo{person}{Linjun Shou}, \bibinfo{person}{Bing Qin},
  \bibinfo{person}{Ting Liu}, \bibinfo{person}{Daxin Jiang}, {et~al\mbox{.}}}
  \bibinfo{year}{2020}\natexlab{}.
\newblock \showarticletitle{Codebert: A pre-trained model for programming and
  natural languages}.
\newblock \bibinfo{journal}{\emph{arXiv preprint arXiv:2002.08155}}
  (\bibinfo{year}{2020}).
\newblock


\bibitem[Foundation(Year)]%
        {TheApacheSoftwareFoundation}
\bibfield{author}{\bibinfo{person}{Apache~Software Foundation}.}
  \bibinfo{year}{Year}\natexlab{}.
\newblock \bibinfo{title}{ASF PROJECT SECURITY FOR COMMITTERS.}
\newblock
  \bibinfo{howpublished}{\url{https://www.apache.org/security/committers.html}}.
\newblock


\bibitem[Francis et~al\mbox{.}(2010)]%
        {francis2010adequate}
\bibfield{author}{\bibinfo{person}{Jill~J Francis}, \bibinfo{person}{Marie
  Johnston}, \bibinfo{person}{Clare Robertson}, \bibinfo{person}{Liz
  Glidewell}, \bibinfo{person}{Vikki Entwistle}, \bibinfo{person}{Martin~P
  Eccles}, {and} \bibinfo{person}{Jeremy~M Grimshaw}.}
  \bibinfo{year}{2010}\natexlab{}.
\newblock \showarticletitle{What is an adequate sample size? Operationalising
  data saturation for theory-based interview studies}.
\newblock \bibinfo{journal}{\emph{Psychology and health}} \bibinfo{volume}{25},
  \bibinfo{number}{10} (\bibinfo{year}{2010}), \bibinfo{pages}{1229--1245}.
\newblock


\bibitem[Fu and Tantithamthavorn(2022)]%
        {fu2022linevul}
\bibfield{author}{\bibinfo{person}{Michael Fu} {and} \bibinfo{person}{Chakkrit
  Tantithamthavorn}.} \bibinfo{year}{2022}\natexlab{}.
\newblock \showarticletitle{Linevul: A transformer-based line-level
  vulnerability prediction}. In \bibinfo{booktitle}{\emph{Proceedings of the
  19th International Conference on Mining Software Repositories}}.
  \bibinfo{pages}{608--620}.
\newblock


\bibitem[Fu et~al\mbox{.}(2023)]%
        {fu2023chatgpt}
\bibfield{author}{\bibinfo{person}{Michael Fu}, \bibinfo{person}{Chakkrit~Kla
  Tantithamthavorn}, \bibinfo{person}{Van Nguyen}, {and} \bibinfo{person}{Trung
  Le}.} \bibinfo{year}{2023}\natexlab{}.
\newblock \showarticletitle{Chatgpt for vulnerability detection,
  classification, and repair: How far are we?}. In
  \bibinfo{booktitle}{\emph{2023 30th Asia-Pacific Software Engineering
  Conference (APSEC)}}. IEEE, \bibinfo{pages}{632--636}.
\newblock


\bibitem[GitHub(2023)]%
        {githubAdvisories}
\bibfield{author}{\bibinfo{person}{GitHub}.} \bibinfo{year}{2023}\natexlab{}.
\newblock \bibinfo{title}{GitHub Security Advisories}.
\newblock
\newblock
\newblock
\shownote{Available at
  \url{https://docs.github.com/en/code-security/security-advisories}}.


\bibitem[GitHub(Year)]%
        {DisclosureGitHubDocs}
\bibfield{author}{\bibinfo{person}{GitHub}.} \bibinfo{year}{Year}\natexlab{}.
\newblock \bibinfo{title}{About coordinated disclosure of security
  vulnerabilities}.
\newblock \bibinfo{howpublished}{\url{https://shorturl.at/wJO01}}.
\newblock


\bibitem[{GitHub, Inc.}(2024)]%
        {GitHubRESTIssuesTimeline}
\bibfield{author}{\bibinfo{person}{{GitHub, Inc.}}}
  \bibinfo{year}{2024}\natexlab{}.
\newblock \bibinfo{title}{GitHub REST API v3: Issues Timeline}.
\newblock
  \bibinfo{howpublished}{\url{https://docs.github.com/en/rest/issues/timeline}}.
\newblock
\newblock
\shownote{Accessed: 2024-04-01}.


\bibitem[Google(Yeara)]%
        {GoogleOSSVuln}
\bibfield{author}{\bibinfo{person}{Google}.} \bibinfo{year}{Year}\natexlab{a}.
\newblock \bibinfo{title}{A guide on coordinated vulnerability disclosure for
  open source projects. includes templates for security policies (security.md)
  and disclosure notifications.}
\newblock
  \bibinfo{howpublished}{\url{https://github.com/google/oss-vulnerability-guide}}.
\newblock


\bibitem[Google(Yearb)]%
        {SecuritytechniquesVulnerabilitydisclosure2018}
\bibfield{author}{\bibinfo{person}{Google}.} \bibinfo{year}{Year}\natexlab{b}.
\newblock \bibinfo{title}{Security techniques - Vulnerability disclosure.}
\newblock
  \bibinfo{howpublished}{\url{https://www.iso.org/standard/72311.html}}.
\newblock


\bibitem[Hin et~al\mbox{.}(2022)]%
        {hin2022linevd}
\bibfield{author}{\bibinfo{person}{David Hin}, \bibinfo{person}{Andrey Kan},
  \bibinfo{person}{Huaming Chen}, {and} \bibinfo{person}{M~Ali Babar}.}
  \bibinfo{year}{2022}\natexlab{}.
\newblock \showarticletitle{Linevd: Statement-level vulnerability detection
  using graph neural networks}. In \bibinfo{booktitle}{\emph{Proceedings of the
  19th international conference on mining software repositories}}.
  \bibinfo{pages}{596--607}.
\newblock


\bibitem[Hirao et~al\mbox{.}(2019)]%
        {hirao2019review}
\bibfield{author}{\bibinfo{person}{Toshiki Hirao}, \bibinfo{person}{Shane
  McIntosh}, \bibinfo{person}{Akinori Ihara}, {and} \bibinfo{person}{Kenichi
  Matsumoto}.} \bibinfo{year}{2019}\natexlab{}.
\newblock \showarticletitle{The review linkage graph for code review analytics:
  A recovery approach and empirical study}. In
  \bibinfo{booktitle}{\emph{Proceedings of the 2019 27th ACM joint meeting on
  European software engineering conference and symposium on the foundations of
  software engineering}}. \bibinfo{pages}{578--589}.
\newblock


\bibitem[Horawalavithana et~al\mbox{.}(2019)]%
        {horawalavithana2019mentions}
\bibfield{author}{\bibinfo{person}{Sameera Horawalavithana},
  \bibinfo{person}{Abhishek Bhattacharjee}, \bibinfo{person}{Renhao Liu},
  \bibinfo{person}{Nazim Choudhury}, \bibinfo{person}{Lawrence O.~Hall}, {and}
  \bibinfo{person}{Adriana Iamnitchi}.} \bibinfo{year}{2019}\natexlab{}.
\newblock \showarticletitle{Mentions of security vulnerabilities on reddit,
  twitter and github}. In \bibinfo{booktitle}{\emph{IEEE/WIC/ACM International
  Conference on Web Intelligence}}. \bibinfo{pages}{200--207}.
\newblock


\bibitem[Householder et~al\mbox{.}(2017)]%
        {householder2017cert}
\bibfield{author}{\bibinfo{person}{Allen~D Householder},
  \bibinfo{person}{Garret Wassermann}, \bibinfo{person}{Art Manion}, {and}
  \bibinfo{person}{Chris King}.} \bibinfo{year}{2017}\natexlab{}.
\newblock \showarticletitle{The cert guide to coordinated vulnerability
  disclosure}.
\newblock \bibinfo{journal}{\emph{Software Engineering Institute, Pittsburgh,
  PA}} (\bibinfo{year}{2017}).
\newblock


\bibitem[Imtiaz et~al\mbox{.}(2022)]%
        {imtiaz2022open}
\bibfield{author}{\bibinfo{person}{Nasif Imtiaz}, \bibinfo{person}{Aniqa
  Khanom}, {and} \bibinfo{person}{Laurie Williams}.}
  \bibinfo{year}{2022}\natexlab{}.
\newblock \showarticletitle{Open or sneaky? fast or slow? light or heavy?:
  Investigating security releases of open source packages}.
\newblock \bibinfo{journal}{\emph{IEEE Transactions on Software Engineering}}
  \bibinfo{volume}{49}, \bibinfo{number}{4} (\bibinfo{year}{2022}),
  \bibinfo{pages}{1540--1560}.
\newblock


\bibitem[Izquierdo et~al\mbox{.}(2015)]%
        {izquierdo2015gila}
\bibfield{author}{\bibinfo{person}{Javier Luis~C{\'a}novas Izquierdo},
  \bibinfo{person}{Valerio Cosentino}, \bibinfo{person}{Bel{\'e}n Rolandi},
  \bibinfo{person}{Alexandre Bergel}, {and} \bibinfo{person}{Jordi Cabot}.}
  \bibinfo{year}{2015}\natexlab{}.
\newblock \showarticletitle{GiLA: GitHub label analyzer}. In
  \bibinfo{booktitle}{\emph{2015 IEEE 22nd International Conference on Software
  Analysis, Evolution, and Reengineering (SANER)}}. IEEE,
  \bibinfo{pages}{479--483}.
\newblock


\bibitem[Kallis et~al\mbox{.}(2021)]%
        {kallis2021predicting}
\bibfield{author}{\bibinfo{person}{Rafael Kallis}, \bibinfo{person}{Andrea
  Di~Sorbo}, \bibinfo{person}{Gerardo Canfora}, {and}
  \bibinfo{person}{Sebastiano Panichella}.} \bibinfo{year}{2021}\natexlab{}.
\newblock \showarticletitle{Predicting issue types on GitHub}.
\newblock \bibinfo{journal}{\emph{Science of Computer Programming}}
  \bibinfo{volume}{205} (\bibinfo{year}{2021}), \bibinfo{pages}{102598}.
\newblock


\bibitem[KATHLEEN~METRICK(Year)]%
        {MandiantOSS}
\bibfield{author}{\bibinfo{person}{SHAMBAVI~SADAYAPPAN KATHLEEN~METRICK,
  JARED~SEMRAU}.} \bibinfo{year}{Year}\natexlab{}.
\newblock \bibinfo{title}{A guide on coordinated vulnerability disclosure for
  open source projects. includes templates for security policies (security.md)
  and disclosure notifications.}
\newblock
  \bibinfo{howpublished}{\url{https://www.mandiant.com/resources/blog/time-between-disclosure-patch-release-and-vulnerability-exploitation}}.
\newblock


\bibitem[Kumar and Rashid(2018)]%
        {kumar2018efficient}
\bibfield{author}{\bibinfo{person}{Madhup Kumar} {and} \bibinfo{person}{Ekbal
  Rashid}.} \bibinfo{year}{2018}\natexlab{}.
\newblock \showarticletitle{An efficient software development life cycle model
  for developing software project}.
\newblock \bibinfo{journal}{\emph{International Journal of Education and
  Management Engineering}} \bibinfo{volume}{8}, \bibinfo{number}{6}
  (\bibinfo{year}{2018}), \bibinfo{pages}{59--68}.
\newblock


\bibitem[Li and Paxson(2017)]%
        {li2017large}
\bibfield{author}{\bibinfo{person}{Frank Li} {and} \bibinfo{person}{Vern
  Paxson}.} \bibinfo{year}{2017}\natexlab{}.
\newblock \showarticletitle{A large-scale empirical study of security patches}.
  In \bibinfo{booktitle}{\emph{Proceedings of the 2017 ACM SIGSAC Conference on
  Computer and Communications Security}}. \bibinfo{pages}{2201--2215}.
\newblock


\bibitem[Li et~al\mbox{.}(2018a)]%
        {li2018issue}
\bibfield{author}{\bibinfo{person}{Lisha Li}, \bibinfo{person}{Zhilei Ren},
  \bibinfo{person}{Xiaochen Li}, \bibinfo{person}{Weiqin Zou}, {and}
  \bibinfo{person}{He Jiang}.} \bibinfo{year}{2018}\natexlab{a}.
\newblock \showarticletitle{How are issue units linked? empirical study on the
  linking behavior in github}. In \bibinfo{booktitle}{\emph{2018 25th
  Asia-Pacific Software Engineering Conference (APSEC)}}. IEEE,
  \bibinfo{pages}{386--395}.
\newblock


\bibitem[Li et~al\mbox{.}(2024)]%
        {li2024effectiveness}
\bibfield{author}{\bibinfo{person}{Zhen Li}, \bibinfo{person}{Ning Wang},
  \bibinfo{person}{Deqing Zou}, \bibinfo{person}{Yating Li},
  \bibinfo{person}{Ruqian Zhang}, \bibinfo{person}{Shouhuai Xu},
  \bibinfo{person}{Chao Zhang}, {and} \bibinfo{person}{Hai Jin}.}
  \bibinfo{year}{2024}\natexlab{}.
\newblock \showarticletitle{On the Effectiveness of Function-Level
  Vulnerability Detectors for Inter-Procedural Vulnerabilities}. In
  \bibinfo{booktitle}{\emph{Proceedings of the IEEE/ACM 46th International
  Conference on Software Engineering}}. \bibinfo{pages}{1--12}.
\newblock


\bibitem[Li et~al\mbox{.}(2016)]%
        {li2016vulpecker}
\bibfield{author}{\bibinfo{person}{Zhen Li}, \bibinfo{person}{Deqing Zou},
  \bibinfo{person}{Shouhuai Xu}, \bibinfo{person}{Hai Jin},
  \bibinfo{person}{Hanchao Qi}, {and} \bibinfo{person}{Jie Hu}.}
  \bibinfo{year}{2016}\natexlab{}.
\newblock \showarticletitle{Vulpecker: an automated vulnerability detection
  system based on code similarity analysis}. In
  \bibinfo{booktitle}{\emph{Proceedings of the 32nd annual conference on
  computer security applications}}. \bibinfo{pages}{201--213}.
\newblock


\bibitem[Li et~al\mbox{.}(2021)]%
        {li2021sysevr}
\bibfield{author}{\bibinfo{person}{Zhen Li}, \bibinfo{person}{Deqing Zou},
  \bibinfo{person}{Shouhuai Xu}, \bibinfo{person}{Hai Jin},
  \bibinfo{person}{Yawei Zhu}, {and} \bibinfo{person}{Zhaoxuan Chen}.}
  \bibinfo{year}{2021}\natexlab{}.
\newblock \showarticletitle{Sysevr: A framework for using deep learning to
  detect software vulnerabilities}.
\newblock \bibinfo{journal}{\emph{IEEE Transactions on Dependable and Secure
  Computing}} \bibinfo{volume}{19}, \bibinfo{number}{4} (\bibinfo{year}{2021}),
  \bibinfo{pages}{2244--2258}.
\newblock


\bibitem[Li et~al\mbox{.}(2018b)]%
        {li2018vuldeepecker}
\bibfield{author}{\bibinfo{person}{Zhen Li}, \bibinfo{person}{Deqing Zou},
  \bibinfo{person}{Shouhuai Xu}, \bibinfo{person}{Xinyu Ou},
  \bibinfo{person}{Hai Jin}, \bibinfo{person}{Sujuan Wang},
  \bibinfo{person}{Zhijun Deng}, {and} \bibinfo{person}{Yuyi Zhong}.}
  \bibinfo{year}{2018}\natexlab{b}.
\newblock \showarticletitle{Vuldeepecker: A deep learning-based system for
  vulnerability detection}.
\newblock \bibinfo{journal}{\emph{arXiv preprint arXiv:1801.01681}}
  (\bibinfo{year}{2018}).
\newblock


\bibitem[Liu et~al\mbox{.}(2019)]%
        {liu2019roberta}
\bibfield{author}{\bibinfo{person}{Yinhan Liu}, \bibinfo{person}{Myle Ott},
  \bibinfo{person}{Naman Goyal}, \bibinfo{person}{Jingfei Du},
  \bibinfo{person}{Mandar Joshi}, \bibinfo{person}{Danqi Chen},
  \bibinfo{person}{Omer Levy}, \bibinfo{person}{Mike Lewis},
  \bibinfo{person}{Luke Zettlemoyer}, {and} \bibinfo{person}{Veselin
  Stoyanov}.} \bibinfo{year}{2019}\natexlab{}.
\newblock \showarticletitle{Roberta: A robustly optimized bert pretraining
  approach}.
\newblock \bibinfo{journal}{\emph{arXiv preprint arXiv:1907.11692}}
  (\bibinfo{year}{2019}).
\newblock


\bibitem[McClanahan et~al\mbox{.}(2024)]%
        {mcclanahan2024chatgpt}
\bibfield{author}{\bibinfo{person}{Kylie McClanahan}, \bibinfo{person}{Sky
  Elder}, \bibinfo{person}{Marie~Louise Uwibambe}, \bibinfo{person}{Yaling
  Liu}, \bibinfo{person}{Rithyka Heng}, {and} \bibinfo{person}{Qinghua Li}.}
  \bibinfo{year}{2024}\natexlab{}.
\newblock \showarticletitle{When ChatGPT Meets Vulnerability Management: the
  Good, the Bad, and the Ugly}. In \bibinfo{booktitle}{\emph{IEEE Int’l Conf.
  on Computing, Networking and Communications (ICNC)}}.
\newblock


\bibitem[Microsoft(Year)]%
        {MicrosoftOSSguide}
\bibfield{author}{\bibinfo{person}{Microsoft}.}
  \bibinfo{year}{Year}\natexlab{}.
\newblock \bibinfo{title}{Microsoft's Approach to Coordinated Vulnerability
  Disclosure}.
\newblock
  \bibinfo{howpublished}{\url{https://www.microsoft.com/en-us/msrc/cvd}}.
\newblock


\bibitem[Nappa et~al\mbox{.}(2015)]%
        {nappa2015attack}
\bibfield{author}{\bibinfo{person}{Antonio Nappa}, \bibinfo{person}{Richard
  Johnson}, \bibinfo{person}{Leyla Bilge}, \bibinfo{person}{Juan Caballero},
  {and} \bibinfo{person}{Tudor Dumitras}.} \bibinfo{year}{2015}\natexlab{}.
\newblock \showarticletitle{The attack of the clones: A study of the impact of
  shared code on vulnerability patching}. In \bibinfo{booktitle}{\emph{2015
  IEEE symposium on security and privacy}}. IEEE, \bibinfo{pages}{692--708}.
\newblock


\bibitem[Nguyen et~al\mbox{.}(2022)]%
        {nguyen2022vulcurator}
\bibfield{author}{\bibinfo{person}{Truong~Giang Nguyen}, \bibinfo{person}{Thanh
  Le-Cong}, \bibinfo{person}{Hong~Jin Kang}, \bibinfo{person}{Xuan-Bach~D Le},
  {and} \bibinfo{person}{David Lo}.} \bibinfo{year}{2022}\natexlab{}.
\newblock \showarticletitle{Vulcurator: a vulnerability-fixing commit
  detector}. In \bibinfo{booktitle}{\emph{Proceedings of the 30th ACM Joint
  European Software Engineering Conference and Symposium on the Foundations of
  Software Engineering}}. \bibinfo{pages}{1726--1730}.
\newblock


\bibitem[Nguyen-Truong et~al\mbox{.}(2022)]%
        {nguyen2022hermes}
\bibfield{author}{\bibinfo{person}{Giang Nguyen-Truong},
  \bibinfo{person}{Hong~Jin Kang}, \bibinfo{person}{David Lo},
  \bibinfo{person}{Abhishek Sharma}, \bibinfo{person}{Andrew~E Santosa},
  \bibinfo{person}{Asankhaya Sharma}, {and} \bibinfo{person}{Ming~Yi Ang}.}
  \bibinfo{year}{2022}\natexlab{}.
\newblock \showarticletitle{Hermes: Using commit-issue linking to detect
  vulnerability-fixing commits}. In \bibinfo{booktitle}{\emph{2022 IEEE
  International Conference on Software Analysis, Evolution and Reengineering
  (SANER)}}. IEEE, \bibinfo{pages}{51--62}.
\newblock


\bibitem[OpenAI(2024)]%
        {openai2024gpt4o}
\bibfield{author}{\bibinfo{person}{OpenAI}.} \bibinfo{year}{2024}\natexlab{}.
\newblock \bibinfo{title}{GPT-4o}.
\newblock \bibinfo{howpublished}{\url{https://openai.com/index/hello-gpt-4o/}}.
\newblock
\newblock
\shownote{Accessed: 2025-02-01}.


\bibitem[Ott et~al\mbox{.}(2018)]%
        {ott2018deep}
\bibfield{author}{\bibinfo{person}{Jordan Ott}, \bibinfo{person}{Abigail
  Atchison}, \bibinfo{person}{Paul Harnack}, \bibinfo{person}{Adrienne Bergh},
  {and} \bibinfo{person}{Erik Linstead}.} \bibinfo{year}{2018}\natexlab{}.
\newblock \showarticletitle{A deep learning approach to identifying source code
  in images and video}. In \bibinfo{booktitle}{\emph{Proceedings of the 15th
  International Conference on Mining Software Repositories}}.
  \bibinfo{pages}{376--386}.
\newblock


\bibitem[Palacio et~al\mbox{.}(2019)]%
        {palacio2019learning}
\bibfield{author}{\bibinfo{person}{David~N Palacio}, \bibinfo{person}{Daniel
  McCrystal}, \bibinfo{person}{Kevin Moran}, \bibinfo{person}{Carlos
  Bernal-C{\'a}rdenas}, \bibinfo{person}{Denys Poshyvanyk}, {and}
  \bibinfo{person}{Chris Shenefiel}.} \bibinfo{year}{2019}\natexlab{}.
\newblock \showarticletitle{Learning to identify security-related issues using
  convolutional neural networks}. In \bibinfo{booktitle}{\emph{2019 IEEE
  International conference on software maintenance and evolution (ICSME)}}.
  IEEE, \bibinfo{pages}{140--144}.
\newblock


\bibitem[Pan et~al\mbox{.}(2022)]%
        {pan2022automated}
\bibfield{author}{\bibinfo{person}{Shengyi Pan}, \bibinfo{person}{Jiayuan
  Zhou}, \bibinfo{person}{Filipe~Roseiro Cogo}, \bibinfo{person}{Xin Xia},
  \bibinfo{person}{Lingfeng Bao}, \bibinfo{person}{Xing Hu},
  \bibinfo{person}{Shanping Li}, {and} \bibinfo{person}{Ahmed~E Hassan}.}
  \bibinfo{year}{2022}\natexlab{}.
\newblock \showarticletitle{Automated unearthing of dangerous issue reports}.
  In \bibinfo{booktitle}{\emph{Proceedings of the 30th ACM Joint European
  Software Engineering Conference and Symposium on the Foundations of Software
  Engineering}}. \bibinfo{pages}{834--846}.
\newblock


\bibitem[Parente et~al\mbox{.}(2025)]%
        {parente2025frape}
\bibfield{author}{\bibinfo{person}{FR Parente}, \bibinfo{person}{Emanuel~B
  Rodrigues}, {and} \bibinfo{person}{C{\'e}sar~LC Mattos}.}
  \bibinfo{year}{2025}\natexlab{}.
\newblock \showarticletitle{FRAPE: A Framework for Risk Assessment,
  Prioritization and Explainability of vulnerabilities in cybersecurity}.
\newblock \bibinfo{journal}{\emph{Journal of Information Security and
  Applications}}  \bibinfo{volume}{89} (\bibinfo{year}{2025}),
  \bibinfo{pages}{103971}.
\newblock


\bibitem[Parsons(2014)]%
        {parsons2014stratified}
\bibfield{author}{\bibinfo{person}{Van~L Parsons}.}
  \bibinfo{year}{2014}\natexlab{}.
\newblock \showarticletitle{Stratified sampling}.
\newblock \bibinfo{journal}{\emph{Wiley StatsRef: Statistics Reference Online}}
  (\bibinfo{year}{2014}), \bibinfo{pages}{1--11}.
\newblock


\bibitem[Prodan and Pelican(2023)]%
        {prodan2023prompt}
\bibfield{author}{\bibinfo{person}{George Prodan} {and} \bibinfo{person}{Elena
  Pelican}.} \bibinfo{year}{2023}\natexlab{}.
\newblock \showarticletitle{Prompt scoring system for dialogue summarization
  using GPT-3}.
\newblock \bibinfo{journal}{\emph{Authorea Preprints}} (\bibinfo{year}{2023}).
\newblock


\bibitem[Program(2023)]%
        {cveservices}
\bibfield{author}{\bibinfo{person}{CVE Program}.}
  \bibinfo{year}{2023}\natexlab{}.
\newblock \bibinfo{title}{CVE Services API}.
\newblock
\newblock
\newblock
\shownote{Available at \url{https://github.com/CVEProject/cve-services}}.


\bibitem[Reddy et~al\mbox{.}(2023)]%
        {reddy2023sumren}
\bibfield{author}{\bibinfo{person}{Revanth~Gangi Reddy}, \bibinfo{person}{Heba
  Elfardy}, \bibinfo{person}{Hou~Pong Chan}, \bibinfo{person}{Kevin Small},
  {and} \bibinfo{person}{Heng Ji}.} \bibinfo{year}{2023}\natexlab{}.
\newblock \showarticletitle{Sumren: Summarizing reported speech about events in
  news}. In \bibinfo{booktitle}{\emph{Proceedings of the AAAI Conference on
  Artificial Intelligence}}, Vol.~\bibinfo{volume}{37}.
  \bibinfo{pages}{12808--12817}.
\newblock


\bibitem[Rodriguez et~al\mbox{.}(2018)]%
        {rodriguez2018analysis}
\bibfield{author}{\bibinfo{person}{Luis Gustavo~Araujo Rodriguez},
  \bibinfo{person}{Julia~Selvatici Trazzi}, \bibinfo{person}{Victor Fossaluza},
  \bibinfo{person}{Rodrigo Campiolo}, {and} \bibinfo{person}{Daniel~Mac{\^e}do
  Batista}.} \bibinfo{year}{2018}\natexlab{}.
\newblock \showarticletitle{Analysis of vulnerability disclosure delays from
  the national vulnerability database}. In \bibinfo{booktitle}{\emph{Workshop
  de Seguran{\c{c}}a Cibern{\'e}tica em Dispositivos Conectados (WSCDC)}}. SBC.
\newblock


\bibitem[Sabetta and Bezzi(2018)]%
        {sabetta2018practical}
\bibfield{author}{\bibinfo{person}{Antonino Sabetta} {and}
  \bibinfo{person}{Michele Bezzi}.} \bibinfo{year}{2018}\natexlab{}.
\newblock \showarticletitle{A practical approach to the automatic
  classification of security-relevant commits}. In
  \bibinfo{booktitle}{\emph{2018 IEEE International conference on software
  maintenance and evolution (ICSME)}}. IEEE, \bibinfo{pages}{579--582}.
\newblock


\bibitem[Sauerwein et~al\mbox{.}(2018)]%
        {sauerwein2018tweet}
\bibfield{author}{\bibinfo{person}{Clemens Sauerwein},
  \bibinfo{person}{Christian Sillaber}, \bibinfo{person}{Michael~M Huber},
  \bibinfo{person}{Andrea Mussmann}, {and} \bibinfo{person}{Ruth Breu}.}
  \bibinfo{year}{2018}\natexlab{}.
\newblock \showarticletitle{The tweet advantage: An empirical analysis of 0-day
  vulnerability information shared on twitter}. In
  \bibinfo{booktitle}{\emph{ICT Systems Security and Privacy Protection: 33rd
  IFIP TC 11 International Conference, SEC 2018}}. Springer,
  \bibinfo{pages}{201--215}.
\newblock


\bibitem[Sejfia et~al\mbox{.}(2024)]%
        {sejfia2024toward}
\bibfield{author}{\bibinfo{person}{Adriana Sejfia}, \bibinfo{person}{Satyaki
  Das}, \bibinfo{person}{Saad Shafiq}, {and} \bibinfo{person}{Nenad
  Medvidovi{\'c}}.} \bibinfo{year}{2024}\natexlab{}.
\newblock \showarticletitle{Toward Improved Deep Learning-based Vulnerability
  Detection}. In \bibinfo{booktitle}{\emph{Proceedings of the 46th IEEE/ACM
  International Conference on Software Engineering}}. \bibinfo{pages}{1--12}.
\newblock


\bibitem[Service(2009)]%
        {service2009book}
\bibfield{author}{\bibinfo{person}{Robert~W Service}.}
  \bibinfo{year}{2009}\natexlab{}.
\newblock \showarticletitle{Book Review: Corbin, J., \& Strauss, A.(2008).
  Basics of Qualitative Research: Techniques and Procedures for Developing
  Grounded Theory . Thousand Oaks, CA: Sage}.
\newblock \bibinfo{journal}{\emph{Organizational Research Methods}}
  \bibinfo{volume}{12}, \bibinfo{number}{3} (\bibinfo{year}{2009}),
  \bibinfo{pages}{614--617}.
\newblock


\bibitem[Shahzad et~al\mbox{.}(2012)]%
        {shahzad2012large}
\bibfield{author}{\bibinfo{person}{Muhammad Shahzad},
  \bibinfo{person}{Muhammad~Zubair Shafiq}, {and} \bibinfo{person}{Alex~X
  Liu}.} \bibinfo{year}{2012}\natexlab{}.
\newblock \showarticletitle{A large scale exploratory analysis of software
  vulnerability life cycles}. In \bibinfo{booktitle}{\emph{2012 34th
  International Conference on Software Engineering (ICSE)}}. IEEE,
  \bibinfo{pages}{771--781}.
\newblock


\bibitem[Steenhoek et~al\mbox{.}(2024)]%
        {steenhoek2024dataflow}
\bibfield{author}{\bibinfo{person}{Benjamin Steenhoek},
  \bibinfo{person}{Hongyang Gao}, {and} \bibinfo{person}{Wei Le}.}
  \bibinfo{year}{2024}\natexlab{}.
\newblock \showarticletitle{Dataflow analysis-inspired deep learning for
  efficient vulnerability detection}. In \bibinfo{booktitle}{\emph{Proceedings
  of the 46th ieee/acm international conference on software engineering}}.
  \bibinfo{pages}{1--13}.
\newblock


\bibitem[Steenhoek et~al\mbox{.}(2023)]%
        {steenhoek2023empirical}
\bibfield{author}{\bibinfo{person}{Benjamin Steenhoek},
  \bibinfo{person}{Md~Mahbubur Rahman}, \bibinfo{person}{Richard Jiles}, {and}
  \bibinfo{person}{Wei Le}.} \bibinfo{year}{2023}\natexlab{}.
\newblock \showarticletitle{An empirical study of deep learning models for
  vulnerability detection}. In \bibinfo{booktitle}{\emph{2023 IEEE/ACM 45th
  International Conference on Software Engineering (ICSE)}}. IEEE,
  \bibinfo{pages}{2237--2248}.
\newblock


\bibitem[Wang et~al\mbox{.}(2022)]%
        {wang2022vcmatch}
\bibfield{author}{\bibinfo{person}{Shichao Wang}, \bibinfo{person}{Yun Zhang},
  \bibinfo{person}{Liagfeng Bao}, \bibinfo{person}{Xin Xia}, {and}
  \bibinfo{person}{Minghui Wu}.} \bibinfo{year}{2022}\natexlab{}.
\newblock \showarticletitle{Vcmatch: a ranking-based approach for automatic
  security patches localization for OSS vulnerabilities}. In
  \bibinfo{booktitle}{\emph{2022 IEEE International Conference on Software
  Analysis, Evolution and Reengineering (SANER)}}. IEEE,
  \bibinfo{pages}{589--600}.
\newblock


\bibitem[Wang et~al\mbox{.}(2021)]%
        {wang2021patchrnn}
\bibfield{author}{\bibinfo{person}{Xinda Wang}, \bibinfo{person}{Shu Wang},
  \bibinfo{person}{Pengbin Feng}, \bibinfo{person}{Kun Sun},
  \bibinfo{person}{Sushil Jajodia}, \bibinfo{person}{Sanae Benchaaboun}, {and}
  \bibinfo{person}{Frank Geck}.} \bibinfo{year}{2021}\natexlab{}.
\newblock \showarticletitle{Patchrnn: A deep learning-based system for security
  patch identification}. In \bibinfo{booktitle}{\emph{MILCOM 2021-2021 IEEE
  Military Communications Conference (MILCOM)}}. IEEE,
  \bibinfo{pages}{595--600}.
\newblock


\bibitem[Wicks(2017)]%
        {wicks2017coding}
\bibfield{author}{\bibinfo{person}{David Wicks}.}
  \bibinfo{year}{2017}\natexlab{}.
\newblock \showarticletitle{The coding manual for qualitative researchers}.
\newblock \bibinfo{journal}{\emph{Qualitative research in organizations and
  management: an international journal}} \bibinfo{volume}{12},
  \bibinfo{number}{2} (\bibinfo{year}{2017}), \bibinfo{pages}{169--170}.
\newblock


\bibitem[Yoshimura et~al\mbox{.}(2023)]%
        {yoshimura2023semi}
\bibfield{author}{\bibinfo{person}{Yuki Yoshimura}, \bibinfo{person}{Shun
  Shiramatsu}, {and} \bibinfo{person}{Takeshi Mizumoto}.}
  \bibinfo{year}{2023}\natexlab{}.
\newblock \showarticletitle{Semi-automatic Summarization of Spoken Discourse
  for Recording Ideas using GPT-3}.
\newblock \bibinfo{journal}{\emph{IIAI Letters on Informatics and
  Interdisciplinary Research}}  \bibinfo{volume}{3} (\bibinfo{year}{2023}).
\newblock


\bibitem[Zhang et~al\mbox{.}(2024)]%
        {zhang2024prompt}
\bibfield{author}{\bibinfo{person}{Chenyuan Zhang}, \bibinfo{person}{Hao Liu},
  \bibinfo{person}{Jiutian Zeng}, \bibinfo{person}{Kejing Yang},
  \bibinfo{person}{Yuhong Li}, {and} \bibinfo{person}{Hui Li}.}
  \bibinfo{year}{2024}\natexlab{}.
\newblock \showarticletitle{Prompt-enhanced software vulnerability detection
  using chatgpt}. In \bibinfo{booktitle}{\emph{Proceedings of the 2024 IEEE/ACM
  46th International Conference on Software Engineering: Companion
  Proceedings}}. \bibinfo{pages}{276--277}.
\newblock


\bibitem[Zhou et~al\mbox{.}(2021)]%
        {zhou2021finding}
\bibfield{author}{\bibinfo{person}{Jiayuan Zhou}, \bibinfo{person}{Michael
  Pacheco}, \bibinfo{person}{Zhiyuan Wan}, \bibinfo{person}{Xin Xia},
  \bibinfo{person}{David Lo}, \bibinfo{person}{Yuan Wang}, {and}
  \bibinfo{person}{Ahmed~E Hassan}.} \bibinfo{year}{2021}\natexlab{}.
\newblock \showarticletitle{Finding a needle in a haystack: Automated mining of
  silent vulnerability fixes}. In \bibinfo{booktitle}{\emph{2021 36th IEEE/ACM
  International Conference on Automated Software Engineering (ASE)}}. IEEE,
  \bibinfo{pages}{705--716}.
\newblock


\bibitem[Zhou et~al\mbox{.}(2019)]%
        {zhou2019devign}
\bibfield{author}{\bibinfo{person}{Yaqin Zhou}, \bibinfo{person}{Shangqing
  Liu}, \bibinfo{person}{Jingkai Siow}, \bibinfo{person}{Xiaoning Du}, {and}
  \bibinfo{person}{Yang Liu}.} \bibinfo{year}{2019}\natexlab{}.
\newblock \showarticletitle{Devign: Effective vulnerability identification by
  learning comprehensive program semantics via graph neural networks}.
\newblock \bibinfo{journal}{\emph{Advances in neural information processing
  systems}}  \bibinfo{volume}{32} (\bibinfo{year}{2019}).
\newblock


\end{thebibliography}

\end{document}